\documentclass[sigconf]{acmart}
\makeatletter
\def\@ACM@checkaffil{% Only warnings
    \if@ACM@instpresent\else
    \ClassWarningNoLine{\@classname}{No institution present for an affiliation}%
    \fi
    \if@ACM@citypresent\else
    \ClassWarningNoLine{\@classname}{No city present for an affiliation}%
    \fi
    \if@ACM@countrypresent\else
        \ClassWarningNoLine{\@classname}{No country present for an affiliation}%
    \fi
}
\makeatother
\AtBeginDocument{%
  \providecommand\BibTeX{{%
    \normalfont B\kern-0.5em{\scshape i\kern-0.25em b}\kern-0.8em\TeX}}}

\setcopyright{acmcopyright}
\copyrightyear{2023}
\acmYear{2023}
\acmConference[SIGMOD'24]{ACM SIGMOD International Conference on Management of Data}{June 11--16,
  2024}{Santiago, Chile}

\acmPrice{15.00}
\acmISBN{978-1-4503-XXXX-X/18/06}

\usepackage{diagbox} 
\usepackage{multirow}
\usepackage{adjustbox}
\usepackage[noabbrev,capitalize,nameinlink]{cleveref}
\usepackage{xspace}
\usepackage{url}
\usepackage{graphicx}
\usepackage{subcaption}
\usepackage{enumitem}
\usepackage{geometry}
\usepackage[noend]{algpseudocode} 
\usepackage{algorithm}
\usepackage{makecell}
\usepackage[inline]{aplcomments}

\newcommand{\integernum}{twelve\xspace}

\newcommand{\leco}{LeCo\xspace}

\newcommand{\eliasfano}{Elias-Fano\xspace}
\newcommand{\FORfix}{FOR\xspace}
\newcommand{\deltafix}{Delta-fix\xspace}
\newcommand{\deltavar}{Delta-var\xspace}
\newcommand{\lecofix}{LeCo-fix\xspace}
\newcommand{\lecovar}{LeCo-var\xspace}
\newcommand{\lecopolyfix}{LeCo-Poly-fix\xspace}
\newcommand{\lecopolyvar}{LeCo-Poly-var\xspace}

\newcommand{\lecoPLA}{LeCo-PLA\xspace}
\newcommand{\lecolavec}{LeCo-la-vec\xspace}

\newcommand{\fsst}{FSST\xspace}

\newcommand{\rocksdb}{RocksDB\xspace}
\newcommand{\parquet}{Parquet\xspace}
\newcommand{\arrow}{Arrow\xspace}

\newcommand{\mycaption}[2]{\caption[#1]{\textbf{#1}\textmd{ -- #2} }}

\usepackage{listings}
\usepackage{xcolor}
\definecolor{dkgreen}{rgb}{0,0.6,0}
\definecolor{gray}{rgb}{0.5,0.5,0.5}
\definecolor{mauve}{rgb}{0.58,0,0.82}
\definecolor{mygray}{RGB}{245,245,245}
\begin{document}

\title{\leco: Lightweight Compression via Learning Serial Correlations}


\vspace{0.5cm}
\author{
    Yihao Liu
}
\affiliation{%
  \institution{Tsinghua University}
}
\email{liuyihao21@mails.tsinghua.edu.cn}

\author{
    Xinyu Zeng
}
\affiliation{%
  \institution{Tsinghua University}
}
\email{zeng-xy21@mails.tsinghua.edu.cn}

\author{
    Huanchen Zhang
}
\affiliation{%
  \institution{Tsinghua University}
}
\email{huanchen@tsinghua.edu.cn}

\renewcommand{\shortauthors}{Liu et al.}




% \keywords{compression}


\vspace{0.5cm}
\begin{abstract}
\label{sec:abstract}
Lightweight data compression is a key technique that allows column stores to exhibit superior performance for analytical queries.
Despite a comprehensive study on dictionary-based encodings to approach
Shannon’s entropy, few prior works have systematically exploited the
serial correlation in a column for compression.
In this paper, we propose \leco (i.e., Learned Compression),
a framework that uses machine learning to remove the serial redundancy
in a value sequence automatically to achieve an outstanding compression
ratio and decompression performance simultaneously.
\leco presents a general approach to this end, making existing (ad-hoc)
algorithms such as Frame-of-Reference (FOR), Delta Encoding, and
Run-Length Encoding (RLE) special cases under our framework.
Our microbenchmark with three synthetic and eight real-world data sets
shows that a prototype of \leco achieves a Pareto improvement on
both compression ratio and random access speed over the existing solutions.
When integrating \leco into widely-used applications, we observe 
up to $5.2\times$ speed up in a data analytical query in the Arrow columnar execution engine,
and a $16\%$ increase in \rocksdb’s throughput.
\end{abstract}
\maketitle

\setcounter{section}{0}
\section{Introduction}
\label{sec:intro}

Almost all major database vendors today have adopted a column-oriented design 
for processing analytical queries
\cite{lahiri2015, larson2015real, dageville2016, gupta2015, armenatzoglou2022, raman2013db2, farber2012, lamb2012vertica}.
One of the key benefits of storing values of the same attribute consecutively 
is that the system can apply a variety of lightweight compression algorithms to 
the columns to save space and disk/network bandwidth~\cite{abadi2006, abadi2013design, welton2011}.
These algorithms, such as Run-Length Encoding (RLE)~\cite{abadi2006} and Dictionary Encoding, 
typically involve a single-pass decompression process (hence, lightweight)
to minimize the CPU overhead.
A few of them (e.g., Frame-of-Reference or FOR~\cite{goldstein1998, zukowski2006super})
allow random access to the individual values.
This is a much-preferred feature because it allows the DBMS to avoid full-block
decompression for highly selective queries, which are increasingly common,
especially in hybrid transactional/analytical processing (HTAP)
\cite{pezzini2014hybrid,kemper2011hyper, plattner2009common, lee2017parallel, huang2020tidb, singlestore}
and real-time analytics~\cite{larson2015real, heatwave2021}.

There are two categories of lightweight compression algorithms that
exploit different sources of redundancy in a value sequence.
The first are dictionary-based algorithms, including those that encode substring
patterns (e.g., FSST~\cite{boncz2020fsst}, HOPE~\cite{zhang2020order}).
These algorithms leverage the uneven probability distribution of the values
and have a compression ratio limited by Shannon’s Entropy~\cite{shannon1948}.
On the other hand, integer compression algorithms such as Run-Length Encoding (RLE)~\cite{abadi2006},
FOR, and Delta Encoding~\cite{abadi2006,lemire2015decoding}
exploit the serial correlation between the values in a sequence:
the value of the current position may depend on its preceding values.

\begin{figure}[t!]
\centering
\begin{minipage}[t]{0.22\textwidth}
\centering
\includegraphics[width=\columnwidth]{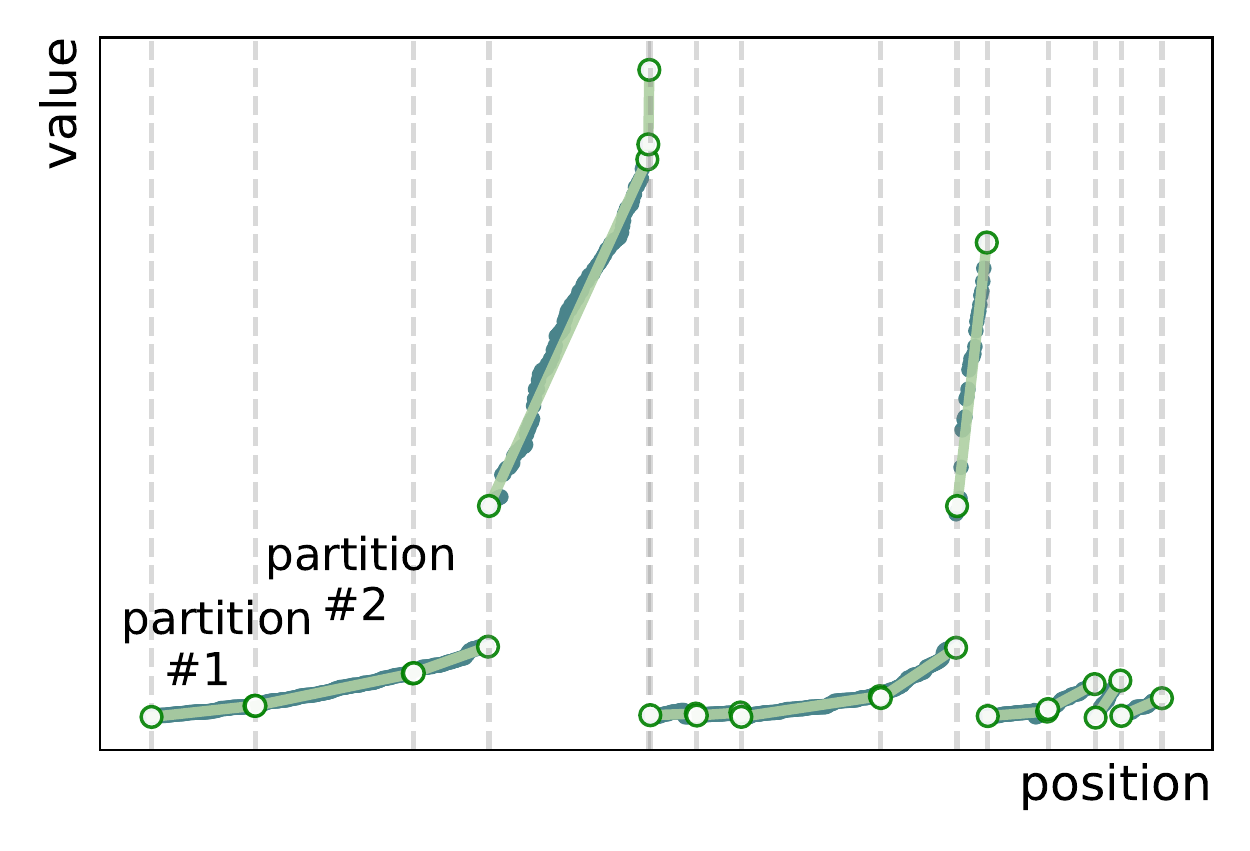}
\mycaption{A Motivating Example.}{On \texttt{movieid} data set.}
\label{fig:movieid}
\end{minipage}
\hspace{0.3cm}
\begin{minipage}[t]{0.23\textwidth}
\centering
\includegraphics[width=\columnwidth]{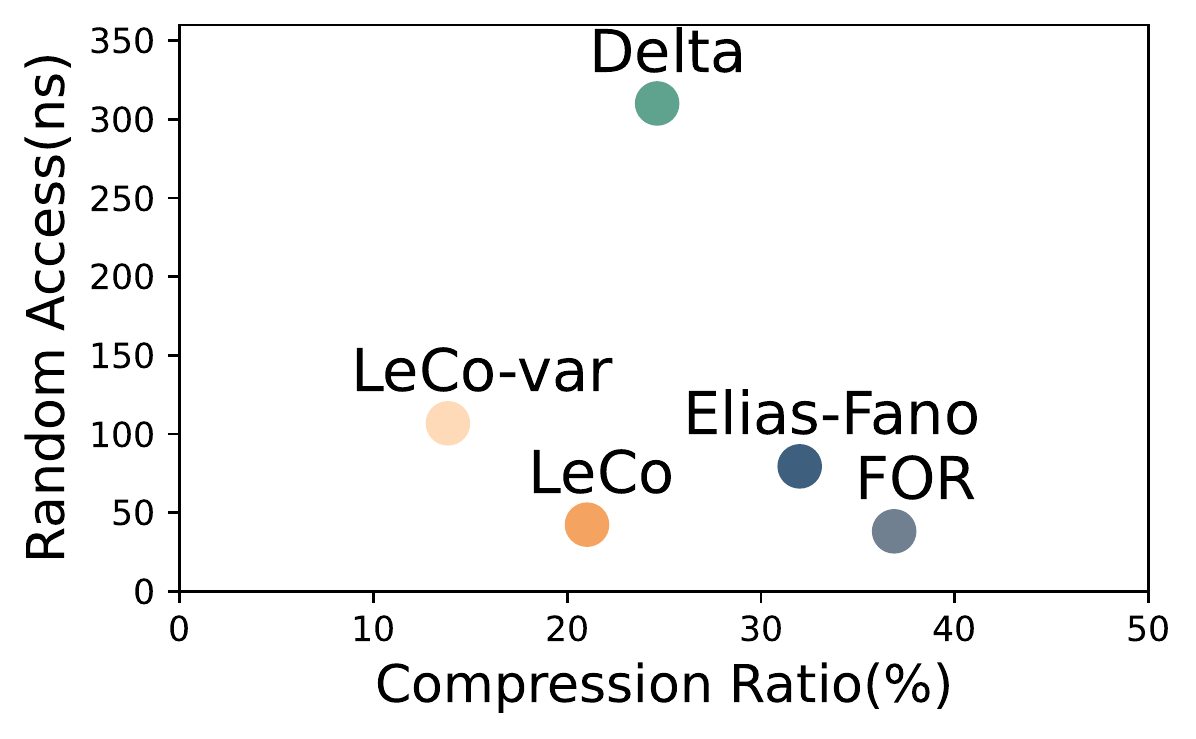}
\caption{Performance-space trade-offs.}
\label{fig:tradeoff}
\end{minipage}
\vspace{-0.4cm}
\end{figure}

However, RLE, FOR, and Delta Encoding are ad-hoc solutions modeling
the simplest serial patterns. For example, Delta adopts a model of a basic step function,
while RLE only works with consecutive repetitions (elaborated in \cref{sec:motivation}).
Consequently, we have missed many opportunities to leverage more sophisticated patterns
such as the piecewise linearity shown in~\cref{fig:movieid} for better compression
in a column store.
Prior studies in time-series data storage~\cite{hung2012evaluation, kitsiossim, luo2015piecewise, elmeleegy2009online, 2015AEichinger, Xie2014Maximum}
have proposed to learn the series distribution and minimize the model sizes to achieve
a \emph{lossy} compression. 
These techniques, however, are not applicable to a general analytical system.
To the best of our knowledge,
none of the existing column stores apply machine learning to improve
the efficiency of their lightweight \emph{lossless} compression systematically.

We, thus, propose a framework called \leco (i.e., Learned Compression) to automatically
learn serial patterns from a sequence and use the models for compression.
Our key insight is that if we can fit such serial patterns with lightweight
machine-learning models, we only need to store the prediction error for each value
to achieve a lossless compression.
Our framework addresses two subproblems.
The first is that given a subsequence of values, how to best fit the data using one model?
This is a classic regression problem.
However, instead of minimizing the sum of the squared errors,
we minimize the maximum error
because we store the deltas (i.e., prediction errors) in a fixed-length array
to support fast random access during query processing.
\leco also includes a Hyperparameter-Advisor to select the regressor
type (e.g., linear vs. higher-order) that would produce the best compression ratios.

The second subproblem is data partitioning: given the type(s) of the regression model,
how to partition the sequence to minimize the overall compression ratio? 
Proactive partitioning is critical to achieving high-prediction accuracy in the regression
tasks above because real-world data sets typically have uneven distributions
\cite{2019SOSD, CARMI}.
The partition schemes introduced by \textit{lossy} time-series compression are not efficient to apply.
They only target minimizing the total size of the model parameters
rather than striking a balance between the model size and the delta array size.
Our evaluation (\cref{sec:partition_compare}) shows that the state-of-the-art
partitioning algorithms~\cite{cameron1966piece, kitsiossim}
are still suboptimal for general \textit{lossless} column compression.

In the \textit{lossless} case, however, 
having smaller partitions might be beneficial for reducing the local max errors,
but it increases the overall model (and metadata) size.
Because optimal partitioning is an NP-hard problem, we developed different heuristic-based algorithms
for different regression models to obtain approximate solutions in a reasonable amount of time.
Another design trade-off is between fixed-length and variable-length partitions.
Variable-length partitions produce a higher compression ratio but are slower in random access.

We implemented a prototype of \leco to show the benefit of using machine learning
to compress columnar data losslessly.
For each partition, we store a pre-trained regression model along with an array of
fixed-length deltas.
Decompressing a value only involves a model inference plus a random access to the delta array.
\leco is highly extensible with built-in support for various model types
and for both fixed-length and variable-length partition schemes.

We compared \leco against state-of-the-art lightweight compression algorithms
including \FORfix, \eliasfano, and Delta Encoding using a microbenchmark consisting of
both synthetic and real-world data sets.
As illustrated in \cref{fig:tradeoff}\footnote{\cref{fig:tradeoff} is based on the weighted average result of twelve data sets in \cref{sec:intbench}.},
\leco achieves a Pareto improvement over these algorithms.
Compared to \FORfix and \eliasfano, \leco improves the compression ratio by up to $91\%$
while retaining a comparable decompression and random access performance.
Compared to Delta Encoding, \leco is an order-of-magnitude faster in random access
with a competitive or better compression ratio.

We further integrated \leco into two widely-used applications to study its benefit
on end-to-end system performance.
We first report \leco's performance on a columnar execution engine, using Apache Arrow~\cite{arrow} and \parquet~\cite{parquet} as the building blocks.
Enabling \leco in this system speeds up a multi-column filter-groupby-aggregation query by up to $5.2\times$ and accelerates single-column bitmap aggregation query up to $11.8\times$ with a $60.5\%$ reduction in memory footprint. 
We also use \leco to compress the index blocks in \rocksdb~\cite{dong2021rocksdb,Rocksdgithub}
and observed a $16\%$ improvement in \rocksdb's throughput compared to its default configuration.

The paper makes four primary contributions.
First, we identify that exploiting the serial correlation between values has a great potential
for efficient column compression.
Second, we make the case for applying machine learning to lightweight lossless
column compression.
Third, we propose the Learned Compression (\leco) framework and implement a prototype
that achieves a Pareto improvement on compression ratio and random access speed over
existing algorithms.
Finally, we integrate \leco into a columnar execution engine and a key-value store
and show that it helps improve the systems' performance and space efficiency simultaneously.

\section{The case for Learned Compression}
\label{sec:motivation}

The performance of persistent storage devices has improved by orders of magnitude
over the last decade~\cite{xu2015performance}. 
Modern NVMe SSDs can achieve ~7GB/s read throughput and over 500,000 IOPS~\cite{samsungssd}. 
The speed of processors, on the other hand, remains stagnant as Moore’s Law fades~\cite{flamm2019measuring}. 
Such a hardware trend is gradually shifting the bottleneck of a data processing system
from storage to computation. 
Hence, pursuing a better compression ratio is no longer the dominating goal
when developing a data compression algorithm. 
Many applications today prefer lightweight compression schemes because decompressing
the data is often on the critical path of query execution. 
Meanwhile, an analytical workload today is often mixed with OLTP-like queries featuring small
range scans or even point accesses~\cite{ozcan2017hybrid,personal-comm}.
To handle such a wide range of selectivity, it is attractive for a data warehouse to
adopt compression algorithms that can support fast random access to the original data
without decompressing the entire block.

Dictionary encoding is perhaps the most widely-used compression scheme in database management systems (DBMSs).
Nonetheless, for a sequence where the values are mostly unique, dictionary encoding does not bring compression
because it assumes independence between the values, and its compression ratio is bounded by Shannon's Entropy~\cite{shannon1948}.
Shannon’s Entropy, however, is not the lower bound for compressing an existing sequence\footnote{The lower bound is known as
the Kolmogorov Complexity. It is the length of the shortest program that can produce the original data \cite{li2008introduction}.
Kolmogorov Complexity is incomputable.}.
In many real-world columns, values often exhibit strong serial correlations (e.g., sorted or clustered)
where the value at a particular position is dependent on the values preceding it.
Unfortunately, to the best of our knowledge, there is no general solution proposed
that can systematically leverage such positional redundancy for compression.

We argue that a learned approach is a natural fit. Extracting serial correlation
is essentially a regression task.
Once the regression model captures the ``common pattern'' of the sequence,
we can use fewer bits to represent the remaining delta for each value.
This Model + Delta framework (a.k.a., \leco) is fundamental for exploiting serial
patterns in a sequence to achieve lossless compression.
For example, Boffa et al. attempted to use linear models for storing
rank$\&$select dictionaries specifically~\cite{boffa2021learned}.
In fact, the widely-used FOR, RLE, and Delta Encoding (Delta)
can be considered special cases under our framework as well.

FOR divides an integer sequence into frames, and for each value $v_i$ in a frame,
it is encoded as $v_i - v_{min}$ where $v_{min}$ is the minimum value of that frame.
From a \leco’s point of view, the regression function for each frame in FOR is
a horizontal line.
Although such a naive model is fast to train and inference, it is usually
suboptimal in terms of compression ratio.
RLE can be considered a special case of FOR, where the values in a frame must be identical.
Delta Encoding achieves compression by only storing the difference between neighboring values.
Specifically, for an integer sequence $v_1, v_2, ..., v_n$, Delta encodes the values as
$v_1, v_2 - v_1, v_3 - v_2, v_n - v_{n-1}$.
Similar to FOR, it uses the horizontal-line function as the model, but each partition/frame
in Delta only contains one item.
The advantage of Delta is that the models can be derived from recovering the previous values
rather than stored explicitly. 
The downside, however, is that accessing any particular value requires a sequential
decompression of the entire sequence.

\leco helps bridge the gap between data compression and data mining.
Discovering and extracting patterns are classic data mining tasks.
Interestingly, these tasks often benefit from preprocessing the data set with entropy compression
tools to reduce “noise” for a more accurate prediction~\cite{taylor2018data}.
As discussed above, these data mining algorithms can inversely boost compression efficiency
by extracting the serial patterns through the \leco framework.
The theoretical foundation of this relationship is previously discussed in~\cite{faloutsos2007data}.
Notice that although we focus on regression in this paper, other data mining techniques, such as
anomaly detection, also reveal serial patterns that can improve compression efficiency~\cite{agrawal2015survey, boniol2021sand}.
The beauty of \leco is that it aligns the goal of sequence compression with that of serial
pattern extraction.
\leco is an extensible framework: it provides a convenient channel to bring related advances
in data mining to the improvement of sequence compression.

Although designed to solve different problems, \leco is related to the recent 
learned indexes~\cite{kraska2018case, ALEX, ferragina2020pgm}
in that they both use machine learning (e.g., regression) to model data distributions.
A learned index tries to fit the cumulative distribution function (CDF) of a sequence
and uses that to predict the quantile (i.e., position) of an input value.
Inversely, \leco takes the position in the sequence as input and tries to predict
the actual value.
\leco's approach is consistent with the mapping direction (i.e., position $\rightarrow$ value)
in classic pattern recognition tasks in data mining.

Moreover, \leco mainly targets immutable columnar formats
such as Arrow~\cite{arrow} and Parquet~\cite{parquet}.
Updating the content requires a complete reconstruction of the files
on which \leco can piggyback its model retraining.
Unlike indexes where incremental updates are the norm,
the retraining overhead introduced by \leco is amortized
because the files in an analytical system typically follow
the pattern of ``compress once and access many times''.

We next present the \leco framework in detail, followed by an extensive microbenchmark
evaluation in \cref{sec:eval}.
We then integrate \leco into two real-world applications and demonstrate their
end-to-end performance in \cref{sec:sys-eval}.

\section{The \leco Framework}
\label{sec:algorithm}
Let us first define the learned compression problem that the \leco framework targets.
Given a data sequence $\vec{v}_{[0, n)} = (v_0,...,v_{n-1})$,
let $P_0 = \vec{v}_{[k_0 = 0, k_1)}, P_1 = \vec{v}_{[k_1, k_2)}, ..., P_{m-1} = \vec{v}_{[k_{m-1}, k_m = n)}$
be a partition assignment $\mathcal{P}$ with $m$ non-overlap segments
where each partition $j$ has a model $\mathcal{F}_j$.
Let $\delta_i = v_i - \mathcal{F}_j(i)$, where $\mathcal{F}_j(i)$ is the model prediction at position $i$, for $v_i \in P_j$.
The goal of learned compression is to find a partition assignment $\mathcal{P}$
and the associated models $\mathcal{F}$
such that the model size plus the delta-array size are minimized:
    $$ \sum_{j=0}^{m-1}(\|\mathcal{F}_j\| + (k_{j+1} - k_j)(\max_{i=k_j}^{k_{j+1}-1}\lceil \log_2\delta_i \rceil))$$
where $\|\mathcal{F}_j\|$ denotes the model size of $\mathcal{F}_j$,
and $\max \lceil \log_2\delta_i \rceil$ is the number of bits required to represent the largest $\delta_i$ in the partition.

\begin{figure}[t!]
\centering
\includegraphics[scale = 0.14]{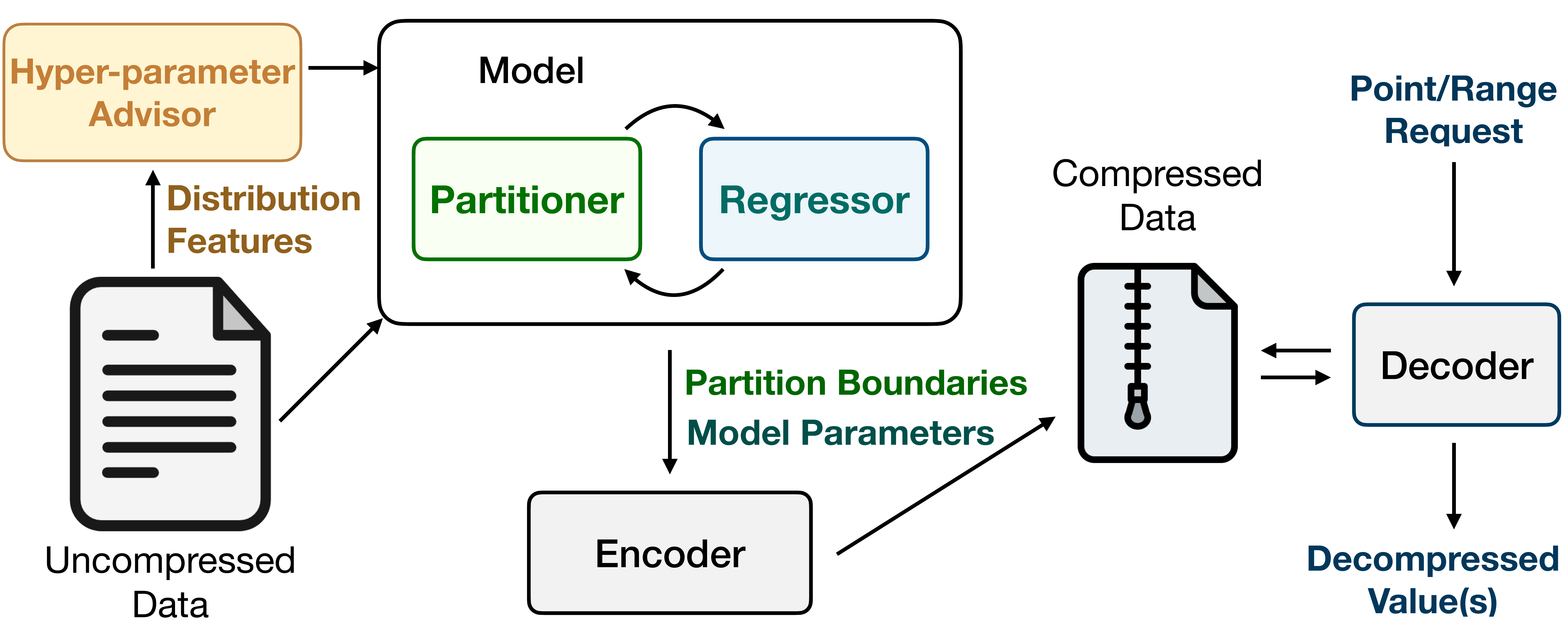}
\mycaption{The \leco Framework}{An overview of the modules and their interactions with each other.}
\vspace{-0.5cm}
\label{fig:leco_overview}
\end{figure}

As shown in \cref{fig:leco_overview}, \leco consists of five modules: 
\texttt{Regressor}, \texttt{Partitioner}, \texttt{Hyperparameter-Advisor}, \texttt{Encoder}, and \texttt{Decoder}. 
The Hyper-parameter Advisor trains a Regressor Selector model offline. 
Given an uncompressed sequence of values at runtime, it extracts features from it for model inference and outputs the recommended Regressor type as well as advises on partitioning strategy.
Then, \leco enters the model learning
phase, where the Regressor and the Partitioner work together to produce a set of
regression models with associated partition boundaries.
The Encoder receives the model parameters as well as the original sequence
and then generates a compact representation of the ``Model + Delta''
(i.e., the compressed sequence) based on a pre-configured format.
The compressed sequence is self-explanatory: all the metadata needed for decoding is embedded in the format.
When a user issues a query by sending one or a range of positions,
the Decoder reads the model of the relevant partition along with
the corresponding locations in the delta array to recover the
requested values.

A design goal of \leco is to make the framework extensible.
We first decouple model learning (i.e., the logical value encoding) from the
physical storage layout because applying common storage-level optimizations
such as bit-packing and null-suppression to a delta sequence is orthogonal to
the modeling algorithms.
We also divide the model learning task into two separate modules.
The Regressor focuses on best fitting the data in a single partition,
while a Partitioner determines how to split the data set into subsequences
to achieve a desirable performance and compression ratio.

Such a modular design facilitates integrating future advances in
serial pattern detection and compressed storage format into \leco.
It also allows us to reason the performance-space trade-off for each component
independently.
We next describe our prototype and the design decisions made for each module
(\cref{sec:regressor} to \cref{sec:enc_dec}),
followed by the extension to handling string data in \cref{sec:string}.

%---------------------------------------------------------------------------------
%---------------------------------------------------------------------------------
% Regressor
%---------------------------------------------------------------------------------
%---------------------------------------------------------------------------------

\subsection{Regressor}
\label{sec:regressor}
The Regressor takes in a sequence of values $v_0, v_1, ..., v_{n-1}$
and outputs a single model that ``best fits'' the sequence.
\leco supports the linear combination of various model types, including constant, linear, polynomial, and more sophisticated models, such as exponential and logarithm.
Given a model $\mathcal{F}(i) = \sum_j (\theta_j \cdot \mathcal{M}_j(i))$ where $\mathcal{M}_j$ denotes different model terms with $\theta_j$ as its linear combination weight and $i$ represents the position in the sequence,  
classic regression methods minimize the sum of the squared errors
$\sum_i (v_i - \mathcal{F}(i))^2$ (i.e., the $l_2$ norm of deltas),
which has a closed-form solution.
If \leco stores deltas in variable lengths, this solution would produce
a delta sequence with minimal size.
As we discussed before, real databases usually avoid
variable-length values because of the parsing overhead during query execution.

\leco, therefore, stores each value in the delta array in fixed length.
Specifically, \leco adopts the bit-packing technique.
Suppose the maximum absolute value in the delta array is $\delta_{maxabs}$, then
each delta occupies a fixed $\phi = \lceil log_2(\delta_{maxabs}) \rceil$ bits.
The storage size of the delta array is thus determined by $\phi$
rather than the expected value of the deltas, and our regression objective
becomes:
\begin{equation*}
\begin{array}{ll@{}ll}
\text{minimize}  & \displaystyle\phi &\\
\text{subject to}& \displaystyle \lceil \log_2 (\lvert \mathcal{F}(i) - \mathbf{v}_i \rvert) \rceil \leq \phi,  & i=0 ,\dots, n-1\\
&\phi \geq 0 \\
\end{array}
\end{equation*}
The constrained optimization problem above can be transformed into a linear programming problem with $2n+1$ constraints where we can get an approximated optimal solution in $O(n)$ time~\cite{seidel1991small}.

We introduce a Regressor Selector (RS) in the Hyperparameter-Advisor to automatically
choose the regressor type (e.g., linear vs. higher-order) for a given sequence partition.
RS takes in features collected from a single pass of the input data and then feeds them to
its classification model (e.g., Classification and Regression Tree or CART).
The model is trained offline using the same features from the training data sets.
We briefly introduce the main features used in the current RS implementation below.

%---------------------------------------------------------------------------------
%---------------------------------------------------------------------------------
% Regressor Selection
%---------------------------------------------------------------------------------
%---------------------------------------------------------------------------------

\texttt{Log-scale data range.} Data range gives an upper bound of the size of the delta array.
A smaller data range prefers simpler models because the model parameters would take a significant portion of the compressed output.

\texttt{Deviation of the $k$th-order deltas.}
Given a data sequence $v_0,...,v_{n-1}$,
we define the first-order delta sequence as $d^0_1 = v_1 - v_0, d^1_1 = v_2 - v_1, ..., d^1_{n-2} = v_{n-1} - v_{n-2}$.
Then, the $k$th-order delta sequence is $\{d_0^k, d_1^k, ..., d_{n-k-1}^k \}$, where $d^k_{i-1} = d^{k-1}_i - d^{k-1}_{i-1}$.
Let $d_{max}^k$, $d_{min}^k$, and $d_{avg}^k$ be the maximum, minimum, and average delta values, respectively.
We then compute the normalized deviation of the $k$th-order deltas as
$\frac{\sum_{i \in [0, n-k)} (d_i^k - d_{avg}^k)}{(n - k)(d_{max}^k - d_{mix}^k)}$.
We use this metric to determine the maximum degree of polynomial needed to fit the data. 
The intuition is that the $k$th-order delta sequence of a $k$th-degree polynomial is constant
(i.e., with minimum deviation).

\texttt{Subrange trend and divergence.}
We first split the data into fixed-length subblocks $\{ \vec{v}_{[i\cdot s, (i+1)\cdot s)} \}_i$,
each containing $s$ records with a data range (i.e., subrange) of $r_i$.
We define the subrange ratio (SR) between adjacent subblocks as $\frac{r_i}{r_{i-1}}$.
The metric ``subrange trend'' $\mathcal{T}$ is the average SR across all subblocks,
while ``subrange divergence'' $\mathcal{D}$ is the difference between the maximum SR and minimum SR.
These two metrics provide a rough sketch of the value-sequence distribution:
$\mathcal{T}$ depicts how fast the values increase on average,
and $\mathcal{D}$ indicates how stable the increasing-trend is.

%---------------------------------------------------------------------------------
%---------------------------------------------------------------------------------
% Partitioner
%---------------------------------------------------------------------------------
%---------------------------------------------------------------------------------

\subsection{Partitioner}
\label{sec:partitioner}
\sloppy Given a Regressor, the Partitioner divides the input sequence
$\vec{v}_{[0, n)} = v_0, v_1, ..., v_{n-1}$ into $m$ consecutive subsequences
(i.e., partitions)
${\vec{v}_{[0, k_1)}, \vec{v}_{[k_1, k_2)}, ..., \vec{v}_{[k_{m-1}, k_m)}}$ 
where a regression model is trained on each partition.
The goal of the Partitioner is to minimize the overall size of the compressed sequences.

Although partitioning increases the number of models to store, it is more likely for
the Regressor to produce a smaller delta array when fitting a shorter subsequence. 
Thus, we require the Partitioner to balance between the model storage overhead and the general model fitting quality.
We can find an optimal partition arrangement by computing the compressed size of each
possible subsequence through dynamic programming~\cite{silvestri2010vsencoding}.
Such an exhaustive search, however, is forbiddingly expensive with time complexity of
$O(n^3)$ and space complexity of $O(n^2)$.

We next propose two practical partitioning schemes developed in \leco that make different
trade-offs between compression ratio and compression/decompression performance.

%---------------------------------------------------------
% Fixed-Length Partitioning
%---------------------------------------------------------

\subsubsection{Fixed-Length Partitioning}
\label{sec:fixedlen}

\begin{figure}[t!]
\centering
\begin{minipage}[t]{0.23\textwidth}
\centering
\includegraphics[width=\columnwidth]{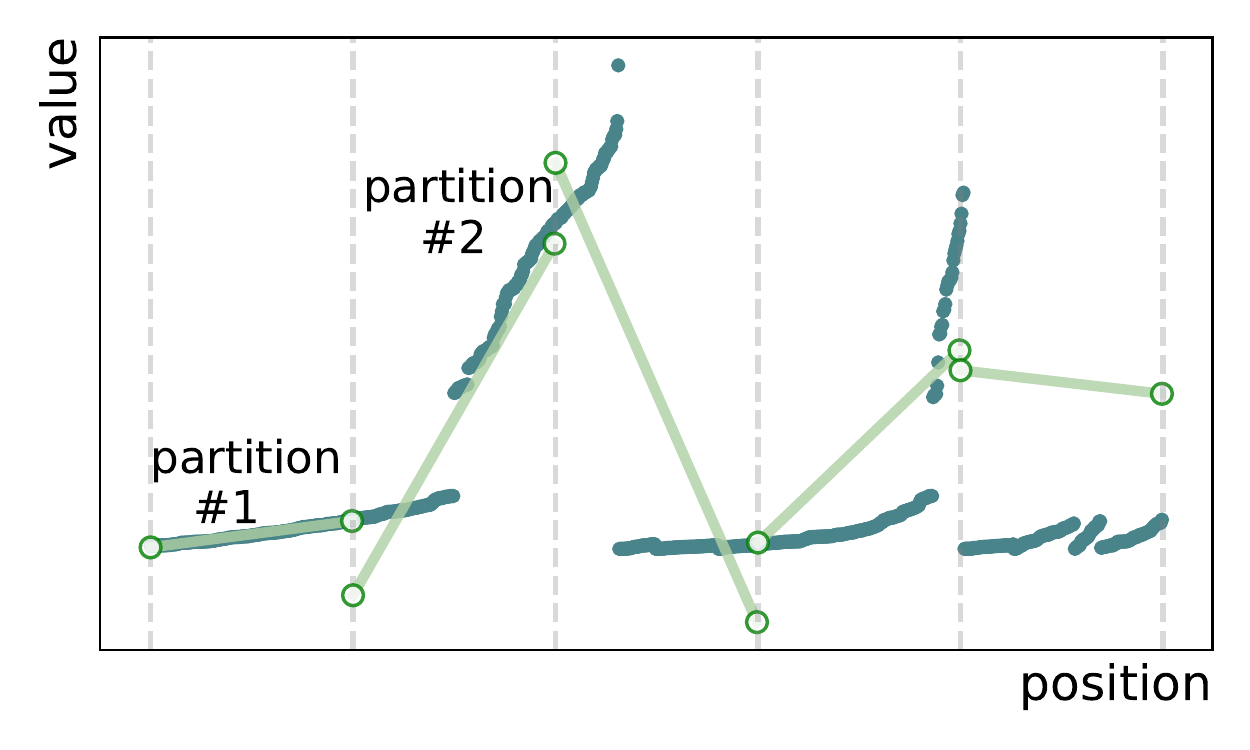}
\caption{Fixed-length Partitioning Example.}
\label{fig:movieid_fix}
\end{minipage}
\hspace{0.3cm}
% \vspace{0.5cm}
\begin{minipage}[t]{0.22\textwidth}
\centering
\includegraphics[width=\columnwidth]{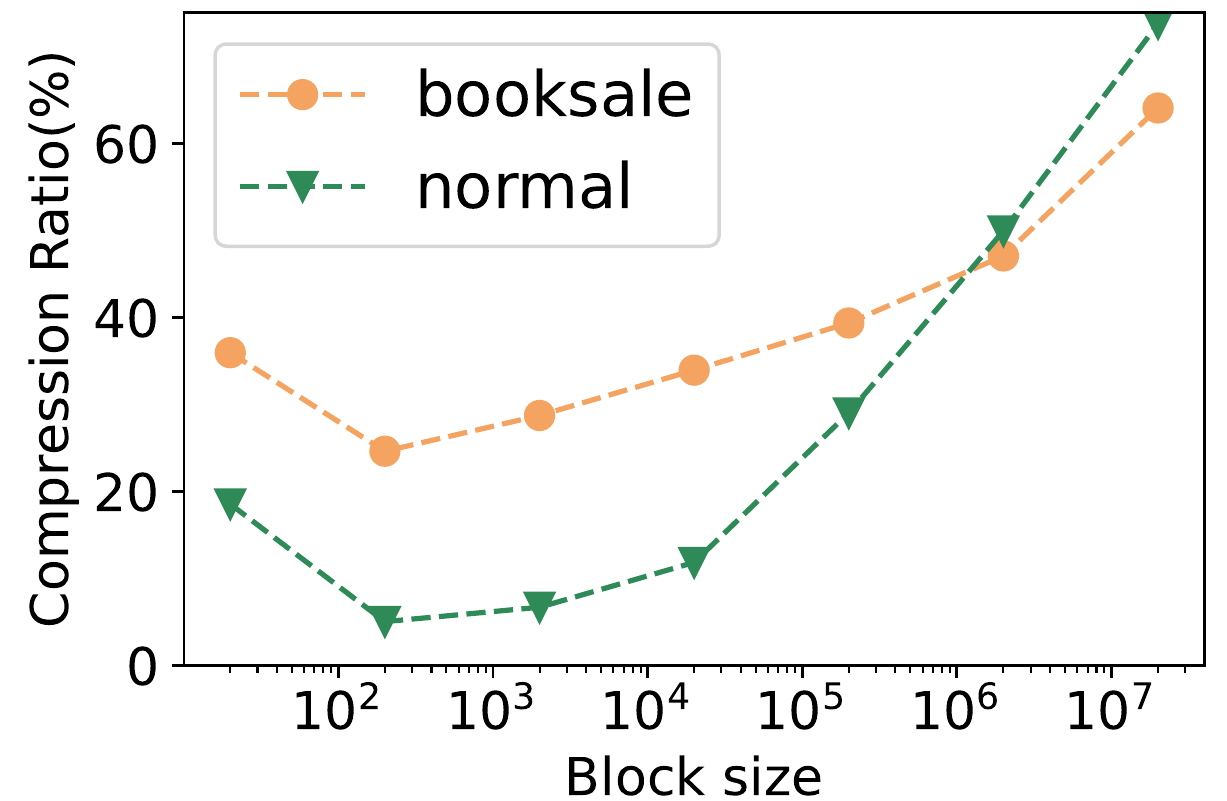}
\mycaption{ompression Ratio Trend.}{Sweeping block size.}
\label{fig:ushape}
\end{minipage}
\vspace{-0.3cm}
\end{figure}

The most common strategy is splitting the sequence into
fixed-length partitions. 
This partitioning scheme is easy to implement and is friendly to random accesses.
Because each partition contains a fixed number of items, given a position,
an application can quickly locate the target partition without the need for
a binary search in the metadata.
The downside, however, is that fixed-length partitioning is not flexible enough
to help the Regressor capture the desired patterns. 
For example, as shown in \cref{fig:movieid_fix}, if we divide the Movie ID data set
into fixed-length partitions,
the Regressor would fail to leverage the piecewise linearity in certain ranges.
To find an optimal partition size:
\begin{enumerate}[leftmargin=10pt]
    \item Sample $<1\%$ of the data randomly, consisting of subsequences of length $N$, where $N$ is the maximum partition length in the search space (e.g., $N = 10k$).
    \item Search the (fixed) partition size between $1$ and $N$ that produces the lowest compression ratio on the samples. Because the compression ratio typically has a ``U-shape'' as we vary the partition size (illustrated in \cref{fig:ushape}), we first perform an exponential search to go past the global minimum.
    Then, we search back with smaller steps to approach the optimal partition size.
    \item Stop the search process once the compression ratio converges (with $<0.01\%$ decline between adjacent iterations).
\end{enumerate}

%---------------------------------------------------------
% Variable-Length Partitioning
%---------------------------------------------------------

\subsubsection{Variable-Length Partitioning}
\label{sec:varlen}
Below, we propose a greedy algorithm for variable-length partitioning
for an arbitrary Regressor discussed in \cref{sec:regressor}
to approximate the optimal solution obtained by the dynamic programming approach.

Our greedy algorithm includes two phases: \textbf{split} and \textbf{merge}.
In the split phase, the algorithm groups consecutive data points into
small partitions 
where the Regressor can predict with small errors.
We impose strict constraints to limit the maximum prediction error
produced by the Regressor for each partition.
Because of our aggressive guarantee of prediction errors, the algorithm
tends to generate an excessive number of partitions in the split phase,
where the cumulative model size could dominate the final compressed size.
To compensate for the over-splitting, the algorithm enters the merge phase
where adjacent partitions are merged if such an action can reduce the
final compressed size.

Specifically, in the \textbf{split} phase, we first pick a few starting partitions.
A starting partition contains at least a minimum number of consecutive values for
the Regressor to function meaningfully (e.g., three for a linear Regressor).
Then, we examine the adjacent data point to determine whether to include this point into the partition.
The intuition is that if the space cost of incorporating this data point is less than a pre-defined threshold, the point is added to the partition;
otherwise, a new partition is created.

The splitting threshold is related to the model size $S_M$ of the Regressor.
Suppose the current partition spans from position $i$ to $j-1$: $\vec{v}_{[i, j)}$.
Let $\Delta(\vec{v})$ be a function that takes in a value sequence and
outputs the number of bits required to represent the maximum absolute
prediction error from the Regressor (i.e., $\lceil log_2(\delta_{maxabs}) \rceil$).
Then, the space cost of adding the next data point $v_j$ is
\[ C = (j + 1 - i) \cdot \Delta(\vec{v}_{[i, j + 1)}) - (j - i) \cdot \Delta(\vec{v}_{[i, j)}) \]
We compare $C$ against $\tau S_M$, where $\tau$ is a pre-defined coefficient
between $0$ and $1$ to reflect the ``aggressiveness'' of the split phase:
a smaller $\tau$ leads to more fine-grained partitions with
more accurate models.
If $C \le \tau S_M$, $v_j$ is included to the current partition $\vec{v}_{[i, j)}$.
Otherwise, we create a new partition with $v_j$ as the first value.

\sloppy In the \textbf{merge} phase, we scan through the list of partitions
${\vec{v}_{[0, k_1)}, \vec{v}_{[k_1, k_2)}, ..., \vec{v}_{[k_{m-1}, k_m)}}$
produced in the split phase and merge the adjacent ones if the size of the
merged partition is smaller than the total size of the individual ones.
Suppose the algorithm proceeds at partition $\vec{v}_{[k_{i-1}, k_i)}$.
At each step, we try to merge the partition to its right neighbor
$\vec{v}_{[k_i, k_{i+1})}$.
We run the Regressor on the merged partition $\vec{v}_{[k_{i-1}, k_{i+1})}$
and compare its size
$S_M + (k_{i+1} - k_{i-1}) \cdot \Delta(\vec{v}_{[k_{i-1}, k_{i+1})})$
to the combined size of the original partitions
$2S_M + (k_i - k_{i-1}) \cdot \Delta(\vec{v}_{[k_{i-1}, k_i)}) + (k_{i+1} - k_i) \cdot \Delta(\vec{v}_{[k_i, k_{i+1})})$.
We accept this merge if it results in a size reduction.
We iterate the partition list multiple times until no qualified merge exists.

We summarize our vari-length partitioning algorithm as follows:
\begin{enumerate}[leftmargin=0pt]
    \item[ ] \textbf{[Init Phase]} Scan all data point once. Pick a few ``good'' initial positions to form the starting partitions. 
    \item[ ] \textbf{[Split Phase]} Scan the starting partition set once.
    \begin{itemize}
        \item Try ``growing'' each starting partition by adding adjacent points.
        \item Calculate the inclusion cost and approve the inclusion if it is below the predefined threshold related to the model size. Otherwise, start a new partition with a single point.
        \item Stops after each point belongs to a partition.
    \end{itemize}
    \item[ ] \textbf{[Merge Phase]} Scan the partition sets multiple times. 
    \begin{itemize}
        \item Merge a partition to its right neighbor if the combined one achieves a lower compression ratio.
        \item Stops when no merge can reduce the total space.
    \end{itemize}
\end{enumerate}

We next discuss two critical aspects that largely determine the efficiency of
the above split-merge algorithm.

\textbf{Computing $\Delta(\vec{v}_{[i, j)})$ Efficiently.}
The computational complexity of $\Delta(\vec{v}_{[i, j)})$ dominates the overall algorithm complexity
because the function is invoked at every data point inclusion in the split phase.
For a general $k$-degree \emph{polynomial model} $\sum_{i\in[0,k]} \theta_i \cdot x^i$,
we can use the method introduced in~\cite{seidel1991small} to compute $\Delta(\vec{v}_{[i, j)})$ in linear time.
To further speed up the process for the \emph{linear Regressor} (which is most commonly used),
we propose a much simpler metric
$\widetilde{\Delta}(\vec{v}_{[i, j)})= \log_2 (\max_{k=i+1}^{j-1}(d_k) - \min_{k=i+1}^{j-1}(d_k))$,
where $d_k = v_k - v_{k-1}$
to approximate the functionality of $\Delta(\vec{v}_{[i, j)})$
The intuition is that the proposed metric $\widetilde{\Delta}(\vec{v}_{[i, j)})$
indicates the difficulty of the linear regression task
and has a positive correlation to max bit-width measure $\Delta(\vec{v}_{[i, j)})$.

As discussed in \cref{sec:motivation}, Delta Encoding is considered a specific design point under the \leco framework.
The model in each Delta partition is an implicit step function,
and only the first value in the partition is explicitly stored as the model.
The prediction errors (i.e., the $\delta's$) of Delta Encoding are the differences between each
pair of the adjacent values.
Therefore, $\Delta(\vec{v}_{[i, j)}) = \lceil \log_2(\max_{k=i+1}^{j-1}d_k) \rceil$, where $d_k = v_k - v_{k-1}$.
After adding the next data point $v_j$ to this partition,
we can directly compute $\Delta(\vec{v}_{[0, j+1)}) = \max{\{\Delta(\vec{v}_{[0, j)}), d_j\}}$.

\textbf{Selecting Good Starting Positions.}
Because the algorithms used in both the split and merge phases are greedy,
the quality of the algorithms' starting partitions can significantly impact the partition results, especially for the split phase.
Suppose we start at a ``bumpy'' region $\vec{v}_{[i, j)}$ during splitting.
Because $\Delta(\vec{v}_{[i, j)})$ of this partition is already large,
there is a high probability that it stays the same when including an extra data point in the partition
(i.e., $\Delta(\vec{v}_{[i, j+1)}) = \Delta(\vec{v}_{[i, j)})$).
Therefore, the space cost of adding this point becomes a constant $C = \Delta(\vec{v}_{[i, j)})$.
As long as $C \le \tau S_M$, this ``bad'' partition
would keep absorbing data points, which is destructive to the overall compression.

For a general \emph{polynomial model} of degree $k$, we select segments where the
$(k+1)$th-order deltas (refer to the definition in \cref{sec:regressor}) are minimized as the positions to initiate the partitioning algorithm.
The intuition is that the discrete $(k+1)$th-order deltas approximate the $(k+1)$th-order derivatives
of a continuous function of degree $k$.
If a segment has small $(k+1)$th-order deltas, the underlying function to be learned
is less likely to contain terms with a degree much higher than $k$.

\begin{figure}[t!]
\centering
\includegraphics[width = 0.9\columnwidth]{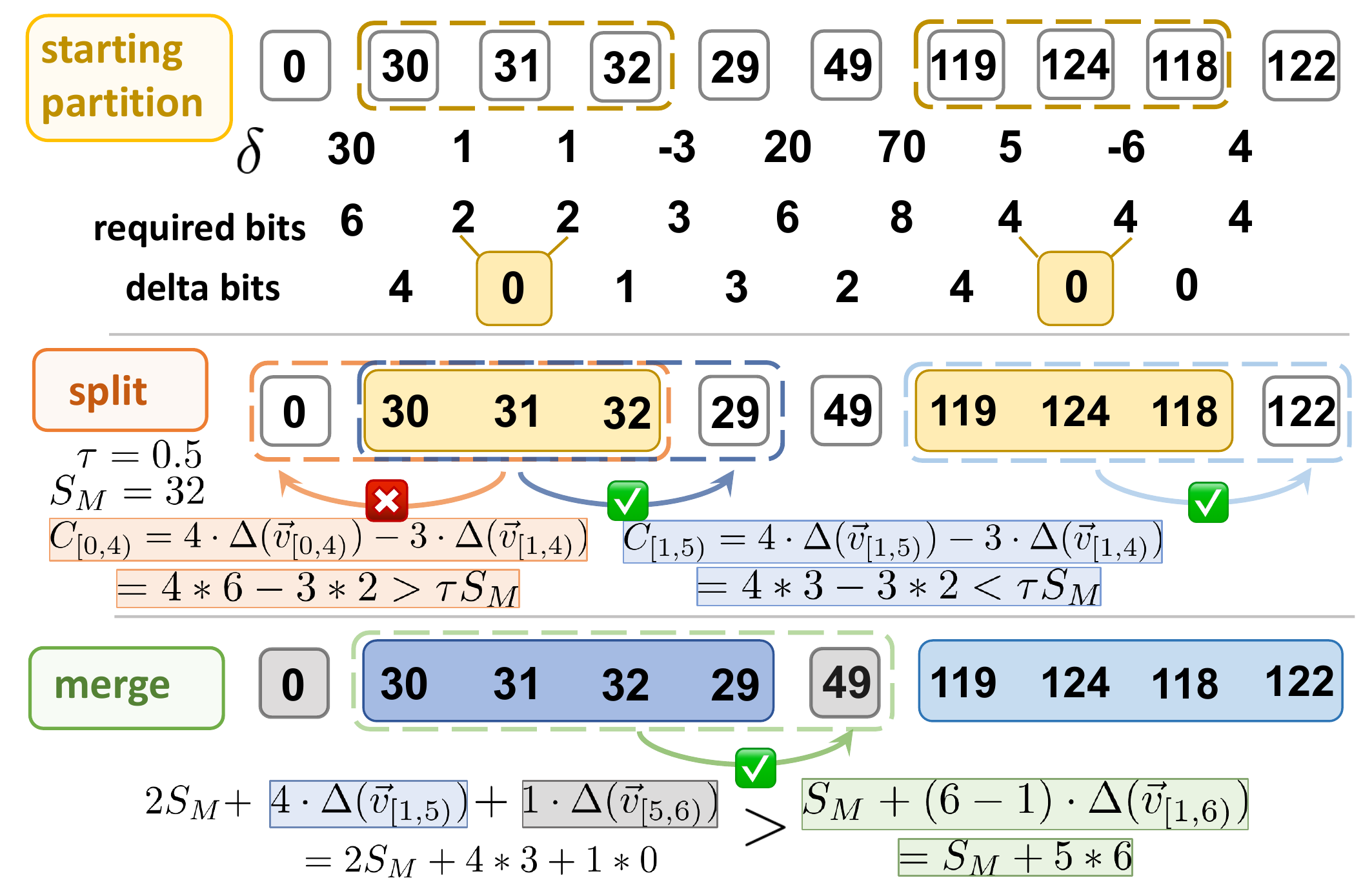}
\vspace{-0.4cm}
\mycaption{Variable-length Partitioning on Delta Encoding}{
Value $29$ is successfully included into segment $\{30,31,32\}$ in the split phase
because its inclusion cost $C_{[1, 5)} = 6$ is less than the pre-defined threshold $\tau S_M = 0.5\cdot 32 = 16$.
In the merge phase, the attempt to merge segment $\{30,31,32,29\}$ and $\{49\}$ succeeds
because the space consumption of the segment formed is smaller than the summation of the two original segments.
}
\vspace{-0.4cm}
\label{fig:partition_alg}
\end{figure}

\begin{figure}[t!]
\centering
\includegraphics[scale = 0.43]{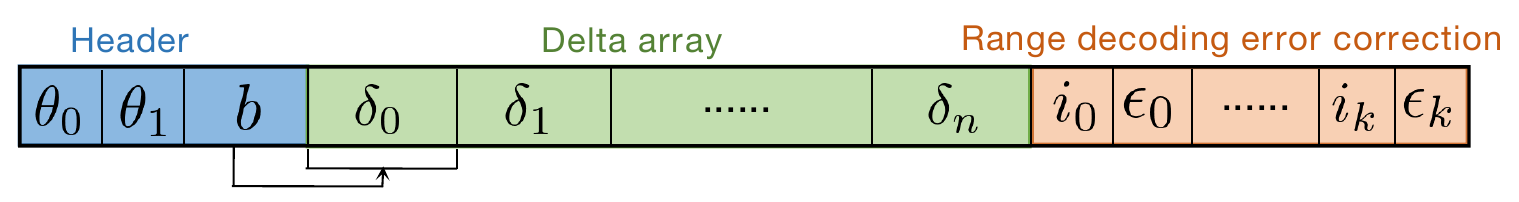}
\vspace{-0.2cm}
\caption{\leco's Storage Format for One Partition}
\vspace{-0.4cm}
\label{fig:storage_format}
\end{figure}

For Delta Encoding, a good starting partition
is when the differences between the neighboring values are small (i.e., a small model prediction error) and
when the neighboring points form roughly an arithmetic progression
(i.e., the partition has the potential to grow larger).
We, therefore, compute the bit-width for each delta in the sequence first
(``required bits'' in \cref{fig:partition_alg}).
We then compute the second-order ``delta bits'' based on those ``required bits''
and pick the positions with the minimum value
(the yellow-boxed zeros in \cref{fig:partition_alg})
as the initial partitions.
The required bits are used as the tie-breaker to determine
the partition growth precedence.

To summarize, we compared the split-merge partitioning algorithm with the linear Regressor
against the optimal partitioning obtained via dynamic programming on real-world data sets
introduced in \cref{sec:dataset}
and found that our greedy algorithm imposes less than $3\%$ overhead on the
final compressed size.

%---------------------------------------------------------------------------------
%---------------------------------------------------------------------------------
% Hyperparameter Tuning
%---------------------------------------------------------------------------------
%---------------------------------------------------------------------------------
\subsubsection{Partitioning Strategy Advising}
\label{sec:hyper_tune}
Compared to fixed-length partitions, variable-length partitions could produce a higher
compression ratio with a cost of slower random access and compression speed.
The choice of the partitioning strategies depends largely on the application's needs.
To facilitate estimating the trade-offs, our Hyperparameter-Advisor provides two scores
to indicate the potential space benefit of adopting the variable-length strategy.

The two scores are inspired by the definitions of ``local hardness'' ($\mathcal{H}_l$)
and ``global hardness'' ($\mathcal{H}_g$) of a data set introduced in~\cite{wongkham2022updatable}.
$\mathcal{H}_l$ captures the local unevenness in the values distribution,
while $\mathcal{H}_g$ depicts the degree of variation of the distribution at a global scale.
Intuitively, if the data set is locally hard (i.e., $\mathcal{H}_l$ is high),
no Regressor would fit the data well regardless of the partitioning strategy.
On the other hand, if the data set is locally easy but globally hard (i.e., $\mathcal{H}_g$ is high),
applying variable-length partitioning could improve the compression ratio significantly
because it is able to catch the ``sharp turns'' in the global trend of the value distribution.

Similar to~\cite{wongkham2022updatable}, we compute $\mathcal{H}_l$ by
running the piece-wise linear approximation (PLA) algorithm with a small error bound (e.g., $\epsilon=7$)
on the data set and count the number of segments generated.
The count is then divided by the data set size to normalize the $\mathcal{H}_l$ score.
For $\mathcal{H}_g$, we run the same PLA algorithm with a much larger error bound (e.g., $\epsilon=4096$).
Instead of counting the number of segments, we use the the average gap\footnote{first value of the latter segment - last value of the former segment}
between adjacent segments and the variance of the segment lengths to estimate
the ``global hardness'' of the value distribution.
$\mathcal{H}_g$ is the summation of these two numbers, with each normalized.

%---------------------------------------------------------------------------------
%---------------------------------------------------------------------------------
% Encoder and Decoder
%---------------------------------------------------------------------------------
%---------------------------------------------------------------------------------

\subsection{Encoder and Decoder}
\label{sec:enc_dec}

The \textbf{Encoder} is responsible for generating the final compressed sequences.
The input to the Encoder is a list of value partitions produced by the
Partitioner, where each partition is associated with a model.
The Encoder computes the delta for each value through model inference
and then stores it in the delta array.

The storage format is shown in \cref{fig:storage_format}.
There is a \texttt{header} and a \texttt{delta array} for each partition.
In the header, we first store the model parameters.
For the default linear Regressor, the parameters are two 64-bit
floating-point numbers: intercept $\theta_0$ and slope $\theta_1$.
Because we bit pack the delta array according to the maximum delta,
we must record the bit-length $b$ for an array item in the header.

For fixed-length partitions, the Encoder stores the partition size
$L$ in the metadata.
If the partitions are variable-length, the Encoder keeps the start index
(in the overall sequence) for each partition so that a random access
can quickly locate the target partition.
We use ALEX~\cite{ALEX} (a learned index) to record those start positions
to speed up the binary search.

To decompress a value given a position $i$, the \textbf{Decoder} first determines
which partition contains the requested value.
If the partitions are fixed-length, the value is located in the
$\lfloor \frac{i}{L} \rfloor$th partition.
Otherwise, the Decoder conducts a ``lower-bound'' search in the metadata to find the partition with the largest start index $\le i$.

After identifying the partition, the Decoder reads the model parameters from the partition header and then performs a model inference using
$i' = i $ $-$ \texttt{start\_index} to get a predicted value $\hat{v}$.
Then, the Decoder fetches the corresponding $\delta_{i'}$ in the delta
array by accessing from the ($b \cdot i'$)th bit to
the $(b \cdot (i' + 1) - 1)$th bit.
Finally, the Decoder returns the decompressed value
$\lfloor \hat{v} \rfloor + \delta_{i'}$.
Decoding a value involves at most two memory accesses,
one for fetching the model (often cached) and the other for fetching the delta.

The basic algorithm for \textbf{range decompression} is to invoke
the above decoding process for each position in the range.
Because of the sequential access pattern, most cache misses are eliminated.
For the default linear regression, the Decoder performs two
floating-point calculations for model inference
(one multiplication and one addition) and an integer addition
for delta correction.

We carry out an optimization to increase the range decompression throughput by $10 - 20\%$.
For position $i$, the model prediction is $\hat{v}_i = \theta_0 + \theta_1 \cdot i$.
We can obtain $\hat{v}_i$ by computing $\hat{v}_{i-1} + \theta_1$,
thus saving the floating-point multiplication. 
However, because of the limited precision in the floating-point representation, the $\theta_1$-accumulation result at certain position $i$ is incorrect (i.e., $\lfloor \theta_0 + \sum_{1}^{i}\theta_1 \rfloor + \delta_i
\ne \lfloor \theta_0 + \theta_1 \cdot i \rfloor + \delta_i$).
Therefore, we append an extra list to the delta array to correct
the deviation at those positions.

%---------------------------------------------------------------------------------
%---------------------------------------------------------------------------------
% Extension to Handling Strings
%---------------------------------------------------------------------------------
%---------------------------------------------------------------------------------
\begin{figure}[t!]
\centering
\includegraphics[scale = 0.45]{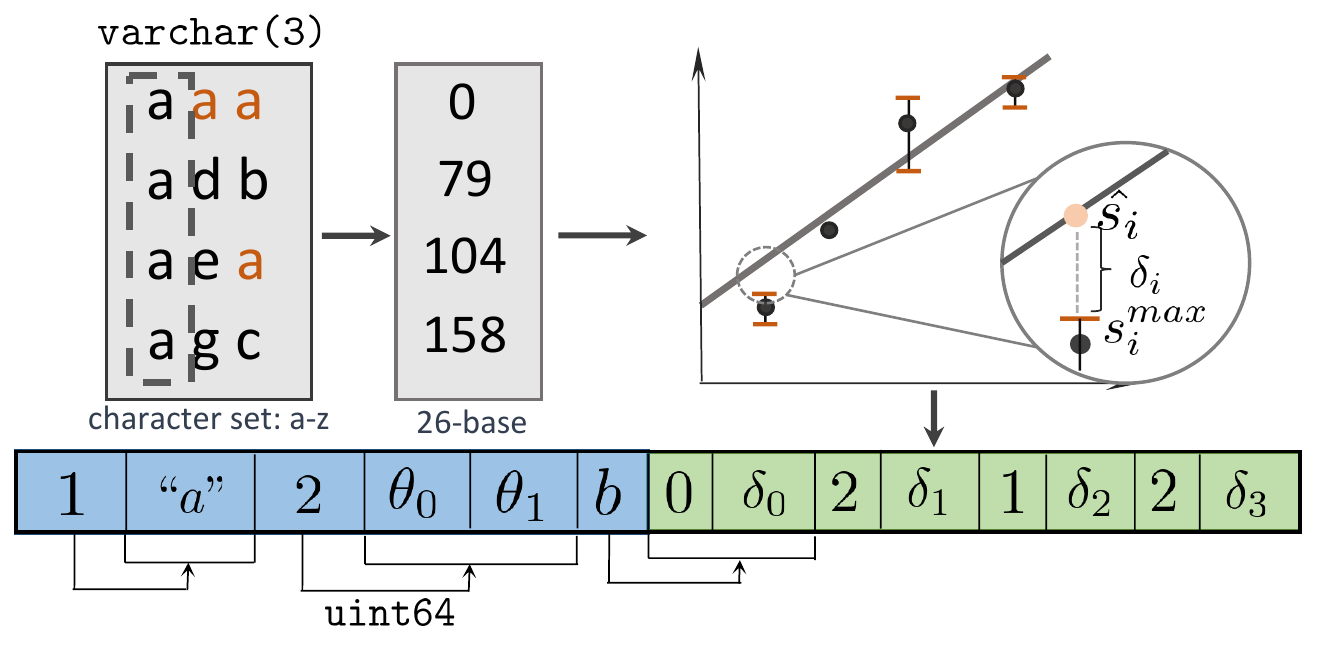}
\vspace{-0.2cm}
\mycaption{\leco String Compression}{An example including algorithm optimizations and storage format modifications.}
\vspace{-0.4cm}
\label{fig:leco_string}
\end{figure}

\subsection{Extension to Handling Strings}
\label{sec:string}
The (integer-based) algorithms discussed so far can already benefit a subset of
the string columns in a relational table where the values are dictionary-encoded.
In this section, we extend our support to mostly unique string values under the \leco framework.
The idea is to create an order-preserving mapping between the strings
and large integers so that they can be fed to the Regressor.

Given a partition of string values, we first extract their common prefix (marked in dashed box in \cref{fig:leco_string})
and store it separately in the partition \texttt{header}.
Then, we shrink the size of the character set if possible.
Because many string data sets refer to a portion of the ASCII table,
we can use a smaller base to perform the string-integer mapping.
For example, we adopt 26-based integers
in \cref{fig:leco_string} with only lower-case letters presenting.

Notice that for an arbitrary $M$-based mapping, the computation required
to recover each character from the integer is expensive.
Given the mapped integer $v$, it requires an integer modulo $v \% M$ to
decode the current character and an integer division $v / M$ to prepare
for decoding the next one.
Both operations take tens of CPU cycles.
To speed up decoding, we set $M$ to its closest power of two ($2^m$) so that
the modulo becomes a left-shift followed by a bit-wise AND ($v \& ((1 << m) - 1)$),
and the division becomes a right-shift ($v >> m$).
For example, for strings that only consist of lower-case characters, we set $M = 32$.

\leco requires strings to be fixed-length.
For a column of \texttt{varchar(3)}, we pad every string to 3 bytes (padding bytes marked with orange ``a'' in \cref{fig:leco_string}).
An interesting observation is that we can leverage the flexibility in choosing
the padding characters to minimize the stored deltas. 
Suppose the string at position $i$ is $s_i$, and the smallest/largest valid
string after padding is $s_i^{min}$/$s_i^{max}$ (i.e., pad each bit position with the smallest/largest character in the character set).
We then choose the padding adaptively based on the predicted value $\hat{s}_i$
from the Regressor to minimize the absolute value of the prediction error.
If $\hat{s}_i < s_i^{min}$, we adopt the minimum padding and store
$\delta_i = s_i^{min} - \hat{s}_i$ in the delta array;
if $\hat{s}_i > s_i^{max}$, we use the maximum padding and produce
$\delta_i = s_i^{max} - \hat{s}_i$;
if $s_i^{min} \le \hat{s}_i \le s_i^{max}$, we choose $\hat{s}_i$ as the
padded string directly and obtain $\delta_i = 0$.

The lower part of \cref{fig:leco_string} shows the updated storage format to accommodate varchars.
Additionally, the \texttt{header} includes the maximum padding length (without prefix) along with the
common prefix of the partition.
We also record the length of each varchar value in the delta array 
(the slot before each delta value) to mark the boundary of the valid bytes from padded bytes in order to decode correctly.
These lengths can be omitted for fixed-length strings.

\section{Microbenchmark Evaluation}
\label{sec:eval}

We evaluate \leco in two steps.
In this section, we compare \leco against state-of-the-art lightweight
compression schemes through a set of microbenchmarks.
We analyze \leco's gains and trade-offs in compression ratio, random
access speed, and range decompression throughput.
In \cref{sec:sys-eval}, we integrate \leco into two widely-used applications
to show the end-to-end performance.

%---------------------------------------------------------------------------------
% Compression Schemes and Data Sets
%---------------------------------------------------------------------------------

\subsection{Compression Schemes and Data Sets}
\label{sec:dataset}

\begin{figure*}
    \begin{subfigure}[b]{0.8\textwidth}%
\hspace{-0.2cm}
\begin{minipage}[t]{0.08\textwidth}
\centering
    \includegraphics[scale=0.138]{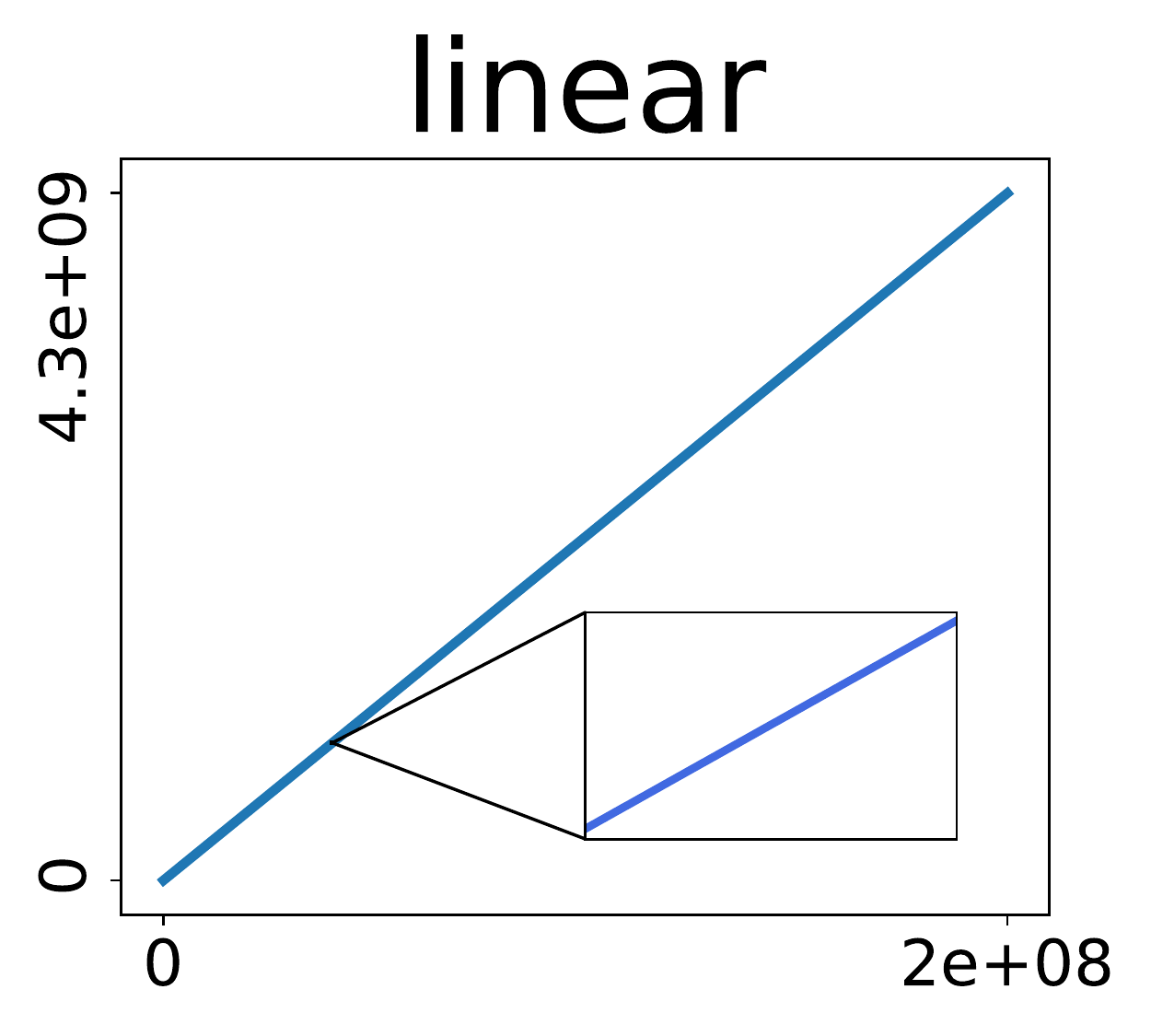}
\end{minipage}
\hspace{0.285cm}
\begin{minipage}[t]{0.08\textwidth}
\centering
    \includegraphics[scale=0.138]{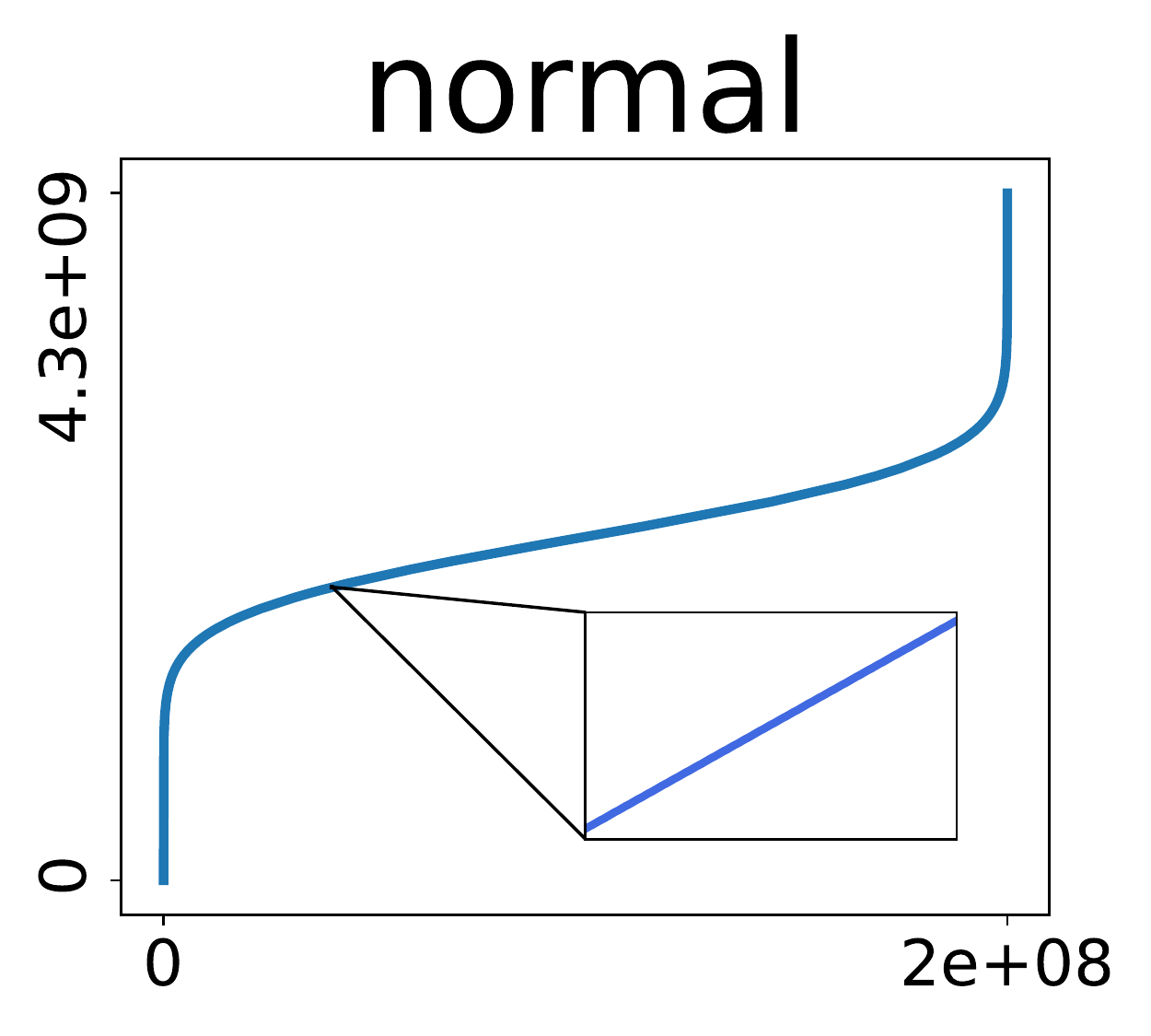}
\end{minipage}
\hspace{0.285cm}
\begin{minipage}[t]{0.08\textwidth}
\centering
    \includegraphics[scale=0.138]{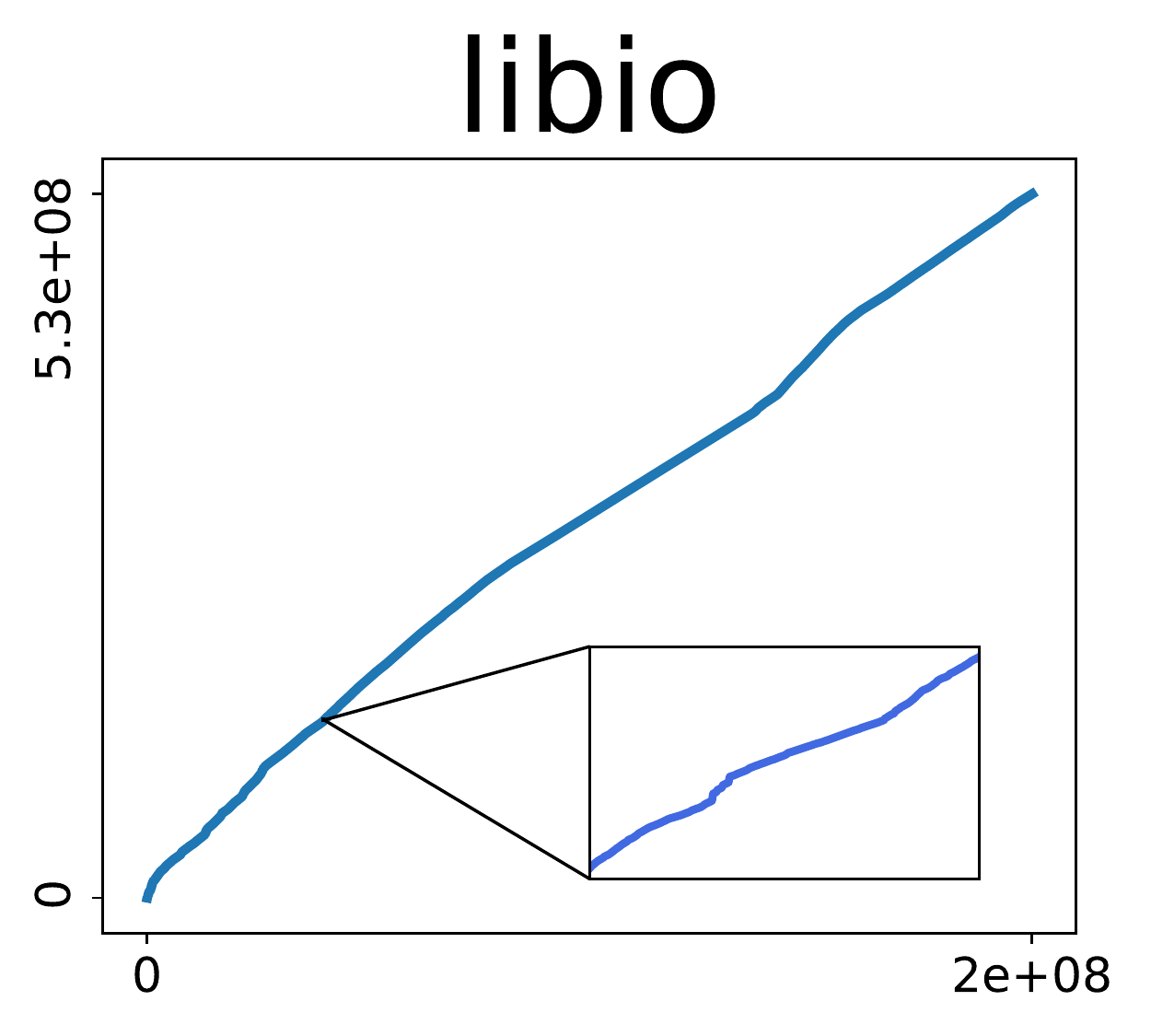}
\end{minipage}
\hspace{0.285cm}
\begin{minipage}[t]{0.08\textwidth}
\centering
    \includegraphics[scale=0.138]{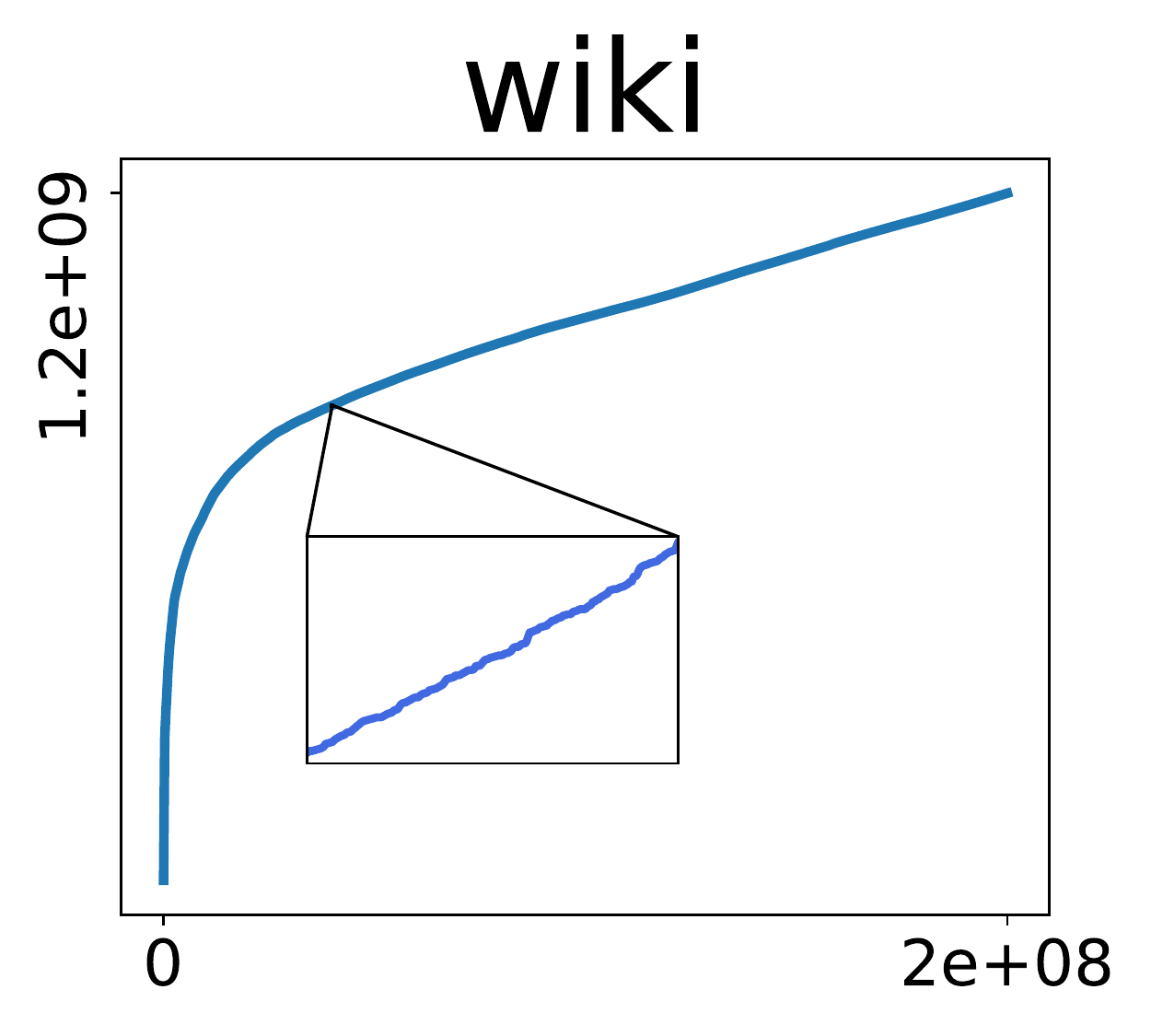}
\end{minipage}
\hspace{0.285cm}
\begin{minipage}[t]{0.08\textwidth}
\centering
    \includegraphics[scale=0.138]{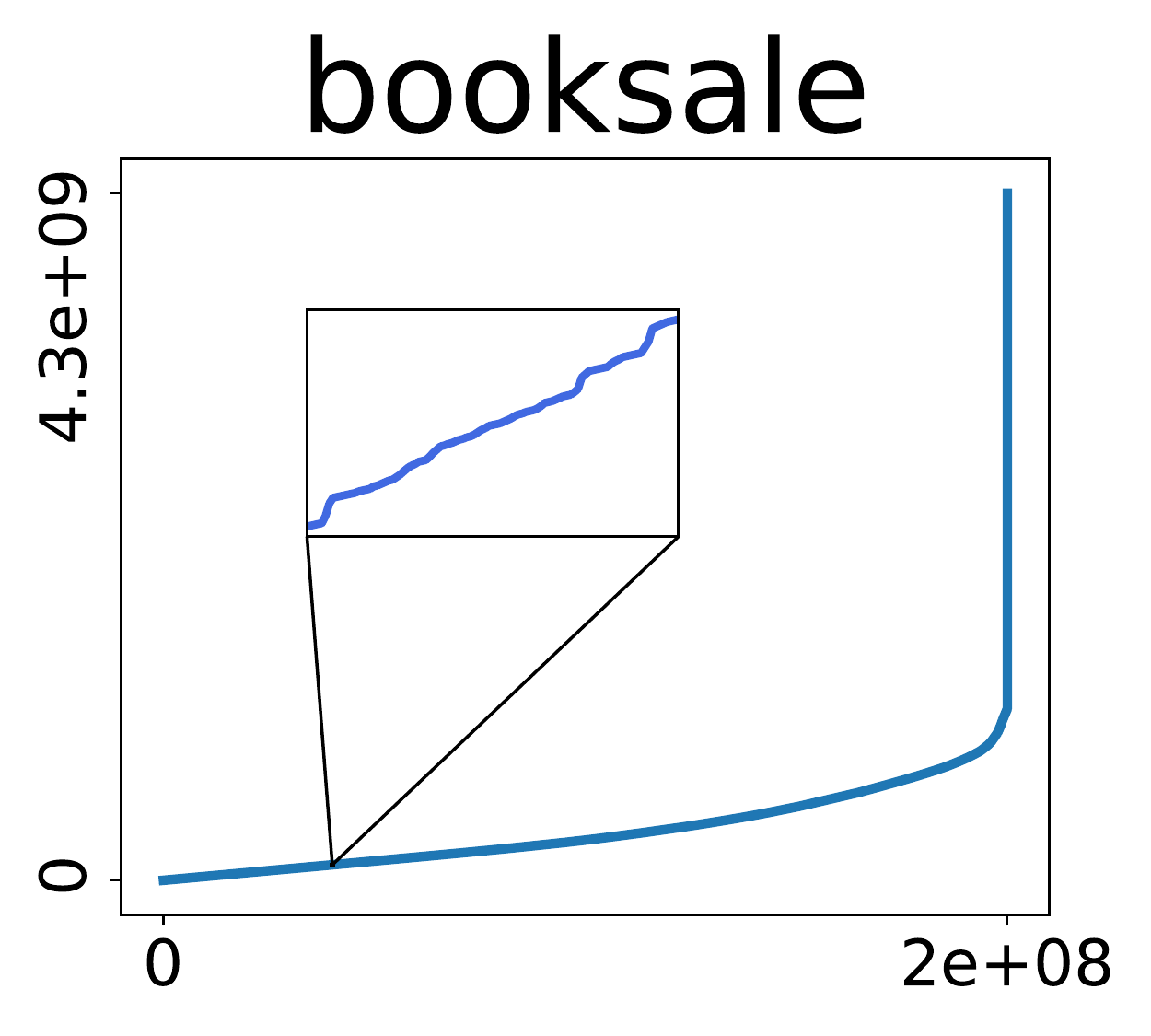}
\end{minipage}
\hspace{0.285cm}
\begin{minipage}[t]{0.08\textwidth}
\centering
    \includegraphics[scale=0.138]{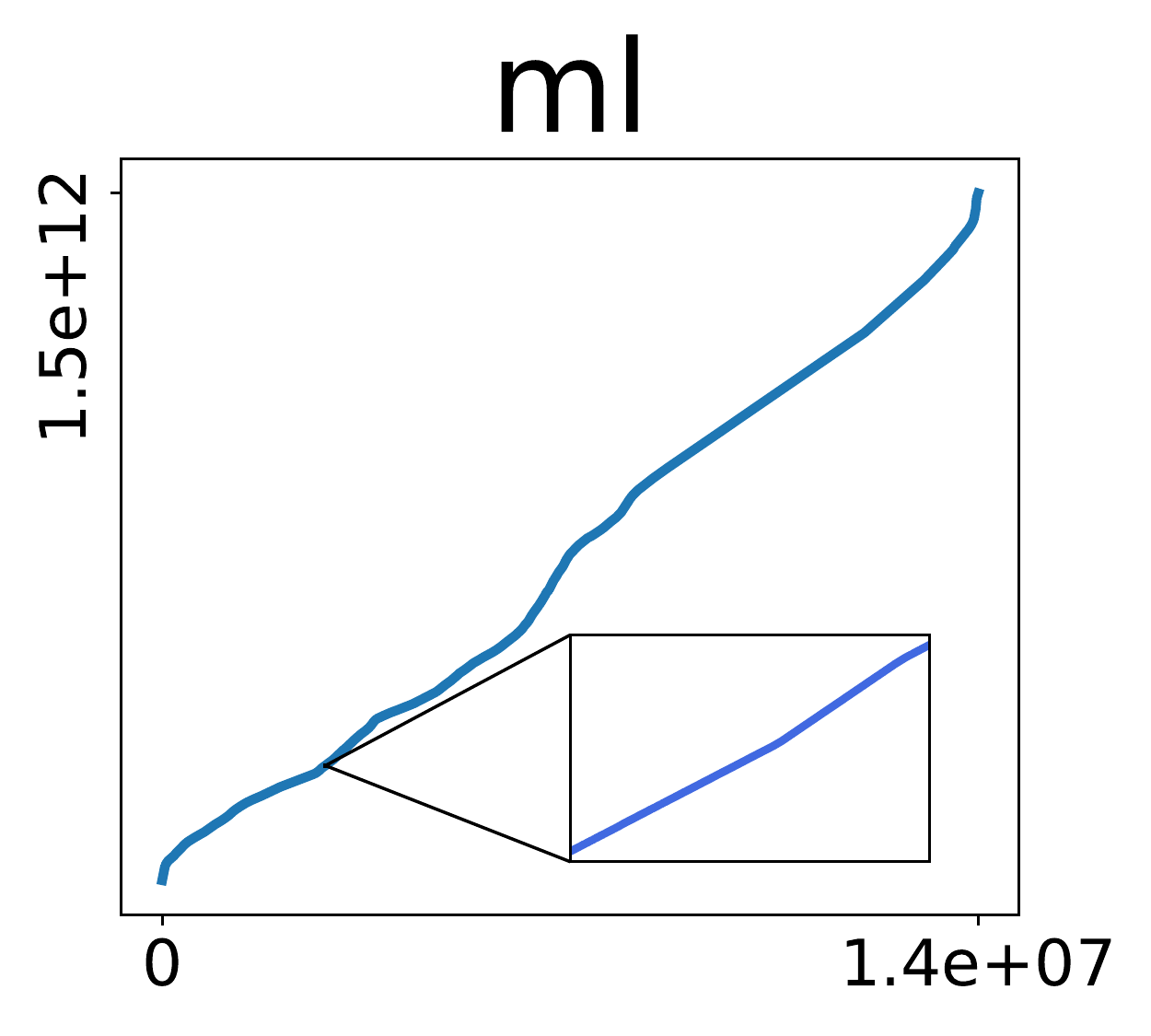}
\end{minipage}
\hspace{0.285cm}
\begin{minipage}[t]{0.08\textwidth}
\centering
    \includegraphics[scale=0.138]{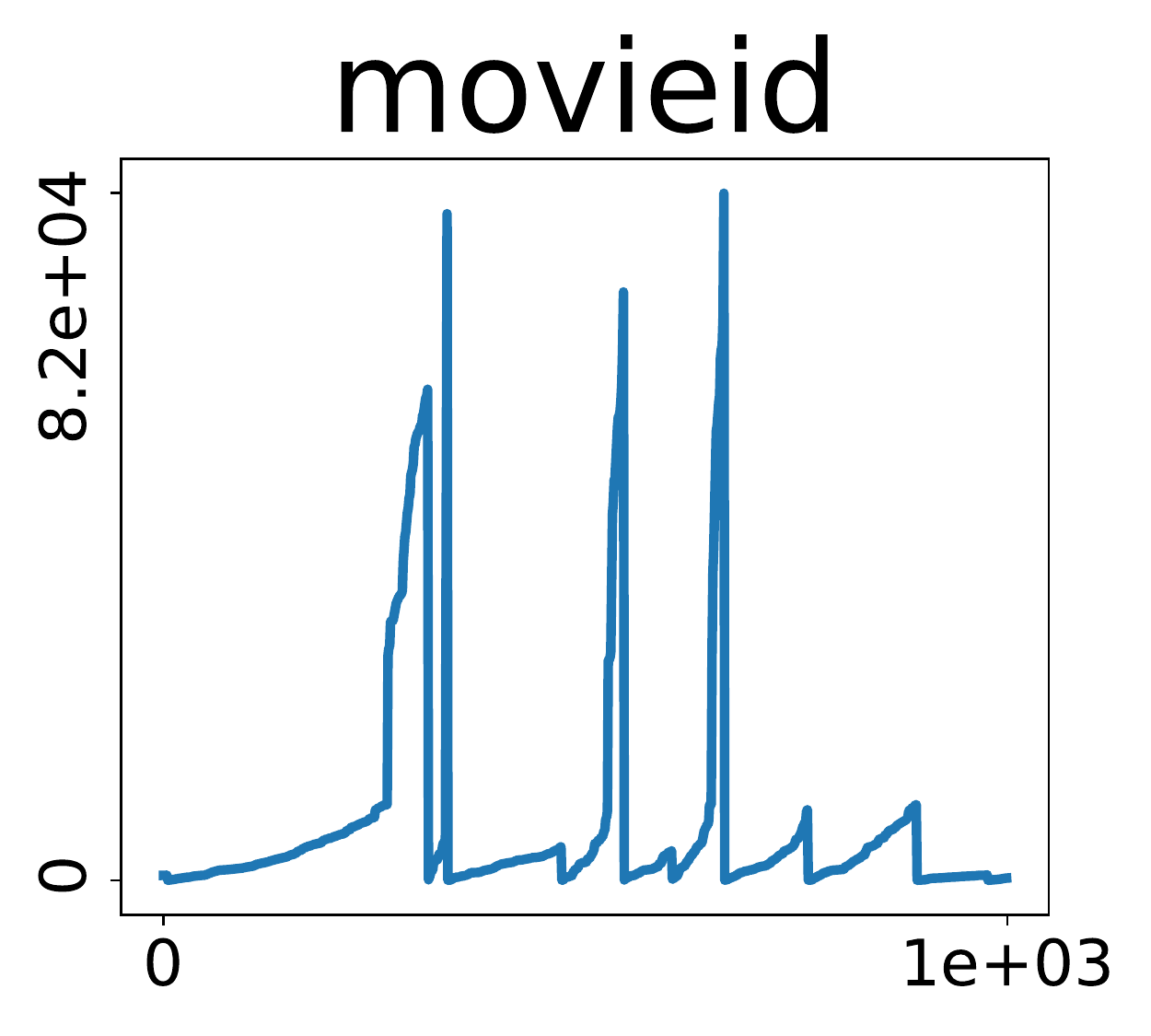}
\end{minipage}
\hspace{0.285cm}
\begin{minipage}[t]{0.08\textwidth}
\centering
    \includegraphics[scale=0.138]{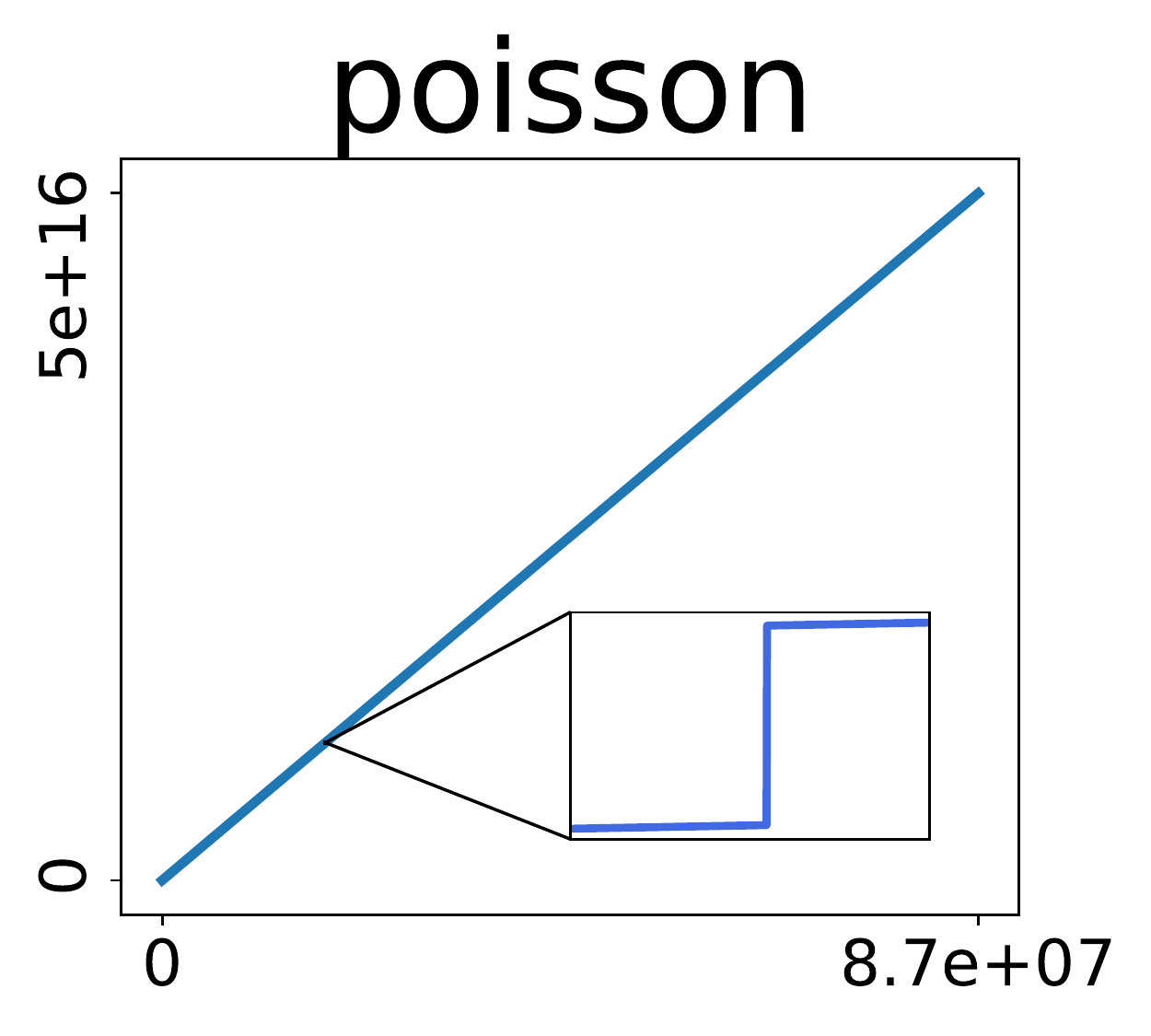}
\end{minipage}
\hspace{0.285cm}
\begin{minipage}[t]{0.08\textwidth}
\centering
    \includegraphics[scale=0.138]{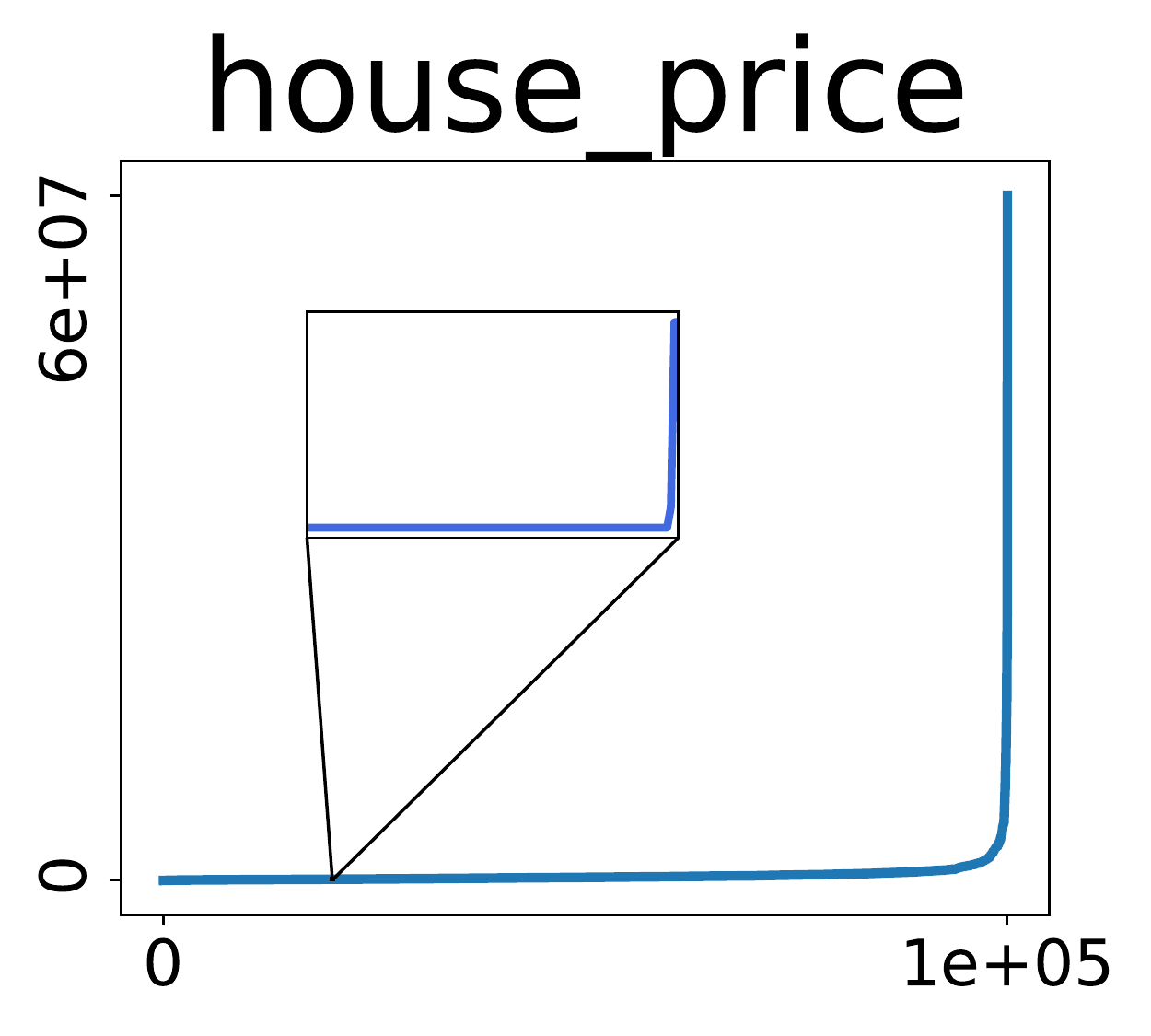}
\end{minipage}

\hspace{-0.2cm}
\begin{minipage}[t]{0.08\textwidth}
\centering
    \includegraphics[scale=0.138]{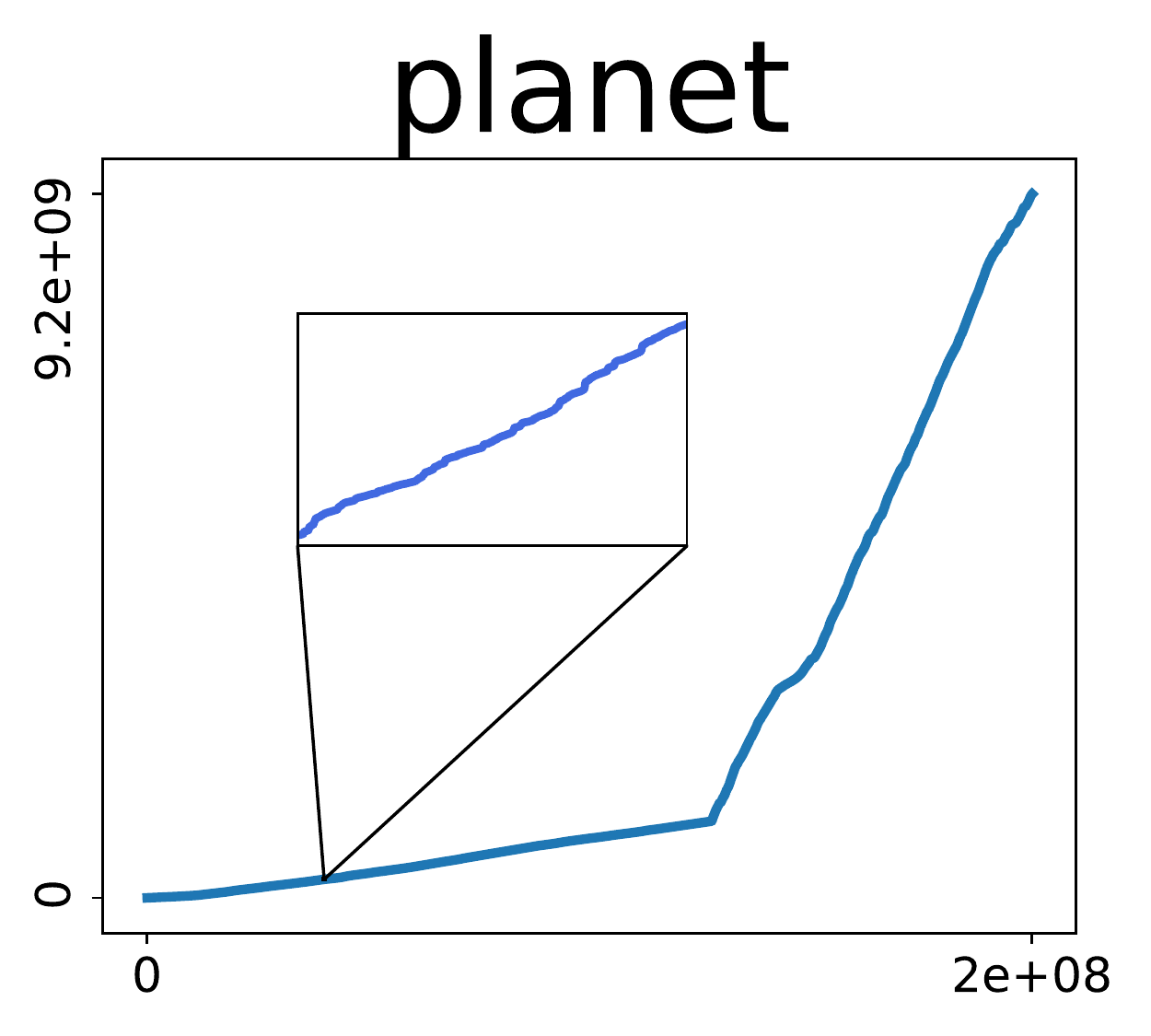}
\end{minipage}
\hspace{0.285cm}
\begin{minipage}[t]{0.08\textwidth}
\centering
    \includegraphics[scale=0.138]{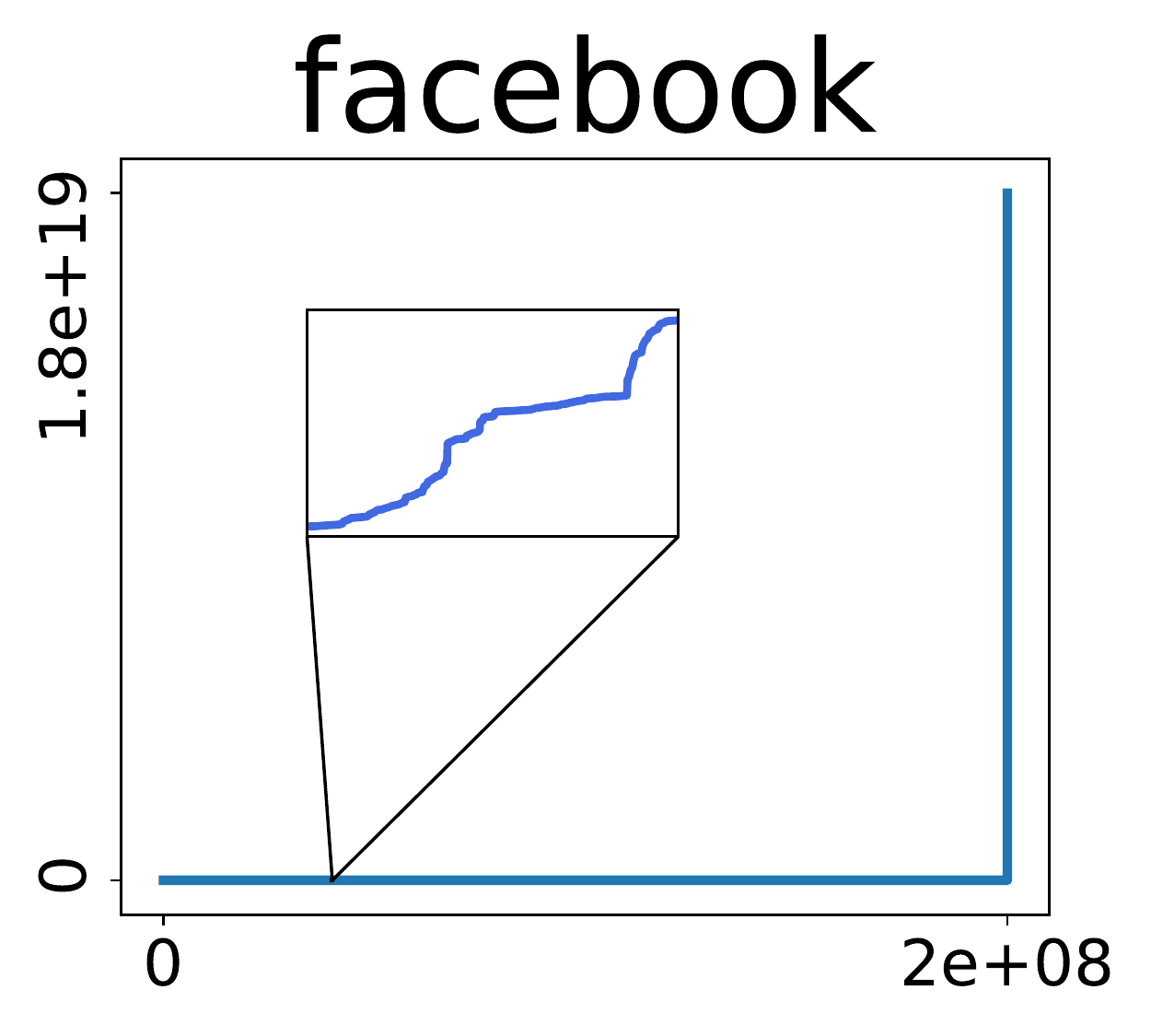}
\end{minipage}
\hspace{0.285cm}
\begin{minipage}[t]{0.08\textwidth}
\centering
    \includegraphics[scale=0.138]{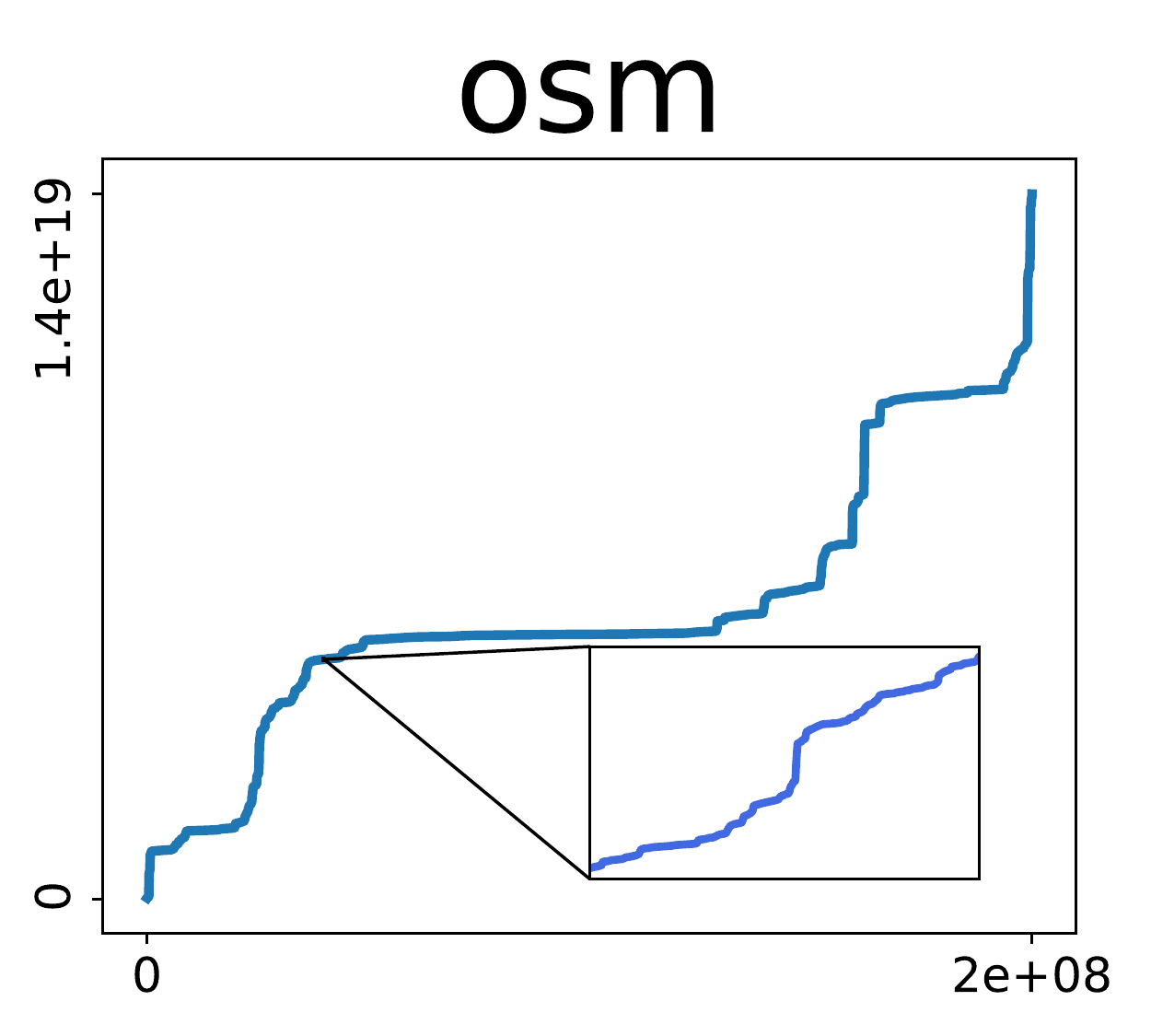}
\end{minipage}
\hspace{0.285cm}
\begin{minipage}[t]{0.08\textwidth}
\centering
    \includegraphics[scale=0.138]{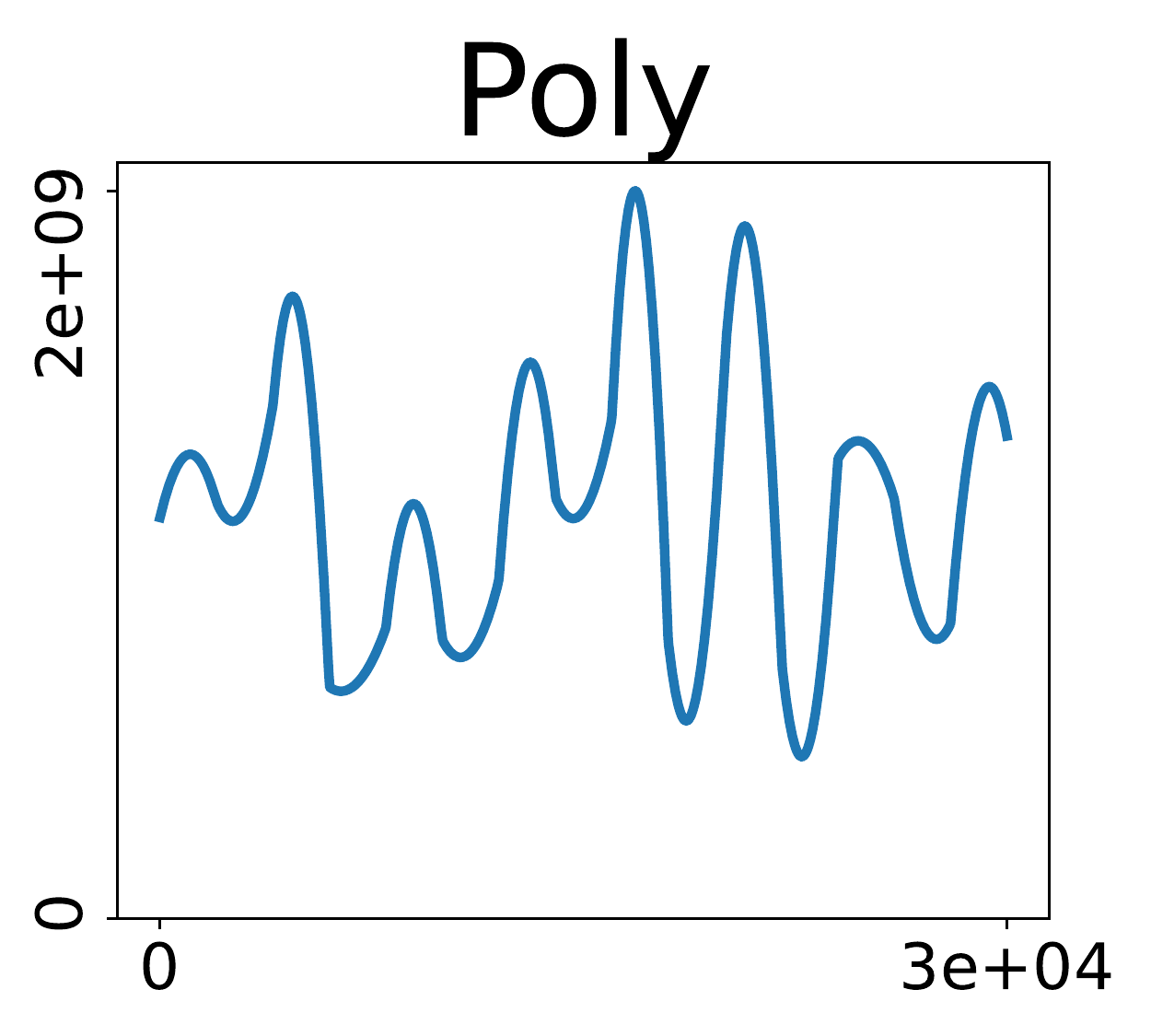}
\end{minipage}
\hspace{0.3cm}
\begin{minipage}[t]{0.08\textwidth}
\centering
    \includegraphics[scale=0.138]{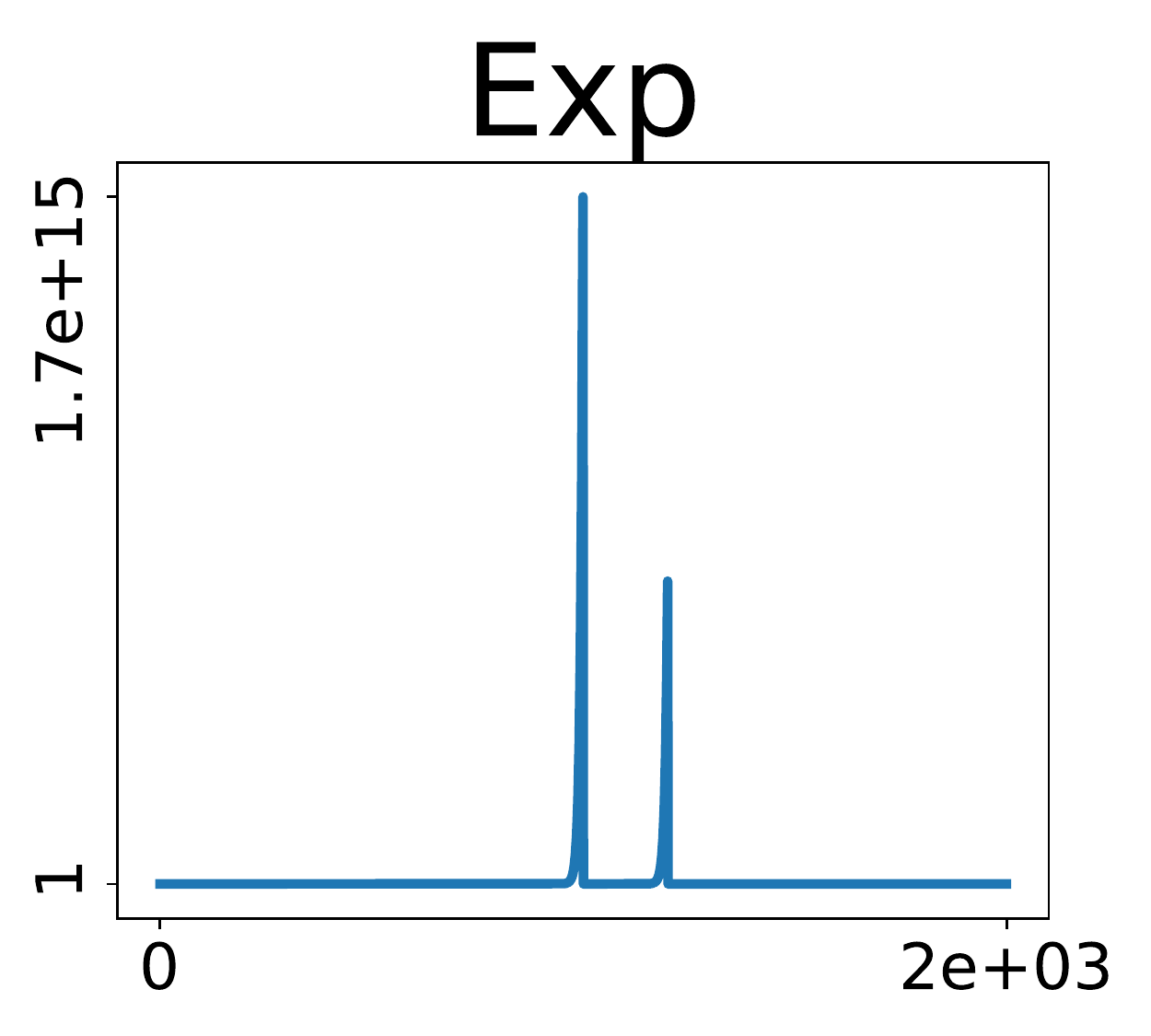}
\end{minipage}
\hspace{0.285cm}
\begin{minipage}[t]{0.08\textwidth}
\centering
    \includegraphics[scale=0.138]{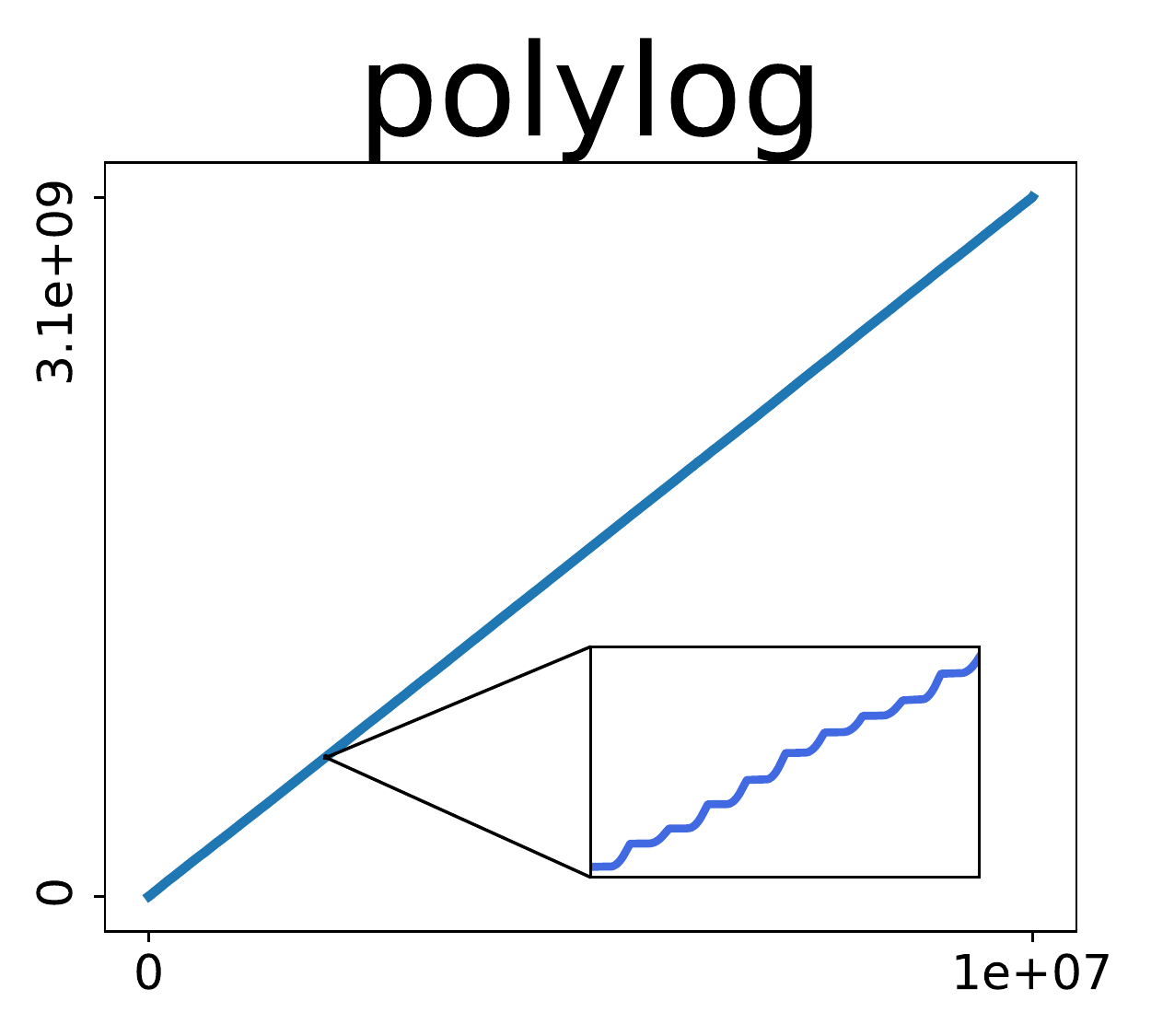}
\end{minipage}
\hspace{0.285cm}
\begin{minipage}[t]{0.08\textwidth}
\centering
    \includegraphics[scale=0.138]{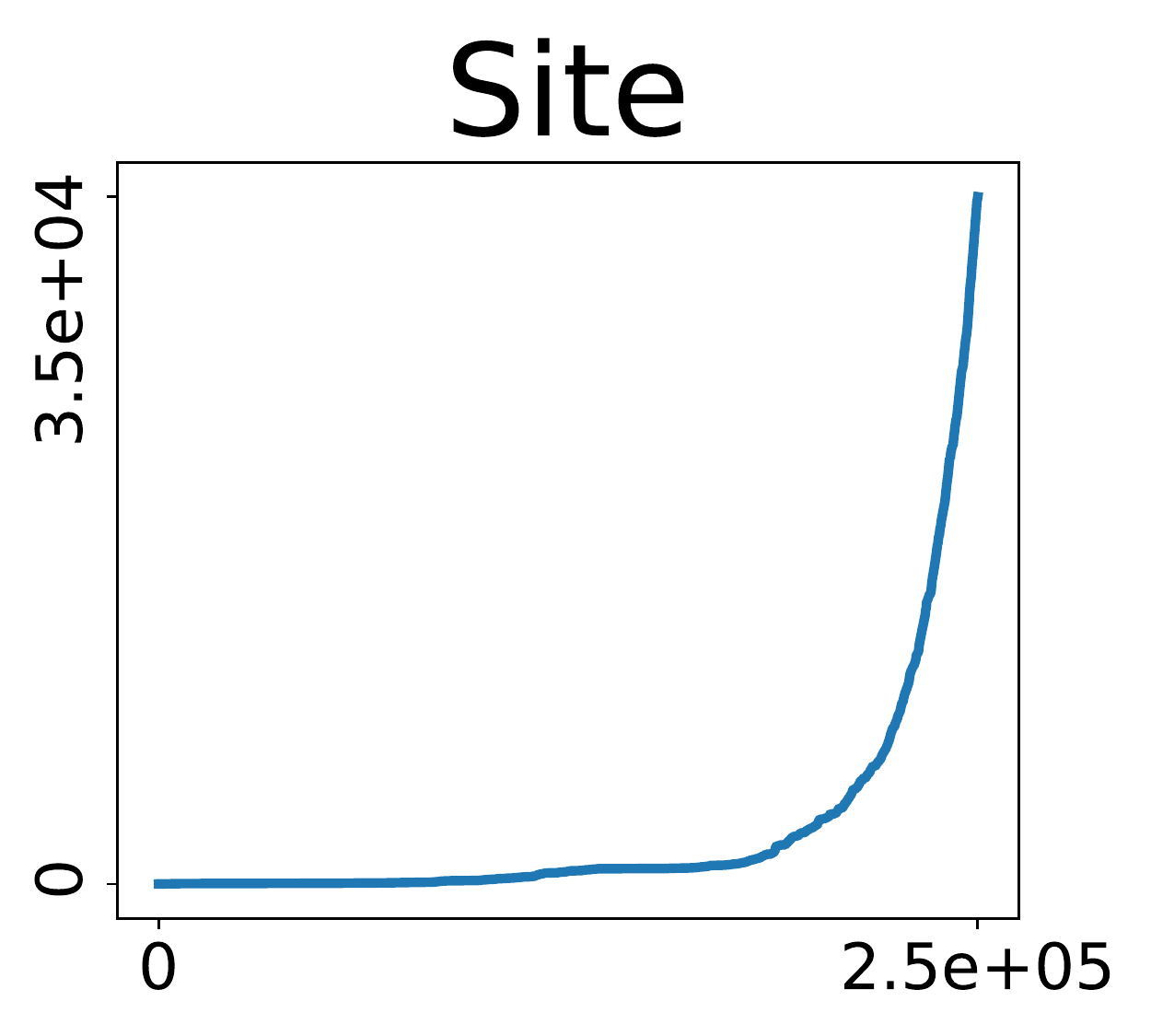}
\end{minipage}
\hspace{0.285cm}
\begin{minipage}[t]{0.08\textwidth}
\centering
    \includegraphics[scale=0.138]{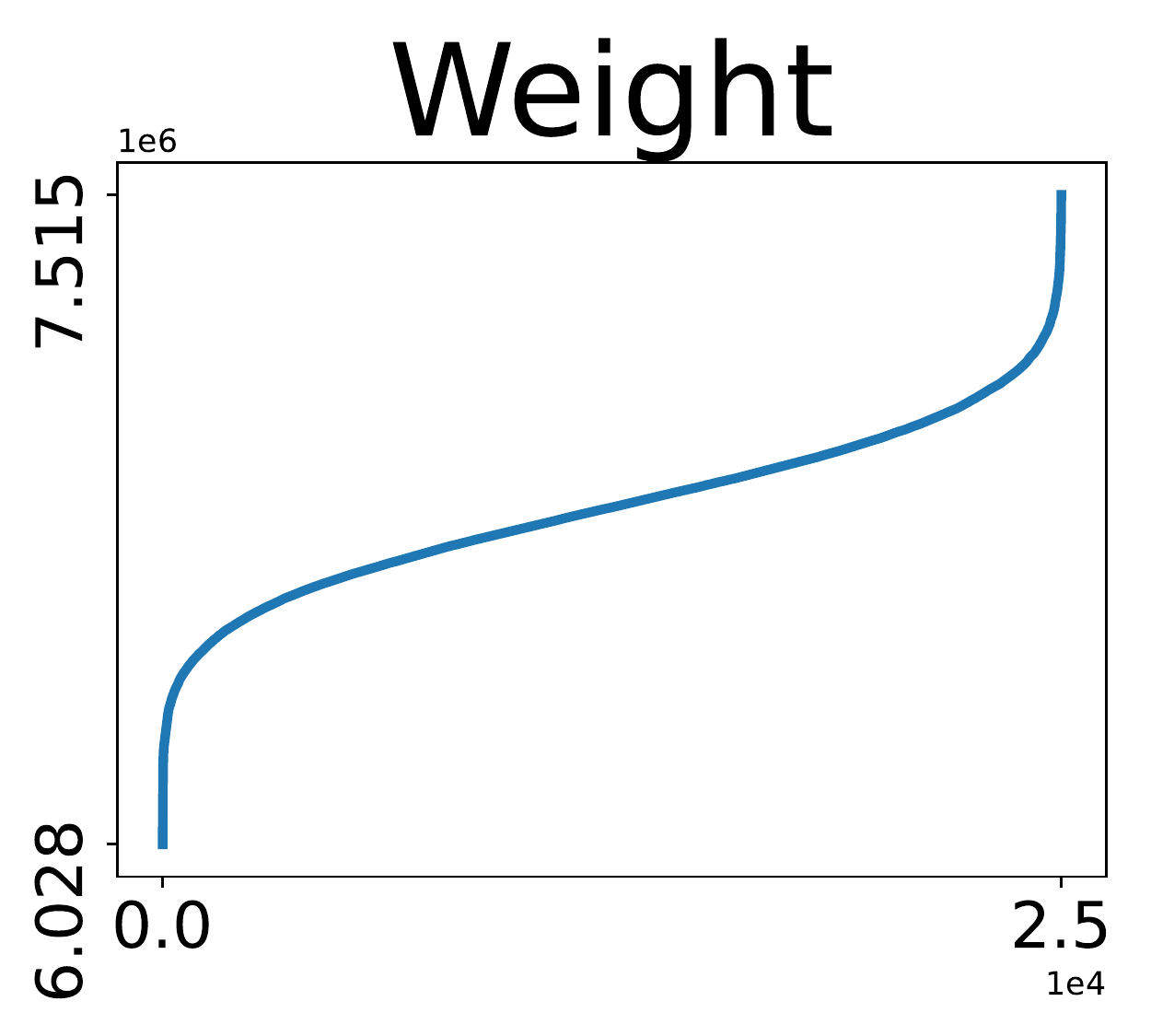}
\end{minipage}
\hspace{0.35cm}
\begin{minipage}[t]{0.08\textwidth}
\centering
    \includegraphics[scale=0.138]{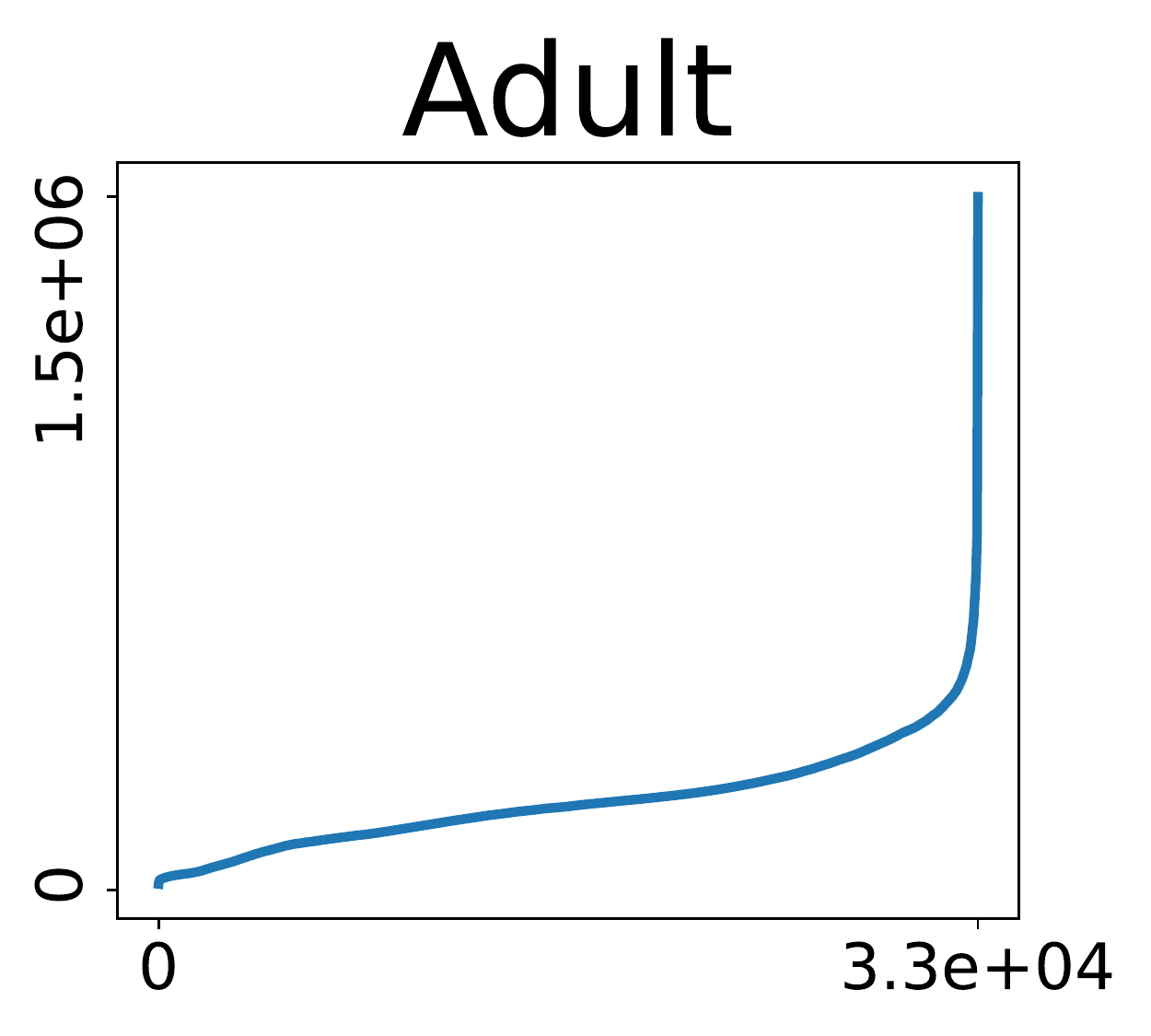}
\end{minipage}
    \vspace{-0.1cm}
\mycaption{Data Distribution Plot.}{The first row presents the nine data sets classified as ``local easy''.}
\label{fig:distribution}
\end{subfigure}%
    \begin{subfigure}[b]{0.2\textwidth}%
        \center
        % \vspace{.25in}
        \hspace{-0.2cm}
        \includegraphics[scale=0.24]{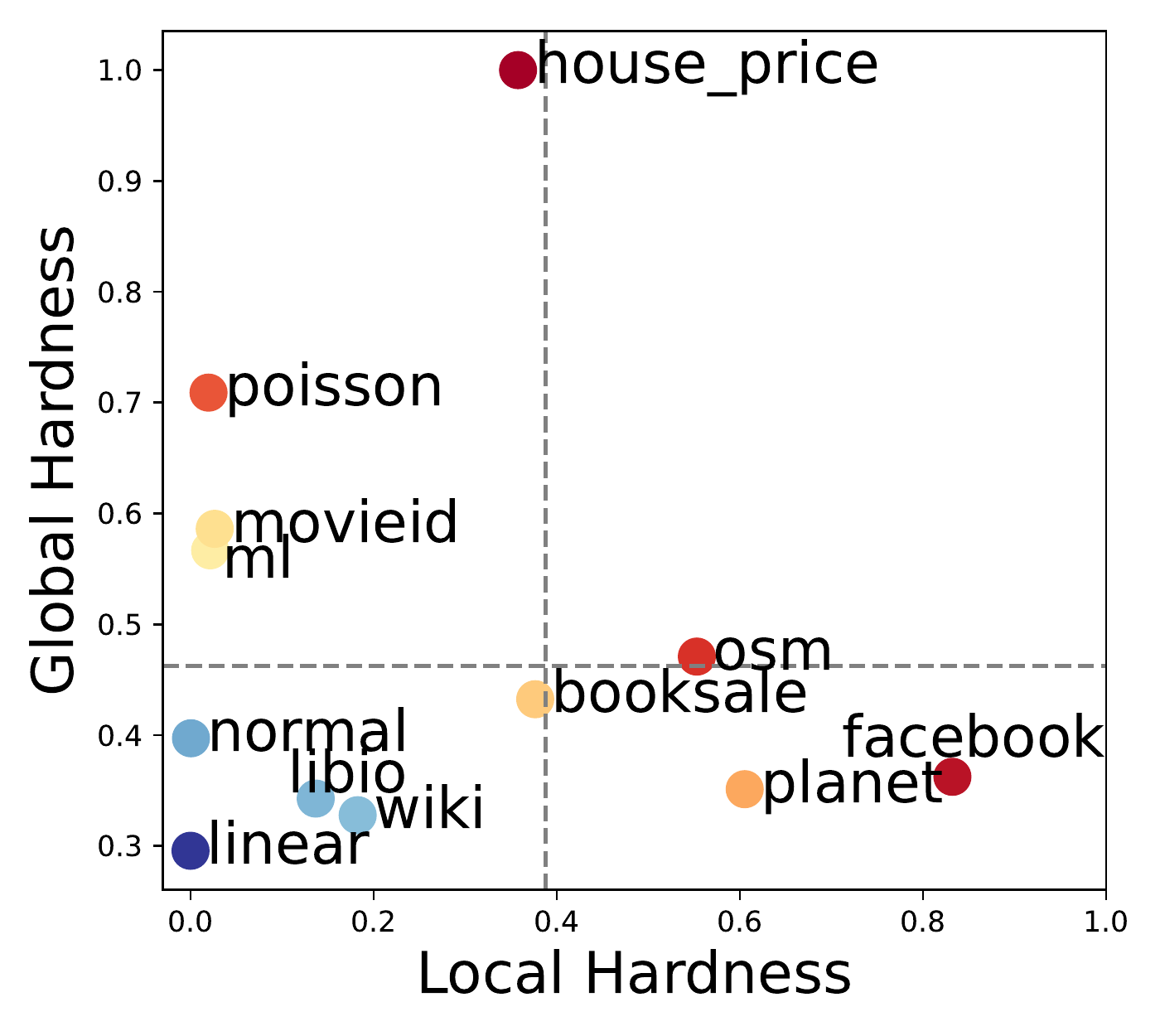}
        \caption{Dataset Hardness.}
        \label{fig:hardness}
    \end{subfigure}%
    \vspace{-0.3cm}
    \caption{Data Distribution with Hardness evaluation.}
\end{figure*}

The baseline compression schemes under evaluation are
\eliasfano~\cite{vigna2013quasi, ottaviano2014partitioned},
Frame-of-Reference (\FORfix)~\cite{goldstein1998, zukowski2006super}, Delta Encoding (Delta)~\cite{abadi2006}, and rANS~\cite{duda2013asymmetric}.
\FORfix and Delta are introduced in \cref{sec:motivation}.
rANS is a variant of arithmetic encoding~\cite{witten1987arithmetic} with a decoding speed similar to Huffman~\cite{huffman1952}.
\eliasfano is an encoding mechanism to compress a sorted list of integers.
Suppose the list has $n$ integers, with $m$ being the difference between the maximum and minimum value of the sequence.
\eliasfano stores the lower $\lceil log_2(\frac{m}{n}) \rceil$ bits for each
value explicitly with bit packing. 
For the remaining higher bits, \eliasfano uses unary coding
to record the number of appearances for each possible higher-bit value.
For example, the binary sequence
00000, 00011, 01101, 10000, 10010, 10011, 11010, 11101
is encoded as
``00 11 01 00 10 11 10 01'' for the lower bits and
``110 0 0 10 1110 0 10 10'' for the higher bits.
\eliasfano is quasi-succinct~\cite{vigna2013quasi} in that it only requires
$(2 + \lceil log_2(\frac{m}{n}) \rceil)$ bits per element.

We evaluate \leco and the baseline solutions extensively on thirteen integer data sets:

\begin{itemize}[leftmargin=10pt]
    \item[--] \textbf{linear, normal}: synthetic data sets with 200M 32-bit sorted integers
        following a clean linear (or normal) distribution.
    \item[--] \textbf{poisson}: 87M 64-bit timestamps following a Poisson distribution
        that models events collected by distributed sensors~\cite{zhang2018surf}.
    \item[--] \textbf{ml}: 14M 64-bit sorted timestamps from the UCI-ML data set~\cite{mltimestamp}.
    \item[--] \textbf{booksale, facebook, wiki, osm}: each with 200M 32-bit or 64-bit sorted integers from the SOSD benchmark~\cite{2019SOSD}.
    \item[--] \textbf{movieid}: 20M 32-bit ``liked'' movie IDs from MovieLens~\cite{movieiddataset}.
    \item[--] \textbf{house\_price}: 100K 32-bit sorted integers representing the distribution
        of house prices in the US~\cite{houseprice}.
    \item[--] \textbf{planet}: 200M 64-bit sorted planet ID from OpenStreetMap~\cite{osm}.
    \item[--] \textbf{libio}: 200M 64-bit sorted repository ID from libraries.io~\cite{libio}.
    \item[--] \textbf{medicare}: (used in \cref{sec:compress_dict}) 1.5 billion augmented 64-bit integers (without order) exported from the public BI benchmark~\cite{publicbi}.
\end{itemize}
seven additional non-linear data sets (used in \cref{sec:nonlinear}):
\begin{itemize}[leftmargin=10pt]
    \item[--] \textbf{cosmos}: 100M 32-bit data simulating a cosmic ray signal\footnote{We use $(\sin\frac{x+10}{60\pi} + \frac{1}{10}\sin\frac{3(x+10)}{60\pi}) \times 10^6 + \mathcal{N}(0,100)$ to construct it.}.
    \item[--] \textbf{polylog}: 10M 64-bit synthetic data of a biological population growth curve\footnote{Constructed by concatenating the polynomial and logarithm distribution, in turn, every 500 records.}.
    \item[--] \textbf{exp, poly}: 200M 64-bit synthetic data, each block follows the exponential or polynomial distribution of different parameters.
    \item[--] \textbf{site, weight, adult}: 250k, 25k and 30k sorted 32-bit integer column exported from the websites\_train\_sessions, weights\_heights, and adult\_train data sets in mlcourse.ai~\cite{mlcourse}.
\end{itemize}
nine tabular data sets, each sorted by its primary key column:
\begin{itemize}[leftmargin=10pt]
    \item[--] \textbf{lineitem, partsupp, orders}: TPC-H~\cite{tpch} tables, scale factor = 1.
    \item[--] \textbf{inventory, catalog\_sales, date\_dim}: from TPC-DS~\cite{tpcds}, sf = 1.
    \item[--] \textbf{geo, stock, course\_info}: real-world tables extracted from geonames~\cite{geoname}, GRXEUR price~\cite{histdata} and Udemy course~\cite{courseinfo}.
\end{itemize}
and three string data sets:
\begin{itemize}[leftmargin=10pt]
    \item[--] \textbf{email}: 30K email addresses (host reversed) with an average string length
        of 15 bytes~\cite{emaildataset}.
    \item[--] \textbf{hex}: 100K sorted hexadecimal strings (up to 8 bytes)~\cite{boncz2020fsst}.
    \item[--] \textbf{word}: 222K English words with an average length of 9 bytes~\cite{englishwords}.
\end{itemize}

\cref{fig:distribution} visualizes the eighteen integer data sets
where noticeable unevenness is observed frequently in real-world data sets.

%---------------------------------------------------------------------------------
% Experiment Setup
%---------------------------------------------------------------------------------

\subsection{Experiment Setup}
\label{sec:setup}
We run the microbenchmark on a machine with Intel\textregistered Xeon\textregistered
(Ice Lake) Platinum 8369B CPU @ 2.70GHz and 32GB DRAM.
The three baselines are labeled as \texttt{\eliasfano}, \texttt{\FORfix}, and \texttt{\deltafix}.
\texttt{\deltavar} represents our improved version of Delta Encoding that uses the
variable-length Partitioner in \leco. %as described in \cref{sec:partitioner}.
\texttt{\lecofix} and \texttt{\lecovar} are \emph{linear-Regressor} \leco prototypes that
adopt fixed-length and variable-length partitioning, respectively.
The corresponding \leco variants with \emph{polynomial Regressor} are labeled \texttt{\lecopolyfix} and \texttt{\lecopolyvar}.

For all the fixed-length partitioning methods, 
the partition size is obtained through a quick sampling-based parameter search described in~\cref{sec:fixedlen}.
For \deltavar, \lecovar, and \lecopolyvar, we set the split-parameter $\tau$
to be small (in the range $[0, 0.15]$) in favor of the compression ratio over
the compression throughput.

Given a data set, an algorithm under test first compresses the whole data set
and reports the compression ratio (i.e., \textit{compressed\_size / uncompressed\_size})
and compression throughput.
Then the algorithm performs $N$ uniformly-random accesses ($N$ is the size of the data set) and reports the average latency.
Finally, the algorithm decodes the entire data set and measures the decompression throughput.
All experiments run on a single thread in the main memory.
We repeat each experiment three times and report the average result for each measurement.

%---------------------------------------------------------------------------------
% Integer Benchmark
%---------------------------------------------------------------------------------
\subsection{Integer Benchmark}
\label{sec:intbench}

\begin{figure*}[t!]
    \centering
    \begin{subfigure}[b]{0.98\textwidth}%
        \center
        \includegraphics[width=\textwidth]{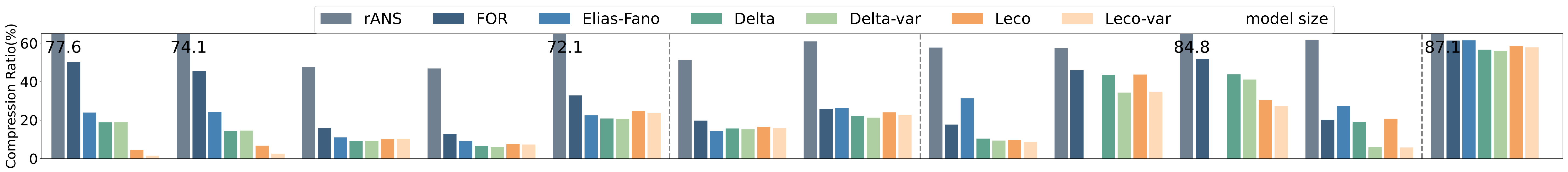}
        \vspace{-.2in}
        % \caption{Compression Ratio}
        \label{fig:CR_result}
    \end{subfigure}%
    
    \begin{subfigure}[b]{0.98\textwidth}%
        \center
        \includegraphics[width=\textwidth]{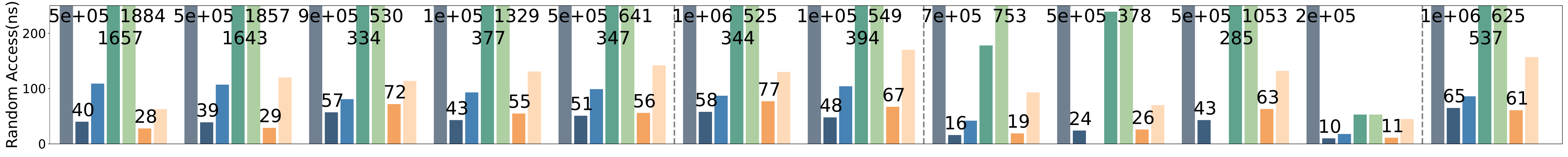}
        \vspace{-.2in}
        \label{fig:RA_lat}
    \end{subfigure}% 
    
    \begin{subfigure}[b]{0.98\textwidth}%
        \center
        \includegraphics[width=\textwidth]{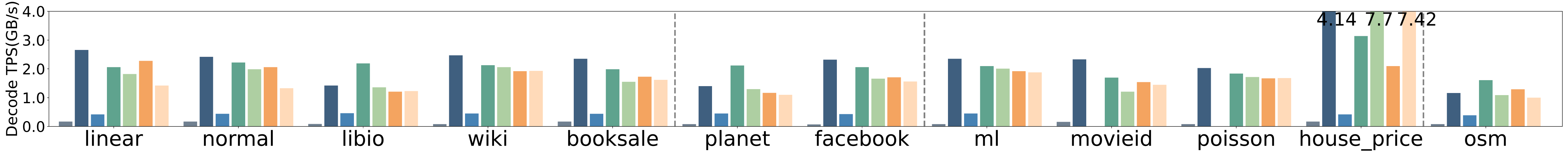}
        \vspace{-.2in}
        \label{fig:DT_result}
    \end{subfigure}% 
    \vspace{-0.1cm}
    \mycaption{Compression Microbenchmark}{Measurement of seven compression schemes on \integernum integer data sets from three aspects: Compression Ratio, Random Access Latency, and Full Decompression Throughput. We break down the compression ratio into model size (marked with the cross pattern) and delta size in the first row. The dashed lines split these data sets into four groups in the order of locally easy - globally easy, locally hard - globally easy, locally easy - globally hard, and locally hard - globally hard according to \cref{fig:hardness}. }
    \vspace{-0.2cm}
    \label{fig:all_results}
\end{figure*}

\cref{fig:all_results} shows the experiment results for compression ratio,
random access latency, and decompression throughput on the \integernum integer data sets.
\eliasfano does not apply to \texttt{poisson} and \texttt{movieid} because these
two data sets are not fully-sorted.

Overall, \leco achieves a Pareto improvement over the existing algorithms.
Compared to \eliasfano and \FORfix, the \leco variants obtain a significantly
better compression ratio while retaining a comparable decompression and
random access speed.
When compared to Delta Encoding, \leco remains competitive in the compression
ratio while outperforming the Delta variants by an order of magnitude in random access.

\subsubsection{\textbf{Compression Ratio}}
\label{sec:cr}
As shown in the first row of \cref{fig:all_results},
the compression ratios from the \leco variants are strictly better than
the corresponding ones from \FORfix.
This is because \FORfix is a special case of \leco:
the output of its Regressor is fixed to a horizontal line (refer to \cref{sec:motivation}).

We further plot the local hardness $\mathcal{H}_l$ and the global hardness $\mathcal{H}_g$
(defined in \cref{sec:hyper_tune}) of the different data sets in \cref{fig:hardness}.
The horizontal/vertical dashed line marks the average global/local hardness among the data sets.
we observe that \leco's compression-ratio advantage over \FORfix
is larger on locally-easy data sets ($40.9\%$ improvement on average)
than the three locally-hard data sets ($9.3\%$ improvement on average).
This is because local unevenness in the distribution makes it
difficult for a regression algorithm to fit well.

\leco also compresses better than \eliasfano
across (almost) all data sets.
Although \eliasfano is proved to be quasi-succinct,
it fails to leverage the embedded serial correlation between the values for further compression.
rANS remains the worst, which indicates that the redundancy embedded in
an integer sequence often comes more from the serial correlation rather than the entropy.

Compared to Delta Encoding, \leco
shows a remarkable improvement in compression ratio for ``smooth'' (synthetic) data sets:
\texttt{linear}, \texttt{normal}, and \texttt{poisson}.
For the remaining (real-world) data sets, however, \leco remains competitive.
This is because many real-world data sets exhibit local unevenness,
as shown in~\cref{fig:distribution}.
The degree of such irregularity is often at the same level as the difference
between adjacent values.

Another observation is that variable-length partitioning is effective
in reducing the compression ratio on real-world data sets that have
rapid slope changes or irregular value gaps
(e.g., \texttt{movieid}, \texttt{house\_price}).
Our variable-length partitioning algorithm proposed in \cref{sec:partitioner}
is able to detect those situations and create partitions accordingly
to avoid oversized partitions caused by unfriendly patterns to the Regressor.
We also notice that \lecovar achieves an additional $28.2\%$ compression
compared to \lecofix on the four locally-easy and globally-hard data sets,
while the improvement drops to $<10\%$ for the remaining data sets\footnote{Except for the ideal cases in \texttt{linear} and \texttt{normal}}.
This indicates that the two metrics used for the partitioning strategy advising
(refer to \cref{sec:hyper_tune}) is effective in identifying data sets that can
potentially benefit from variable-length partitions.

\subsubsection{\textbf{Random Access}}
\label{sec:ra}
The second row of \cref{fig:all_results} presents the average latency of
decoding a single value in memory for each compression scheme.
The random access speed of \lecofix is comparable to that of \FORfix
because they both require only two memory accesses per operation.
\FORfix is often considered the lower bound of the random access latency
for lightweight compression
because it involves minimal computation (i.e., an integer addition).
Compared to \FORfix, \lecofix requires an additional floating-point multiplication.
This overhead, however, is mostly offset by a better cache hit ratio
because \lecofix produces a smaller compressed sequence.

\lecovar is slower because it has to first search the metadata to determine the corresponding partition for a given position.
This index search takes an extra $35 - 90$ ns depending on the total number
of partitions.
The Delta variants are an order of magnitude slower than the others
in most data sets because they must decompress the entire partition
sequentially to perform a random access.

\subsubsection{\textbf{Full Decompression}}
\label{sec:fd}
The third row in \cref{fig:all_results} shows the throughput of each compression
algorithm for decompressing an entire data set.
In general, \lecofix is $14\% - 34\%$\footnote{except for \texttt{house\_price} where the enhancement of \FORfix over \lecofix is 49\% } slower than its fastest competitor
\FORfix because \lecofix involves an extra floating-point operation upon decoding each record.
\deltavar and \lecovar perform exceptionally well on \texttt{house\_price}.
The reason is that part of the data set contains sequences of repetitive values.
\leco's Partitioner would detect them and put them into the same segment, making the decompression task trivial for these partitions.

\begin{table}[!t]
\small
\begin{center}
\begin{tabular}{|c|c|c|c|c|c|}
\hline
\FORfix & \eliasfano & \deltafix & \deltavar & \lecofix & \lecovar\\ 
\hline
0.81$\pm$0.28 & 0.58$\pm$0.17 & 1.04$\pm$0.14 & 0.04$\pm$0.01 & 0.78$\pm$0.11 & 0.02$\pm$0.01\\
\hline 
\end{tabular}
\end{center}
\caption{Compression Throughput (GB/s).}
\label{tab:Compress_throughput}
\vspace{-0.8cm}
\end{table}

\subsubsection{\textbf{Compression throughput}}
\label{sec:ct}
\cref{tab:Compress_throughput} shows the compression throughput for each 
algorithm weighted averaged across all the \integernum data sets
with error bars.
\lecofix has a similar compression speed to the baselines because
our linear Regressor has a low computational overhead.
Algorithms that adopt variable-length partitioning (i.e., \deltavar and \lecovar),
however, are an order of magnitude slower
because the Partitioner needs to perform multiple scans through the data set
and invokes the Regressor (or an approximate function) frequently along the way.
Such a classic trade-off between compression ratio and throughput is often
beneficial to applications that do not allow in-place updates.

\subsection{Cases for higher-order models}
\label{sec:nonlinear}

\begin{figure}[t!]
\centering
\includegraphics[width=0.5\textwidth]{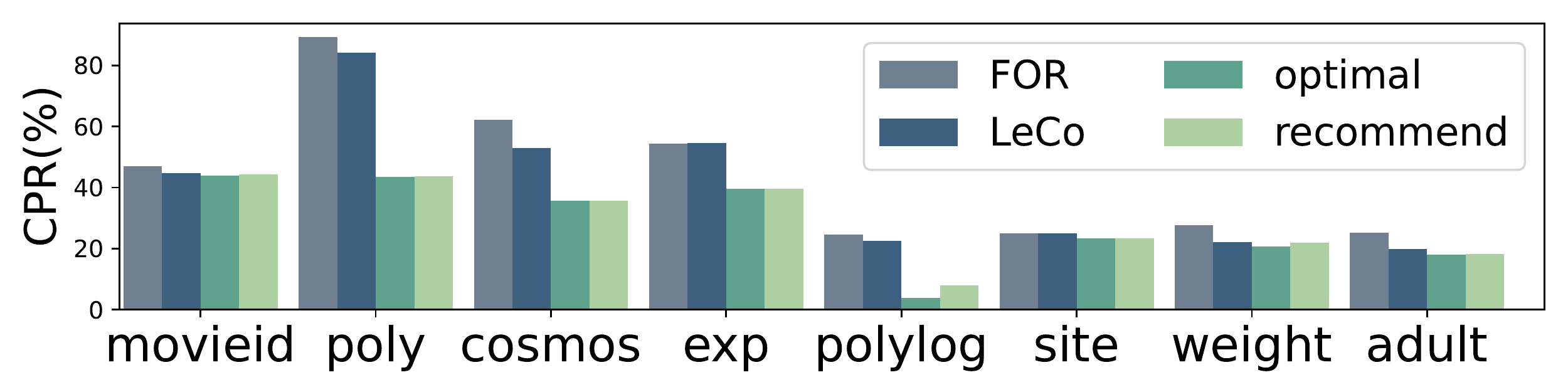}
\vspace{-0.6cm}
\caption{Regressor Selection Result.}
\vspace{-0.5cm}
\label{fig:codec_sel}
\end{figure}

\begin{figure}[t!]
\centering
\includegraphics[width=0.5\textwidth]{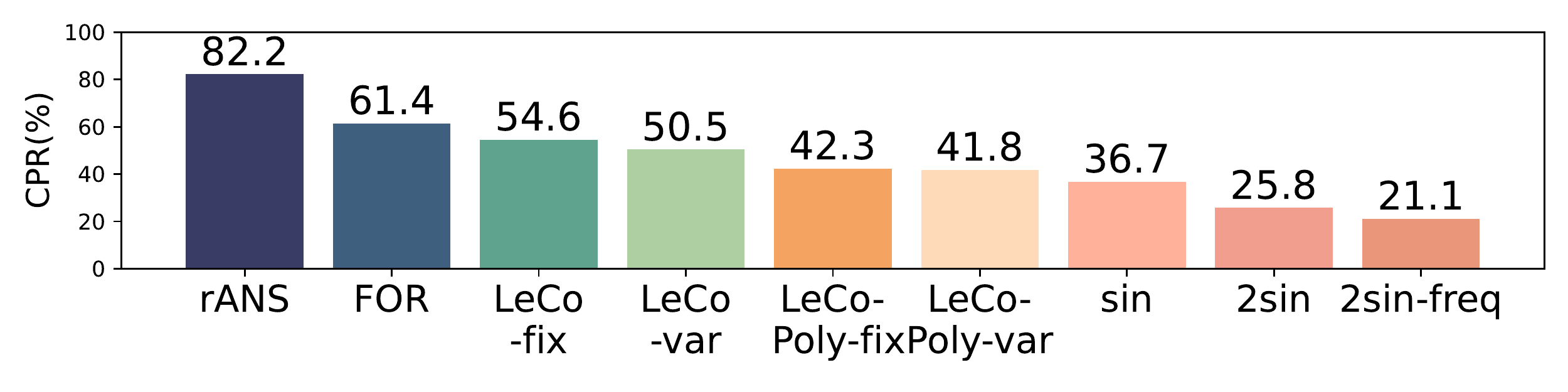}
\vspace{-0.75cm}
\caption{Compression ratio on \texttt{cosmos}.}
\vspace{-0.5cm}
\label{fig:poly_mix_sin}
\end{figure}

Although linear models perform sufficiently well in the above integer benchmark\footnote{Many data sets
in the integer benchmark come from the SOSD benchmark~\cite{2019SOSD}, which favors linear models.},
there are cases where higher-order models shine.
Because our setting is mostly read-only,
it is usually worthwhile to spend more computation to compress the data once
and then benefit from long-term space and query efficiency.

We first verify the effectiveness of our Regressor Selector in the Hyperparameter Advisor
(refer to \cref{sec:hyper_tune}). In this experiment, we consider the following six Regressor types:
constant (FOR), linear, polynomial up to a degree of three, exponential, and logarithm.
We create synthetic data sets (with random noise) for each Regressor type and extract the
features introduced in \cref{sec:hyper_tune} to train the classification model (i.e., CART) offline.

We compare the compression ratios obtained by using our recommended
Regressor per partition (labeled \texttt{recommend}) to those obtained by \FORfix, \lecofix,
and the \texttt{optimal} (i.e., exhaustively search in the candidate Regressor types
and pick the one with the best compression ratio).
\cref{fig:codec_sel} shows the results.
Note that none of the eight tested data sets were used for training.
We observe that \texttt{recommend} achieves a compression ratio close to the \texttt{optimal},
with up to $64.7\%$ improvement over \lecofix (with linear regression only) on data sets that
exhibit higher-order patterns.
For data sets that are mostly linear (e.g. \texttt{movieid}),
the benefit of applying higher-order models is limited, as expected.

One can even extend the \leco framework to leverage domain knowledge easily.
For example, the \texttt{cosmos} data set contains a mixture of two signals (i.e., sine function)
with random noise.
As shown in \cref{fig:poly_mix_sin}, if we include a sine term in the Regressor (labeled \texttt{sin}),
we are able to achieve a better compression ratio ($36.7\%$) compared to the recommended polynomial model ($42.3\%$).
If we include two sine terms (labeled \texttt{2sin}), we are able to extract an additional
$29.7\%$ compression out of the \leco framework compared to \texttt{sin}.
If we further know the approximate frequencies of the two sine terms (labeled \texttt{2sin-freq}),
\leco produces an even better compression ratio, as presented in \cref{fig:poly_mix_sin}.

\subsection{Compressing Dictionaries}
\label{sec:compress_dict}
Building dictionaries that preserve the key ordering is a common technique
to achieve compression and speed up query processing~\cite{zhang2020order,liu2019mostly,binnig2009dictionary}.
Reducing the memory footprint of such dictionaries is an important use case of \leco.
In the following experiment, we perform a hash join with the probe side being dictionary encoded.
Specifically, we use the \texttt{medicare} dataset as the probe-side column,
and we pre-build a hash table of size 84MB in memory, which contains 50\% of the unique values
(i.e., 50\% hash table hit ratio during the join).
The probe side first goes through a filter of selectivity of 1\% and then probes the hash table for the join.
The probe-side values are encoded using an order-preserving dictionary
compressed by \leco (i.e., \lecofix), \FORfix, and Raw (i.e., no compression).
We vary the memory budget from 3GB to 500MB and report the throughput
(defined as the raw data size of the probe side divided by the query execution time)
of executing this query.

\cref{fig:leco_dict} shows that
applying \leco improves the throughput up to $95.7\times$
compared to \FORfix when the memory budget for this query is limited.
This is because \leco compresses the probe-side dictionary from 2.4GB to 5.5MB
(cpr ratio = $0.23\%$) so that it constantly fits in memory.
For comparison, the dictionary size compressed using \FORfix is still 400MB
(cpr ratio = $17\%$). When the available memory is limited,
this larger dictionary causes a significant number of buffer pool misses,
thus hurting the overall query performance.

\subsection{Multi-Column Benchmark}
\label{sec:eval-multi-column}
\begin{figure*}[t!]
    \centering
    \begin{subfigure}[b]{0.98\textwidth}%
        \center
        \includegraphics[width=\textwidth]{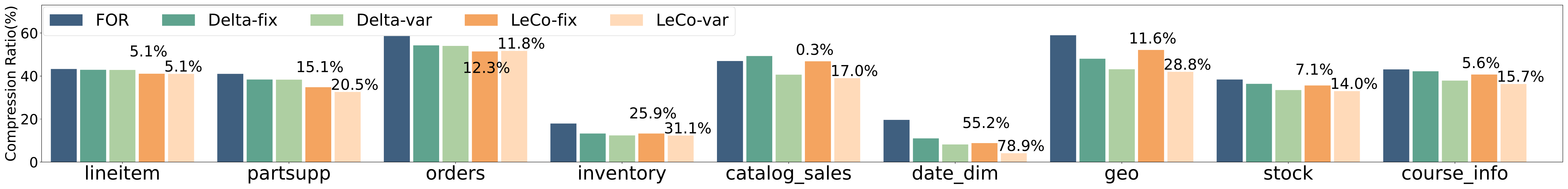}
    \end{subfigure}%
    
    \begin{subfigure}[b]{0.98\textwidth}%
        \center
        \includegraphics[width=\textwidth]{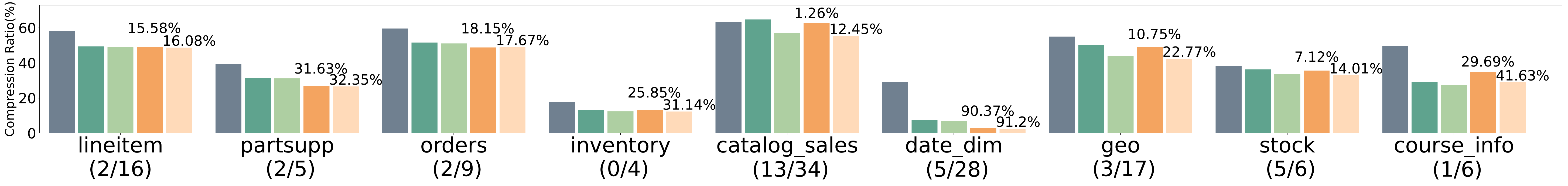}
    \end{subfigure}% 

    \begin{subfigure}[b]{0.62\textwidth}%
        \center
        \includegraphics[width=\textwidth]{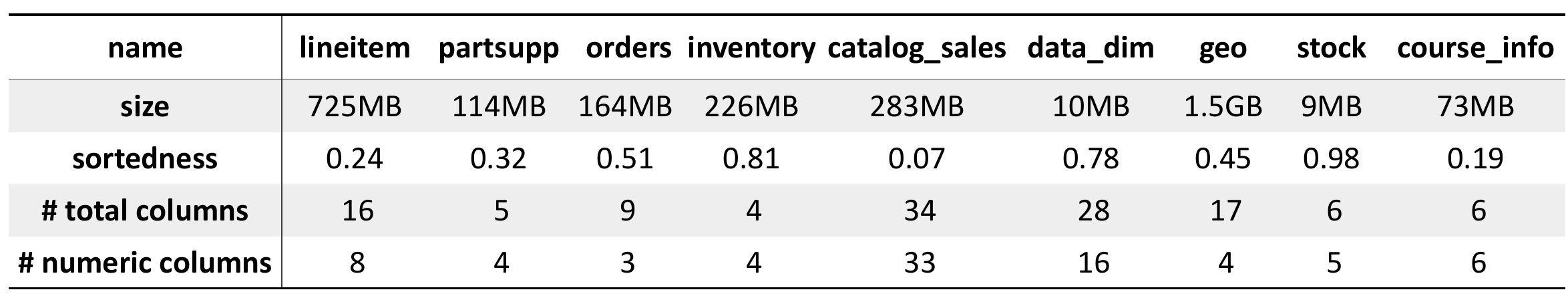}
        \vspace{-.2in}
    \end{subfigure}% 
    \vspace{-.1in}
    \mycaption{Multiple Column}{Compression ratio of five methods on nine tabular data sets.
    The second row of the result only considers columns with cardinality $\geq 10\%$.
    We report the size in bytes, average sortedness (in the range $[0,1]$), total column number, and integer/numerical column number of each table. 
    We mark the enhancement ratio of \leco variants over \FORfix above the bars.}
    \label{fig:mulcol_results}
    \vspace{-0.3cm}
\end{figure*}

In this section, we evaluate the effectiveness of \leco on nine multi-column
tabular data sets\footnote{Elias-Fano is not included as a baseline because most columns are not strictly sorted.}.
As shown in ~\cref{fig:mulcol_results} (bottom right), we compute the ``sortedness'' of a table (in the range [0, 1])
by averaging the sortedness of each column 
using the portion of inverse pairs~\cite{Borroni2013} as the metric.

From~\cref{fig:mulcol_results} (the top row), we observe that \leco
achieves a better compression ratio than \FORfix in all nine tables.
This is because columns in a table are often correlated~\cite{raman2006wring, gao2016squish, ilkhechi2020deepsqueeze}.
Our ``sortedness'' metric indicates that non-primary-key columns have
different degrees of correlation with the primary-key (i.e., sorting) column
across tables, thus partially inheriting the serial patterns.
Tables with high sortedness such as \texttt{inventory} and \texttt{data\_dim}
are more likely to achieve better compression ratios with the \leco variants.

The bottom left of \cref{fig:mulcol_results} presents the compression ratios
of the TPC-H tables\footnote{Due to space limitations, we only present the results of TPC-H.
Results of the other six data sets can be found in our technique report at \cite{techrep}.}
with high-cardinality columns only (i.e., NDV > 10\% \#row).
\leco's has a more noticeable advantage over \FORfix on columns that are likely to
select \FORfix as the compression method.

\subsection{String Benchmark}
\label{sec:stringbench}

\begin{figure*}[t!]
\centering
\hspace{-0.5cm}
\begin{minipage}[t]{0.242\textwidth}
\centering
    \includegraphics[scale=0.357]{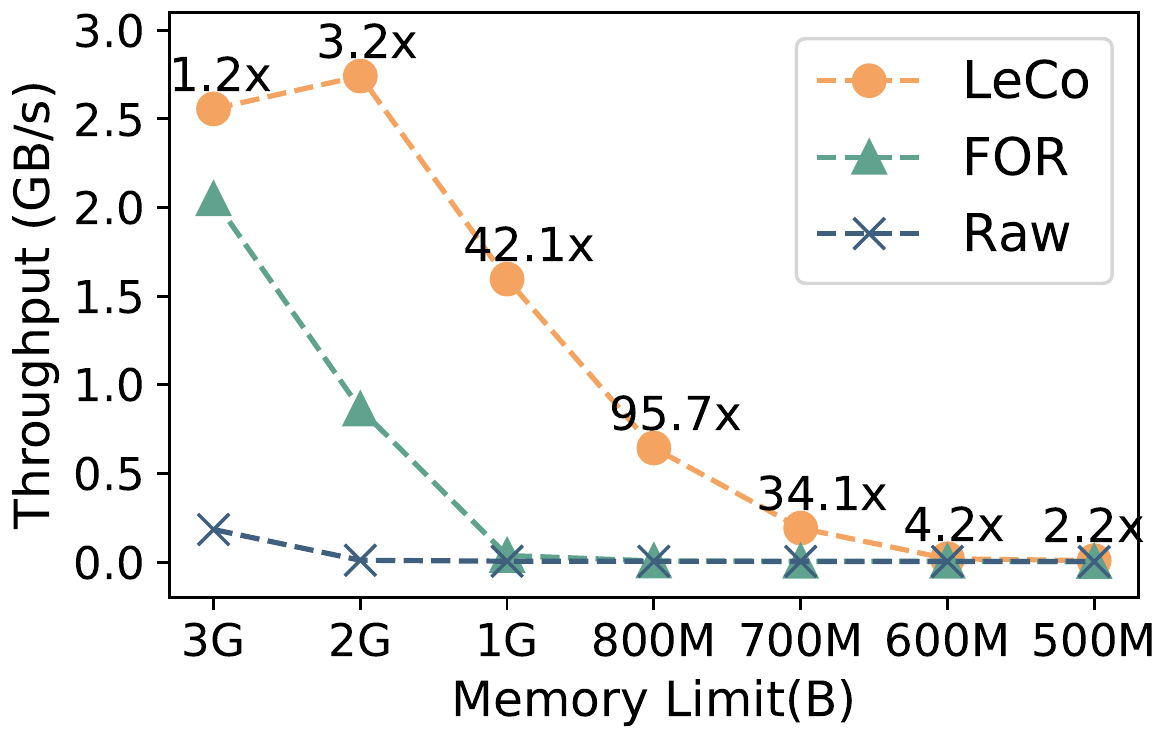}
\vspace{-.28in}
\caption{Hash Probe TPS.}
\label{fig:leco_dict}
\end{minipage}
\begin{minipage}[t]{0.24\textwidth}
    \includegraphics[scale=0.352]{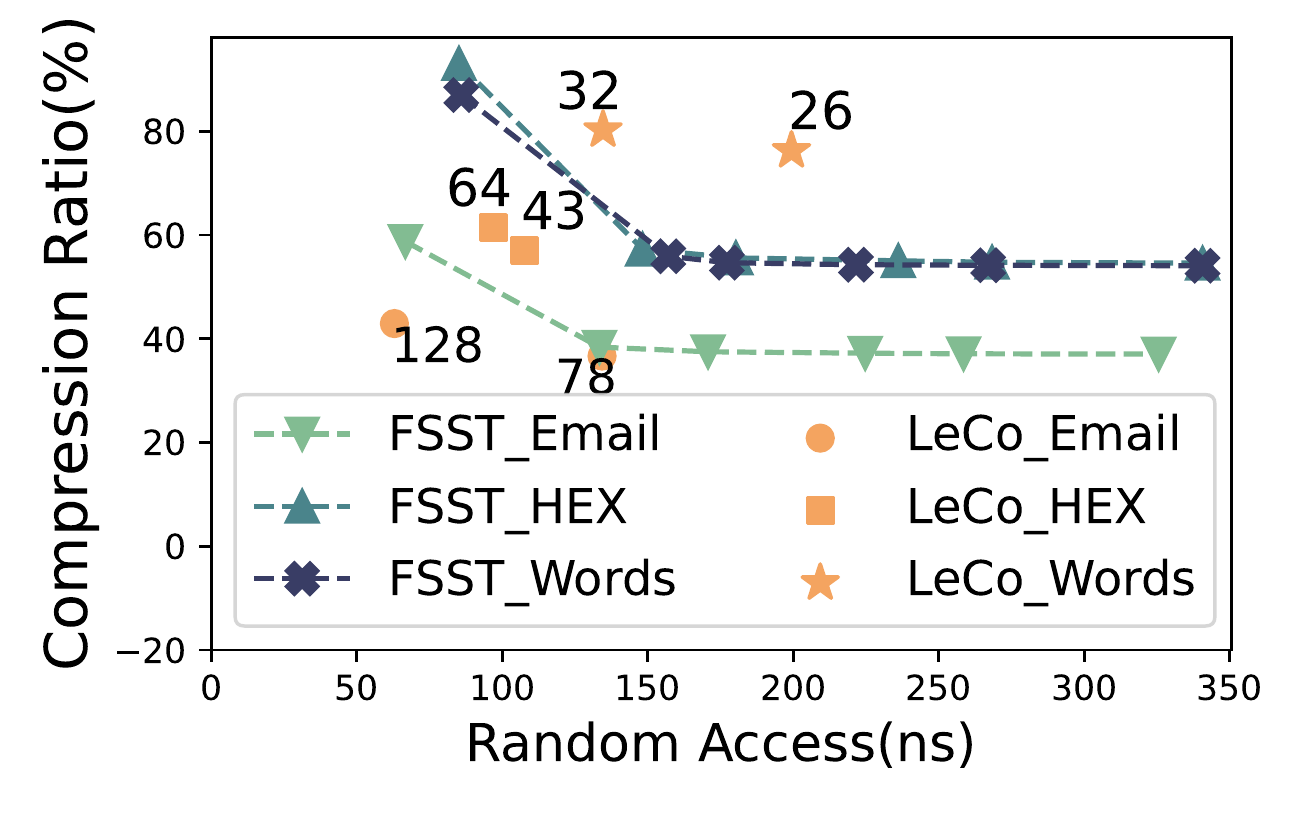}
	\vspace{-.28in}
\caption{String Evaluation.}
\label{fig:string_result}
\end{minipage}
\hspace{0.285cm}
\begin{minipage}[t]{0.24\textwidth}
\centering
    \includegraphics[scale=0.34]{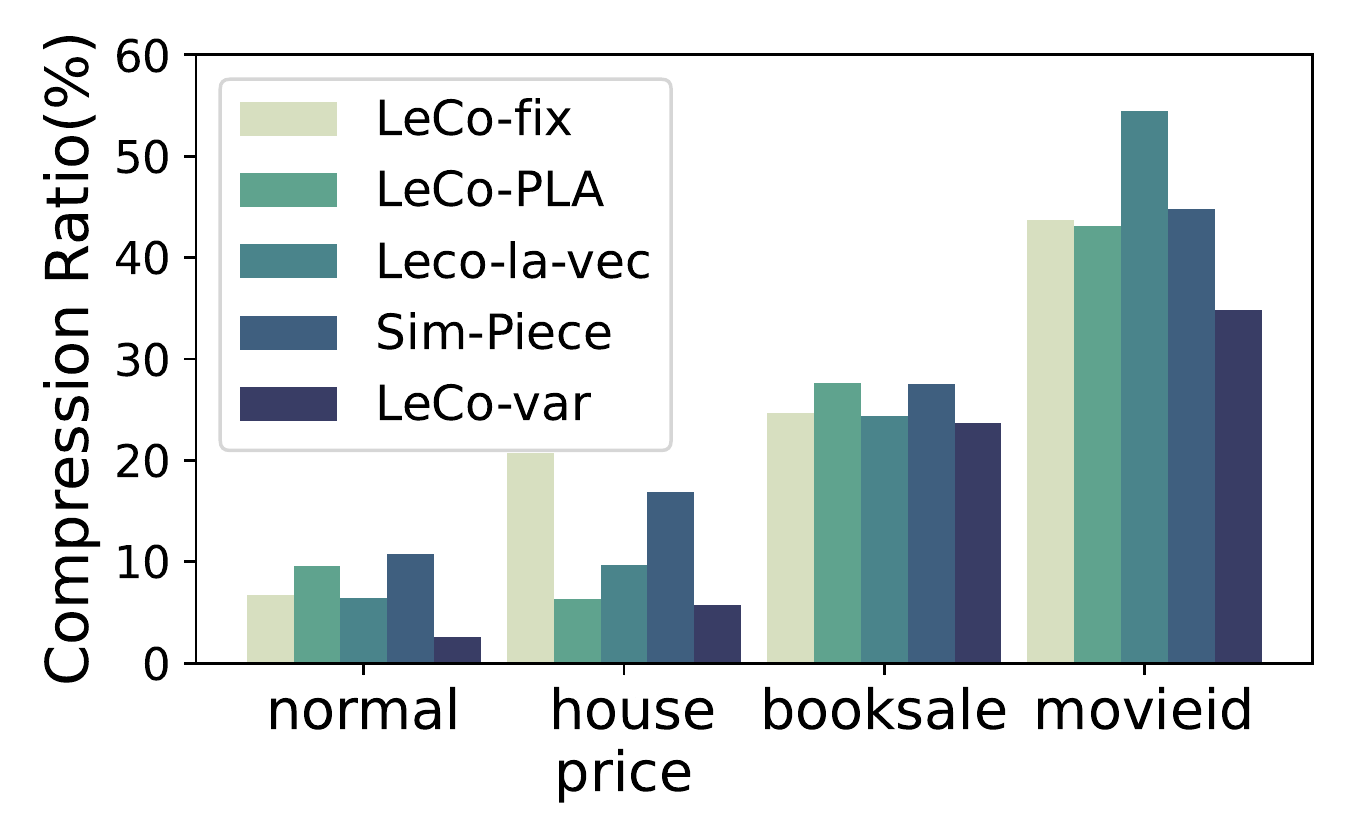}
    \vspace{-.28in}
\caption{Partition efficiency.}
\label{fig:leco_partitioning}
\end{minipage}
\hspace{0.285cm}
\begin{minipage}[t]{0.24\textwidth}
\centering
    \includegraphics[scale=0.41]{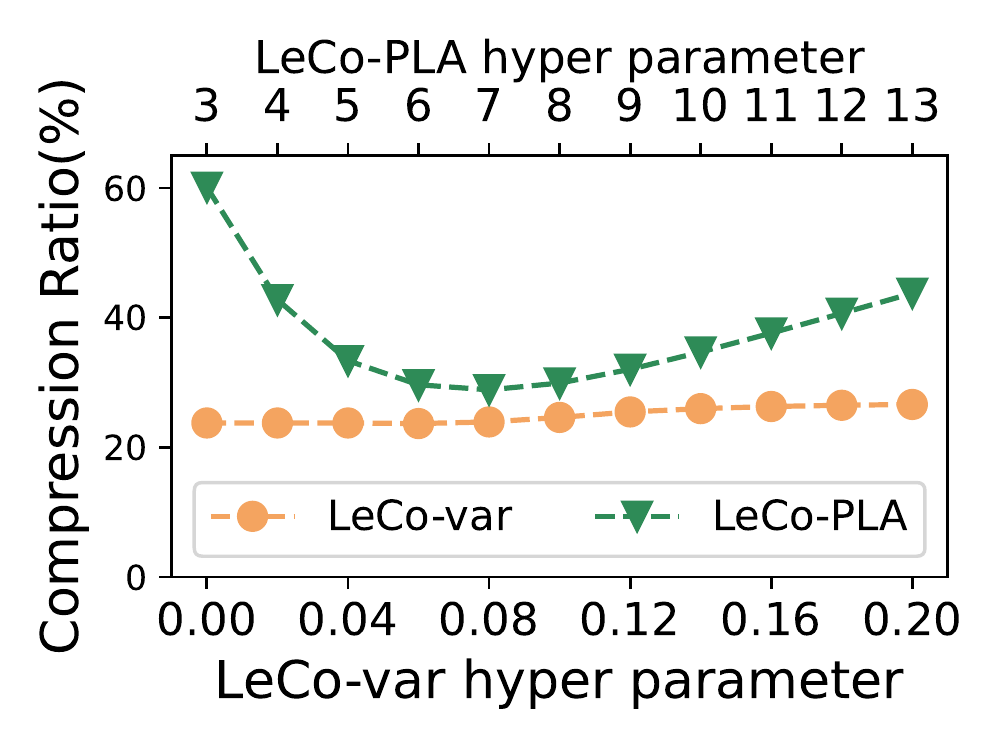}
    \vspace{-.14in}
\caption{Robustness test.}
\label{fig:partition_hyper_search}
\end{minipage}
\hspace{0.285cm}
\vspace{-0.4cm}
\end{figure*}

We compare \leco (i.e., \lecofix) against the state-of-the-art
lightweight string compression algorithm \fsst~\cite{boncz2020fsst}
using three string data sets \texttt{email}, \texttt{hex} and \texttt{words}. 
\fsst adopts a dictionary-based approach by building a fine-grained
static symbol table to map a partial string to a 1-byte code.
Because each compressed string has a variable length, \fsst must store
a byte-offset array to support random access.
An optimization (not mentioned in the \fsst paper) is to
delta-encode this offset array to trade its random access speed for
a better compression ratio.
To perform a fair comparison, we tested six different block sizes
of the delta encoding: 0 (i.e., no delta compression), 20, 40, 60,
80, and 100.
For \leco, we present two data points with different character-set sizes.

\cref{fig:string_result} shows the random access latencies and compression ratios
for different algorithm configurations.
Each \leco point is marked with the base value used to convert strings.
We observed that \leco's string extension provides a higher random access speed
while retaining a competitive compression ratio, compared to \fsst
on both \texttt{email} and \texttt{hex} data sets.
The compression ratio of \leco, however,
is slightly worse than that of \fsst on \texttt{word}.
This is because dictionary-based algorithms are more suitable for human-readable strings
that contain repeating patterns such as common prefixes, roots, and suffixes,
while learned compression is better at leveraging serial patterns between values.

\subsection{Partitioner Efficiency}
\label{sec:partition_compare}
In this section, we compare \leco's default Partitioner (as described in \cref{sec:partitioner})
to state-of-the-art partitioning algorithms, including the PLA algorithm adopted by time-series compression~\cite{luo2015piecewise},
as well as FITing tree~\cite{galakatos2019fiting},
Sim-Piece introduced in~\cite{kitsiossim}, and the la\_vector algorithm proposed in~\cite{boffa2021learned}.
The angle-based PLA predefines a fixed global prediction error bound ($\epsilon$) and determines the partition boundaries greedily in one pass.
Sim-Piece adopts the angle-based PLA as its partitioner and compactly stores linear models with the same intercept together to reduce the overall space.
They sacrifice model parameter precisions to create more segments with the same intercept.
On the other hand, la\_vector translates each data point $v_i$ into a vertex $i$,
where the weight of edge $(i,j)$ is defined as the compression ratio of segment $[v_i, v_j]$.
The optimal partitioning problem is thus converted into finding the shortest path in the above graph $\mathcal{G}$.
la\_vector approximates $\mathcal{G}$ with $\mathcal{G}'$ with fewer edges
and proofs that the best compression ratio achieved on $\mathcal{G}'$ is at most $k\cdot l$ larger than that on $\mathcal{G}$
where $k$ is a constant and $l$ is the shortest path length.

We integrated PLA, Sim-Piece, and la\_vector into the \leco framework (with the linear Regressor denoted by \lecoPLA, Sim-Piece and \lecolavec, respectively)
and repeated the experiments in \cref{sec:intbench} on four representative data sets.
As shown in \cref{fig:leco_partitioning}, all three candidate methods exhibit significantly worse
compression ratios compared to \lecovar.
The globally-fixed error bound in \lecoPLA fails to adapt to data segments with rapidly changing slopes.
We also found that \lecoPLA is more sensitive to its hyperparameter compared to \lecovar,
as shown in~\cref{fig:partition_hyper_search} where we sweep the hyperparameters for
\lecoPLA ($\epsilon$) and \lecovar ($\tau$) on the \texttt{books} data set.
The model compaction in Sim-Piece doesn't take effect because, on mostly sorted data sets, the intercept of each linear model is also increasing.
The precision sacrifice in their implementation results in an even worse compression ratio on \texttt{house\_price} compared to \lecoPLA.
For \lecolavec, although it finds the shortest path in the approximate ``compression-ratio graph'',
it overlooked the length of the shortest path, resulting in an excessive number of models
that dominate the compressed size on data sets such as \texttt{movieid}. 
\vspace{-0.1cm}

\section{System Evaluation}
\label{sec:sys-eval}
To show how \leco can benefit real-world systems, we integrated \leco into
two system applications:
(1) a columnar execution engine implemented using \arrow~\cite{arrow} and \parquet~\cite{parquet}
and (2) \rocksdb~\cite{Rocksdgithub}.
All experiments are conducted on a machine with 4$\times$ Intel\textregistered
Xeon\textregistered~(Cascade Lake) Platinum 8269CY CPU @ 2.50GHz, 32GB DRAM,
and a local NVMe SSD of 447GB with 250k maximum read IOPS.
We use Apache Arrow 8.0.0, \parquet version 2.6.0, and \rocksdb Release version 6.26.1
in the following experiments.

\vspace{-0.2cm}
\subsection{Integration to Arrow and \parquet}
\label{sec:parquet}
We first integrated \leco (as well as FOR and Delta for comparison)
into Apache \arrow (the most widely-used columnar \emph{in-memory} format)
and Apache \parquet (the most widely-used columnar \emph{storage} format),
and built an execution engine prototype using their C++ libraries to demonstrate how
\leco can benefit query processing.

\parquet uses dictionary encoding as the default compression method.
It falls back to plain encoding if the dictionary grows too large.
We refer to this mechanism as \texttt{Default}.
In the following experiments, we set \parquet's row group size to 10M rows
and disable block compression unless specified otherwise.

The primary component of the \arrow format is the \arrow Array
that represents a sequence of values of the same type.
Except for basic dictionary encoding, no compression is applied to \arrow arrays
to guarantee maximum query-processing performance.
We re-implemented the \arrow Array structure using lightweight compression methods
(i.e., \leco, FOR, and Delta) without changing its interface.
We use a consistent lightweight-compressed format for the \arrow Array and
\parquet Column Chunk so that no additional decoding is required when scanning
the data from disk to memory.

The \arrow Compute library implements various basic database operators
(e.g., Take, Filter, GroupBy) on \arrow arrays as compute functions.
Our execution engine uses these compute functions as building blocks.
The engine is implemented using late materialization~\cite{DBLP:conf/cidr/BonczZN05}
where intermediate results are passed between operators as position bitmaps.
We also push down the filters to the storage layer (i.e., \parquet).

\subsubsection{Filter-Groupby-Aggregation}
\label{sec:parquet_filterscan}

\begin{figure}[t!]
\centering
\includegraphics[width=\linewidth]{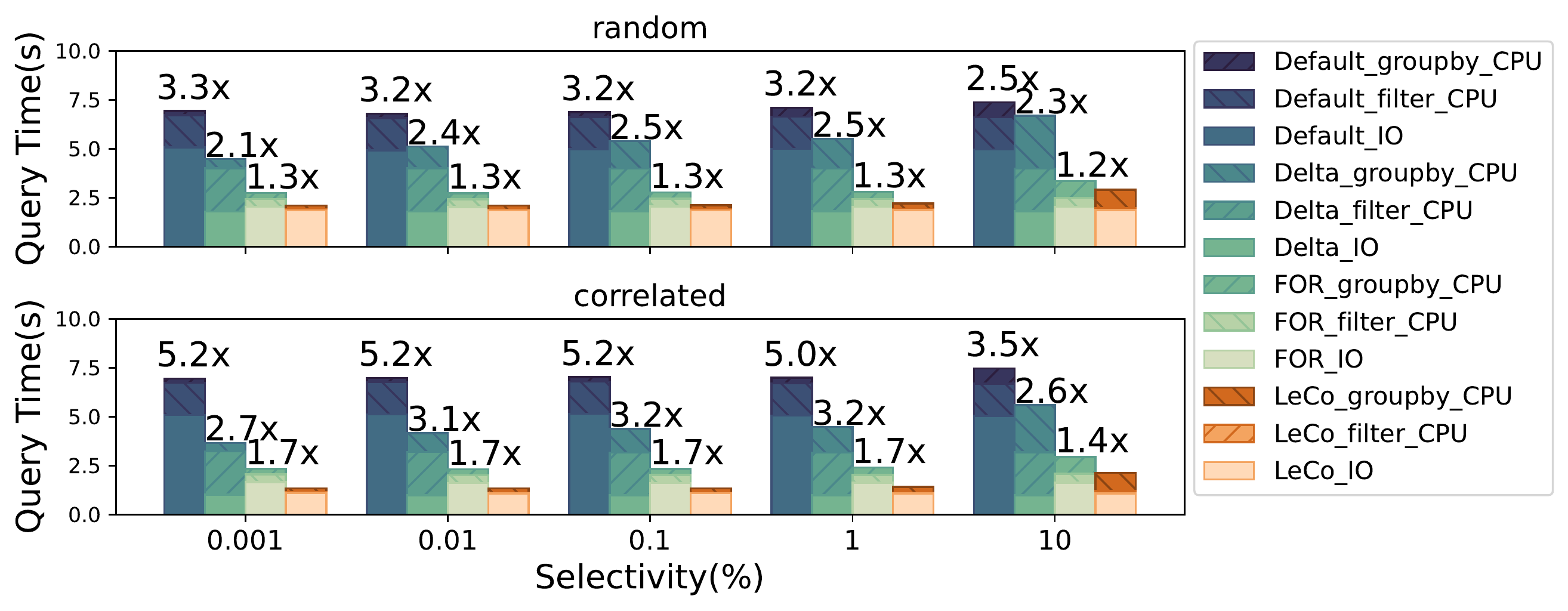}
\vspace{-0.8cm}
\caption{Filter Groupby Aggregation}
\vspace{-0.5cm}
\label{fig:filter_aggsum}
\end{figure}

We create a query template of a typical filter-groupby-aggregation as follows.
Suppose we have 10k sensors recording measurements.
The table \texttt{T} has three columns:
(1) \texttt{ts}, timestamps (in seconds, almost sorted) extracted from the \texttt{ml}~\cite{mltimestamp} data set,
(2) \texttt{id}, 16-bit sensor IDs ranging from 1 to 10k, and
(3) \texttt{val}, 64-bit-integer sensor readings.
To vary the compressibility of the table, we generate two different data distributions
for the \texttt{id} and the \texttt{val} columns:
(1) \texttt{random}: both \texttt{id} and \texttt{val} are randomly generated
and are difficult to compress no matter which algorithm, and
(2) \texttt{correlated}: \texttt{id}s are clustered in groups of 100,
and \texttt{val}s are monotonically increasing across groups
(but random within a group). There are serial patterns in this setting
for lightweight compression algorithms to leverage.

We construct the following query that outputs the average reading
for each sensor within a given time range per day:
\lstinline[language=SQL]{SELECT AVG(val) FROM T WHERE ts\_begin < ts \% val\_2 < ts\_end GROUP BY id}.
We adjust the time range (i.e., \texttt{ts\_end} - \texttt{ts\_begin})
to control the query's selectivity.
When executing this query, our execution engine first pushes down the
filter predicate to \parquet, which outputs a bitmap representing the filtering results.
The engine then scans the \texttt{id} and the \texttt{val} column from \parquet
into \arrow arrays and performs the groupby-aggregation.
Both groupby and aggregation only decode entries that are still valid
according to the filter-bitmap, which involves random accesses to the corresponding \arrow arrays.

We generated four \parquet files with Default, Delta, \FORfix, and \leco
as the encoding algorithms (with a partition size of 10k entries).
In the case of \texttt{random} distribution, 
the resulting file sizes are 3.8GB, 1.3GB, 1.5GB, and 1.4GB, respectively.
For the \texttt{correlated} distribution,
the corresponding file sizes are 3.8GB, 706MB, 1.2GB, and 785MB
(with better compression ratios).
We execute the above query template and repeat each query instance
three times with its average execution time reported.

As shown in \cref{fig:filter_aggsum},
all three lightweight compression algorithms outperform the Default
because of the significant I/O savings proportional to the file size reduction.
Compared to Delta, \leco is much more CPU-efficient because Delta requires
to decode the entire partition to random-access particular entries during
the groupby-aggregation.
Compared to \FORfix, \leco mainly gains its advantage through the I/O reduction
due to a better compression ratio.
This I/O advantage becomes larger with a more compressible data set (i.e., \texttt{correlated}).

Interestingly, \leco is up to $10.5\times$ faster than \FORfix
when performing the filter operation.
Suppose that the model of a partition is $\theta_0 + \theta_1 \cdot i$,
and the bit-length of the delta array is $b$.
For a less-than predicate $v < \alpha$, for example,
once \leco decodes the partition up to position $k$, where
$\theta_0 + \theta_1 \cdot k - 2^{b-1} > \alpha$ (assume $\theta_1 \geq 0$),
we can safely skip the values in the rest of the sequence because
they are guaranteed to be out of range.
\FORfix cannot perform such a computation pruning because the \texttt{ts}
column is not \textit{strictly} sorted.

\begin{figure}[t!]
    \begin{subfigure}[b]{0.30\textwidth}%
        \hspace{-.3in}
        % \center
        \includegraphics[width=0.9\linewidth]{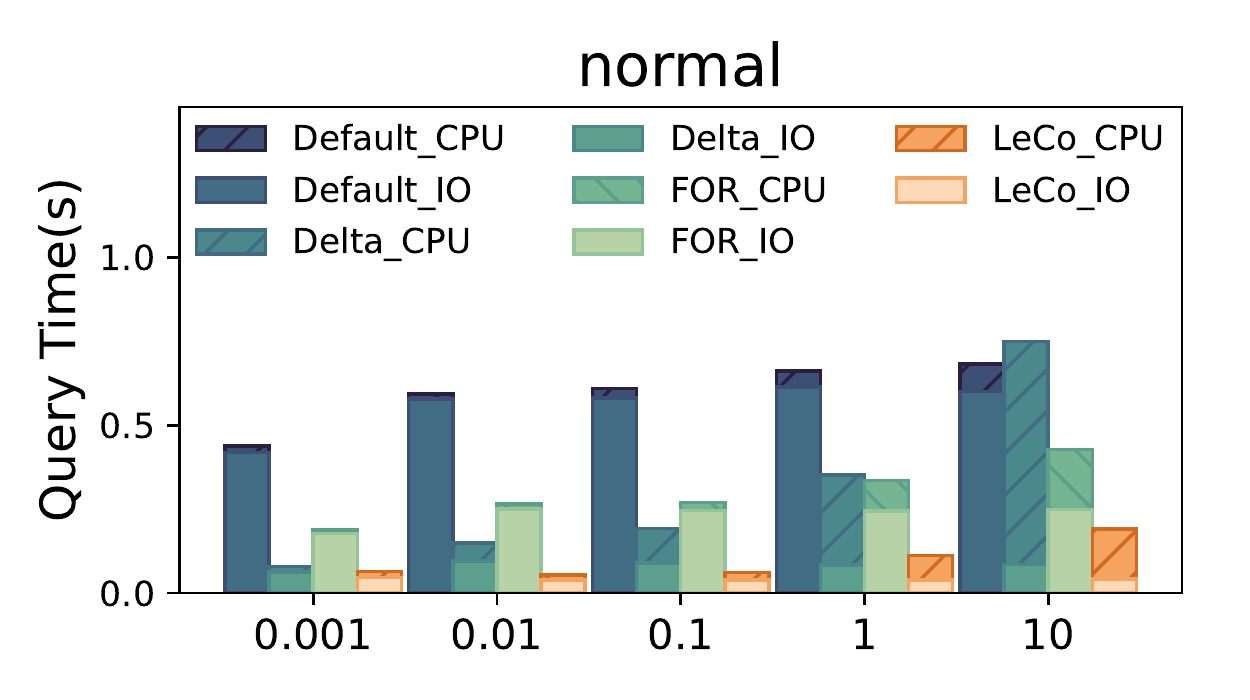}
        \vspace{-.2in}
        \label{fig:parquet_normal}
    \end{subfigure}%
    \begin{subfigure}[b]{0.30\textwidth}%
        % \center
        \hspace{-.54in}
        \includegraphics[width=0.9\linewidth]{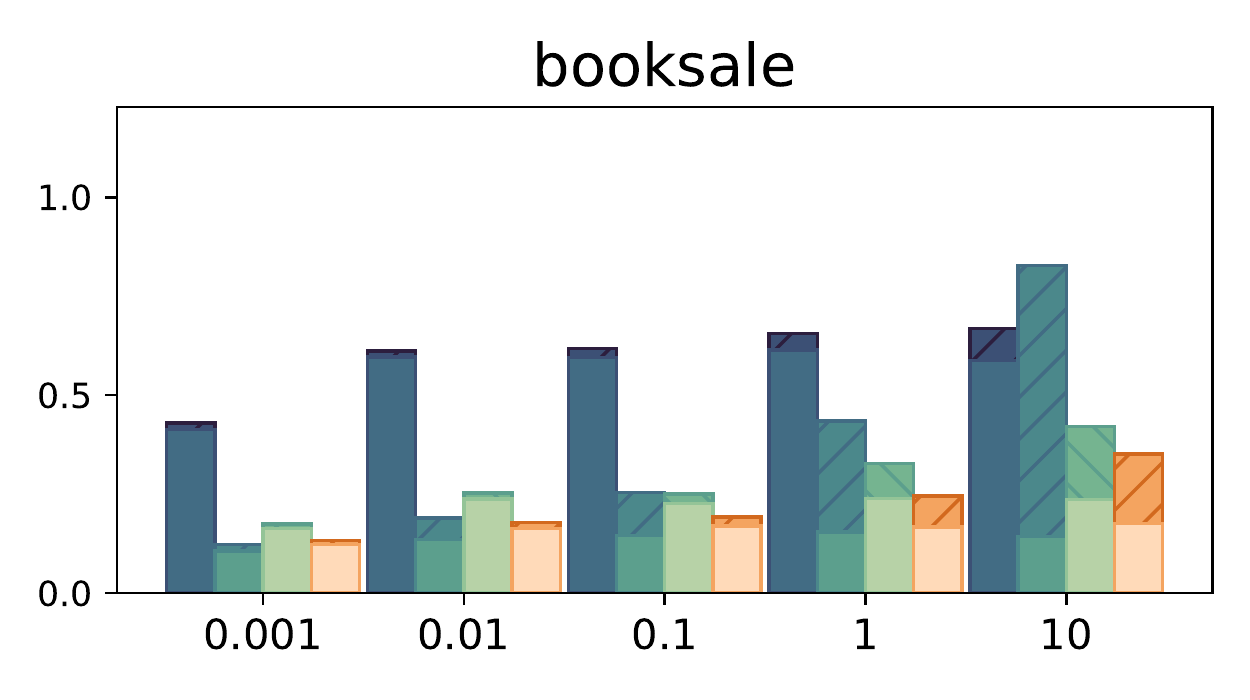}
        \vspace{-.2in}
        \label{fig:parquet_books}
    \end{subfigure}% 
    \vspace{.1in}
    \begin{subfigure}[b]{0.30\textwidth}%
        \hspace{-.3in}
        % \center
        \includegraphics[width=0.9\linewidth]{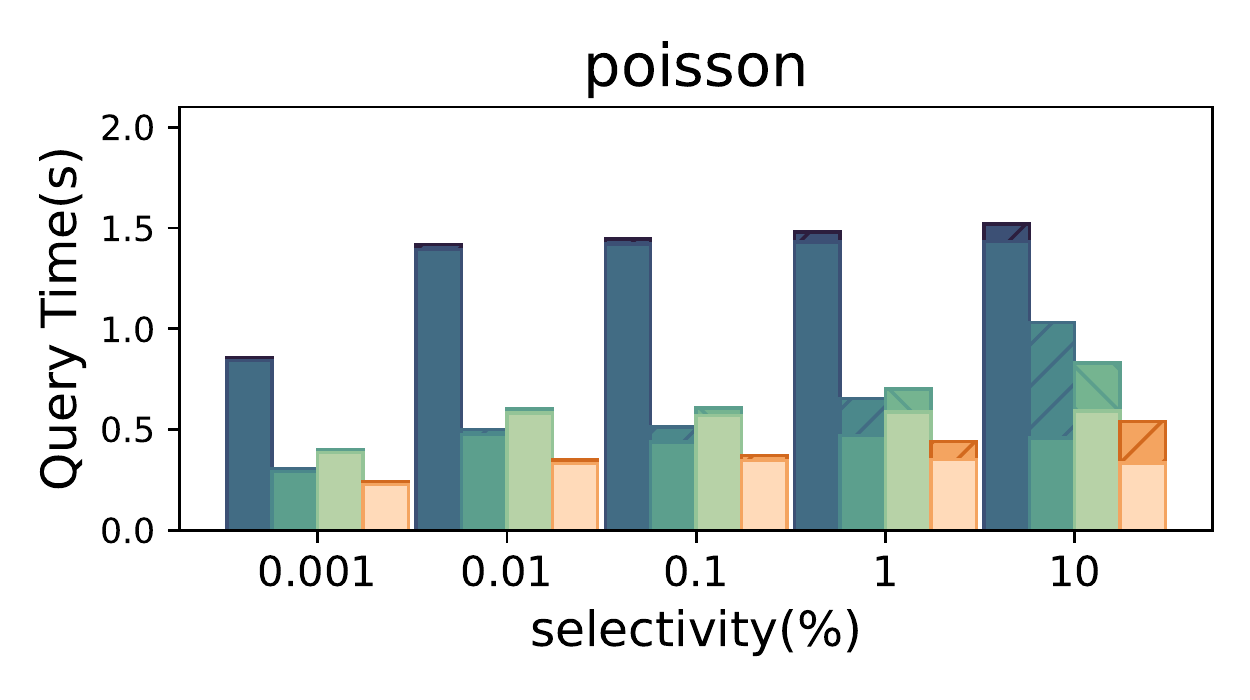}
        \vspace{-.2in}
        \label{fig:parquet_poisson}    
    \end{subfigure}% 
    \begin{subfigure}[b]{0.30\textwidth}%
        \hspace{-.54in}
        % \center
        \includegraphics[width=0.9\linewidth]{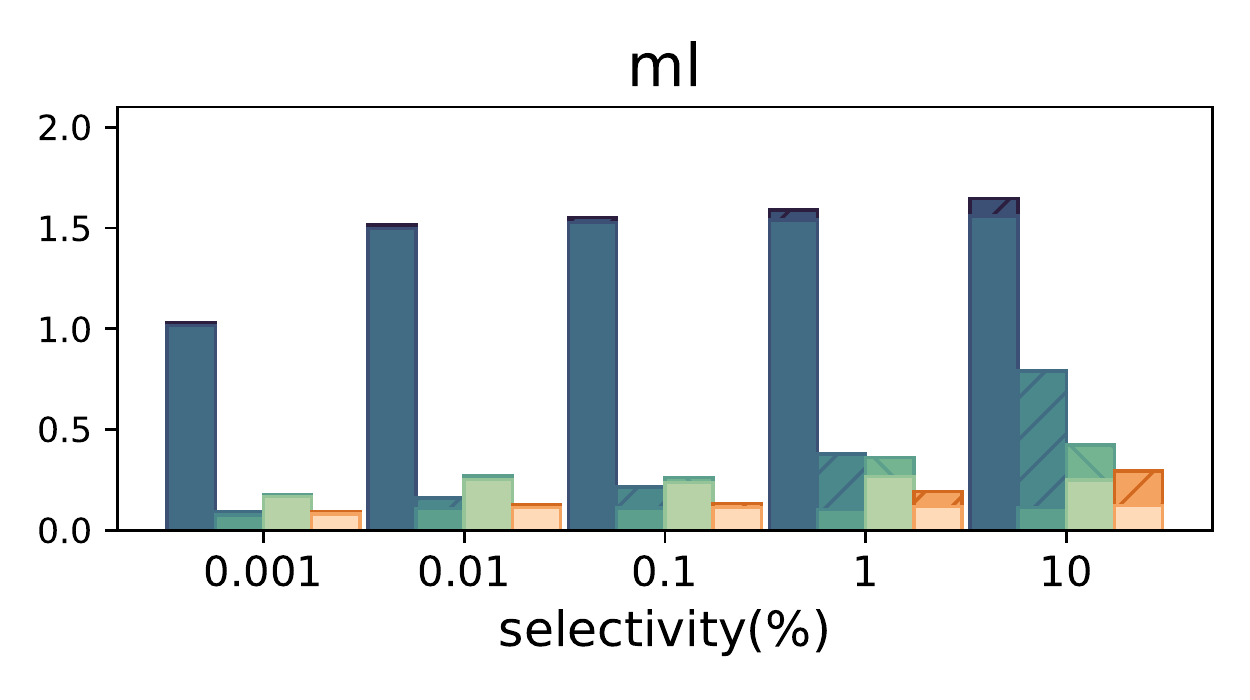}
        \vspace{-.2in}
        \label{fig:parquet_ml}
    \end{subfigure}%
    % \vspace{-0.2cm}
    \caption{Bitmap Aggregation.}
    \label{fig:parquet_agg}
    \vspace{-0.3cm}
\end{figure}

\subsubsection{Bitmap Aggregation}
\label{sec:parquet_bitmap_agg}
In this experiment, we zoom in on the critical bitmap aggregation 
operation of the above end-to-end query and further verify \leco's performance and
space benefits on four different data sets introduced in \cref{sec:dataset}:
\texttt{normal}, \texttt{poisson}, \texttt{booksale},
and \texttt{ml}\footnote{we scale \texttt{ml} to 200M rows while preserving its value distribution.}.
For each data set, we create four \parquet files with different
lightweight compression algorithms (i.e., Default, Delta, \FORfix, and \leco) enabled as above.
The bitmaps used in the experiments include ten set-bit clusters following a Zipf-like distribution
with a varying ratio of ``ones'' (to represent different filter selectivities).
Data is scanned directly into \arrow arrays in a row-group granularity, where a row-group
is skipped if the bits in the corresponding area in the bitmap are all zeros.
We then feed the arrays and the bitmap to the \arrow Compute function to perform the summation.

As shown in \cref{fig:parquet_agg}, \leco consistently outperforms Default (by up to $11.8\times$),
Delta (by up to $3.9\times$), and \FORfix (by up to $5.0\times$).
\leco's speedup comes from both the I/O reduction (due to a better compression ratio)
and the CPU saving (due to fast random access and better caching).
Moreover, we found that \leco consumes less memory during the execution.
The peak memory usage (for processing a \parquet row group) of \leco is
$60.5\%$, $35.3\%$, and $10.0\%$ less compared to Default, \FORfix, and Delta, respectively on average.
This is much preferred for systems with constrained memory budgets.

\subsubsection{Enabling Block Compression}
\label{sec:parquet_blk_cpr}

\begin{figure}[t!]
\centering
\includegraphics[width=0.9\linewidth]{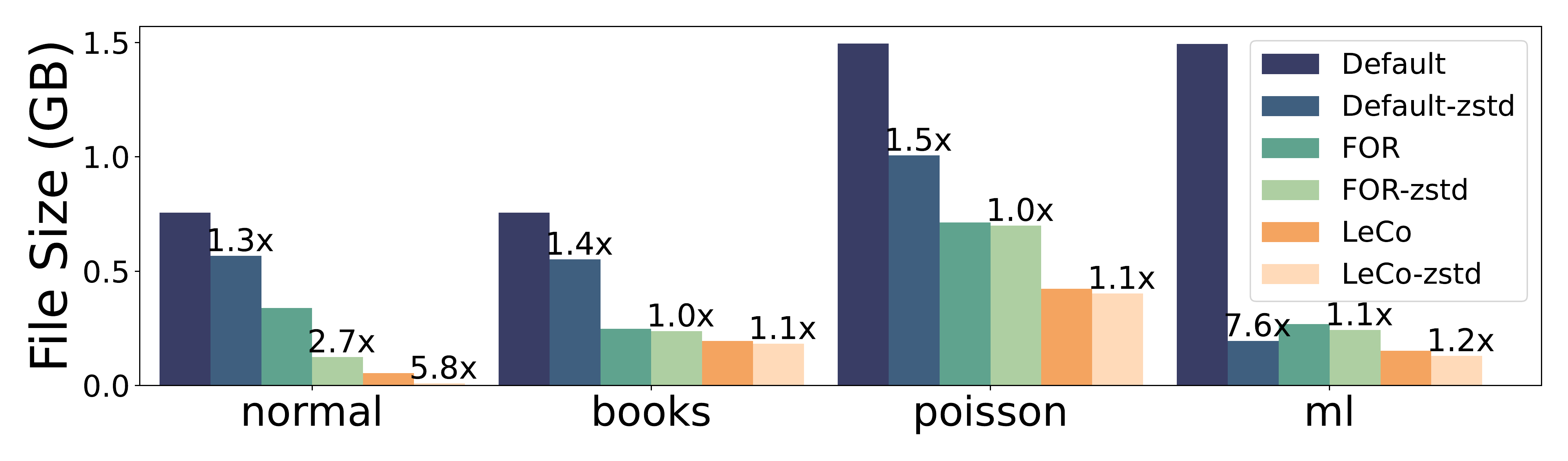}
\vspace{-0.4cm}
\mycaption{Parquet With zstd Compression}{Numbers on bars indicating additional improvement introduced by zstd.}
\vspace{-0.4cm}
\label{fig:parquet_zstd}
\end{figure}

People often enable block compression on columnar storage formats
such as \parquet and ORC~\cite{orc} to further reduce the storage overhead.
We repeat the \parquet loading phase of the above experiments with zstd~\cite{zstd}
enabled to show how block compression algorithms affect the final file sizes.

As shown in \cref{fig:parquet_zstd}, the additional improvement introduced by zstd is marked above each bar.
Applying zstd on top of the lightweight encoding schemes in \parquet
can further reduce the file sizes.
The relative improvement of \leco + zstd over \leco is higher than
that in the case of \FORfix.
This shows that \leco's ability to remove serial redundancy
is complementary to some degree to the general-purpose block compression algorithms.

\begin{figure}[t!]
\centering
% \hspace{-0.3cm}
\begin{minipage}[t]{0.22\textwidth}
\centering
\includegraphics[width=\columnwidth]{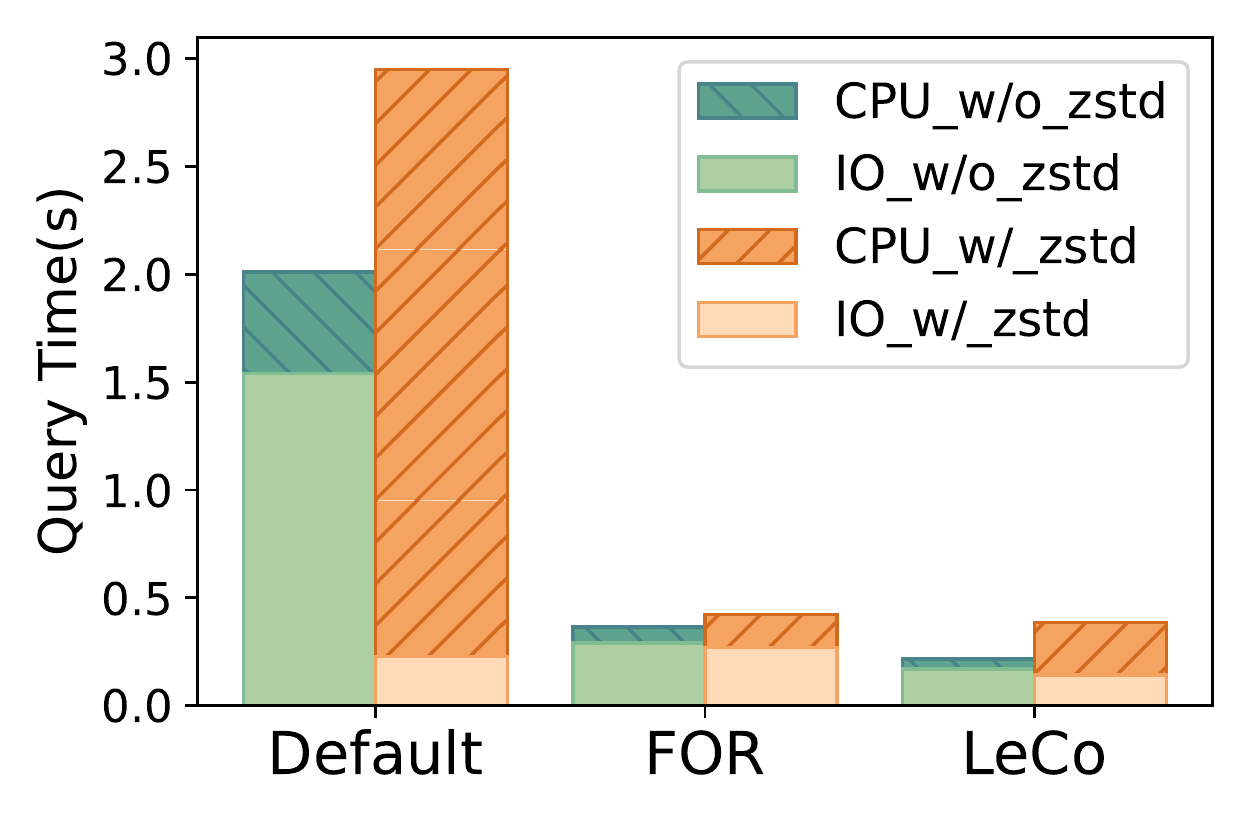}
\vspace{-0.6cm}
\caption{Time breakdown of zstd on \parquet.}
\label{fig:zstd_breakdown}
\end{minipage}
\hspace{0.2cm}
\begin{minipage}[t]{0.23\textwidth}
\centering
\includegraphics[width=\columnwidth]{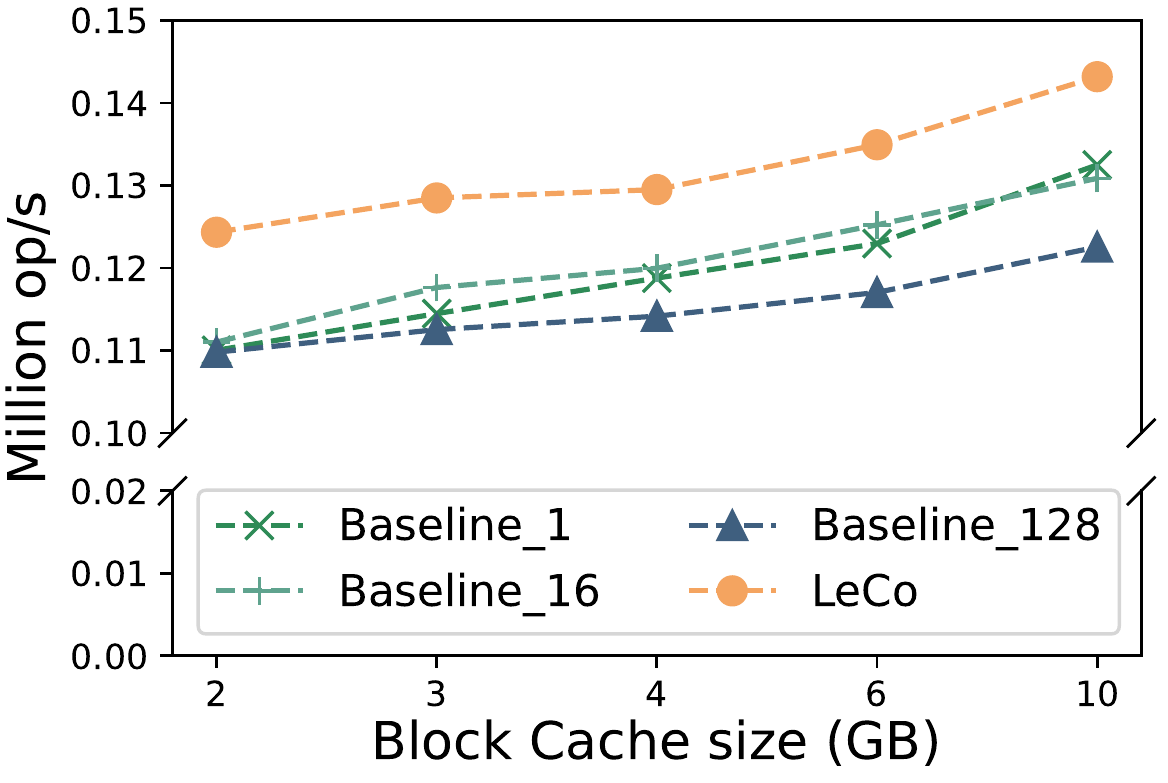}
\vspace{-0.6cm}
\caption{\rocksdb Seek Query Throughput.}
\label{fig:rocksdb_index}
\end{minipage}
\vspace{-0.6cm}
\end{figure}

The decompression overhead of zstd, however, can be significant.
We perform the bitmap selection experiment
with zstd turned on for \parquet.
\cref{fig:zstd_breakdown} shows an example result (\texttt{ml} data set, selectivity = 0.01).
We observe that the I/O savings from zstd are outweighed by its CPU overhead,
leading to an increase in the overall query time.
The result confirms our motivation in \cref{sec:motivation} that
heavyweight compression algorithms are likely to cause CPU bottlenecks
in modern data processing systems.

\subsection{\rocksdb Index Block Compression}
\label{sec:rocksdb}
\rocksdb is a key-value store based on log-structured merge trees.
Each level consists of a sorted run of key-value pairs stored in a sequence of SSTables. 
Each SSTable is divided into multiple data blocks (4KB by default).
\rocksdb builds an index on top of the data blocks.
For each pair of adjacent data blocks $B_{i-1}$ and $B_i$,
an index entry is created where the key is the shortest string
greater than the last key in $B_{i-1}$
and smaller than the first key in $B_i$.
The value of the index entry is a ``block handle'' that records
the byte offset and the size of $B_i$.
To locate a particular key $k$,
\rocksdb performs a binary search in the index block and obtains the entry with the smallest key $\ge k$. 
It then reads the associated ``block handle'' and fetches the corresponding data block that (potentially)
contains $k$.

\rocksdb offers a native compression scheme for the index blocks.
It includes a hyper-parameter called ``restart interval'' (RI) to
make trade-offs between the lookup performance and the index size.
The value of RI determines the size of a compression unit in an index block.
Within each compression unit, \rocksdb applies a variation of Delta Encoding to both the keys and values.
For the index keys, suppose $k_{i-1}$ proceeds $k_i$ in the compressed sequence.
Then $k_i$ is encoded as $(m_i, k_i')$ where $m_i$ denotes the length
of the shared prefix between $k_{i-1}$ and $k_i$,
and $k_i'$ is the remaining suffix.
For the ``block handles'', \rocksdb simply stores the offset of
each block in a delta-encoded sequence.

We use \leco
to compress the keys
and values separately in a \rocksdb index block to shrink its
size and to improve the lookup performance at the same time.
We adopt \lecofix for both key and value sequences.
Because all internal keys in \rocksdb are strings,
we use \leco with the string extension to compress the keys.

We compare \rocksdb with \leco against\footnote{The fixed partition size are set to 64 entries for \leco.} three baseline configurations:
\texttt{Baseline\_1}, \texttt{Baseline\_16}, and \texttt{Baseline\_128}.
The number at the end of each label denotes the value of the RI parameter
(1 is \rocksdb's default).
We configured \rocksdb according to the settings in its Performance
Benchmark~\cite{Rocksdbperf}\footnote{block\_size = 4096B;  pin\_l0\_filter\_and\_index\_blocks\_in \_cache is enabled.}.
We turned on direct I/O to bypass the large OS page cache.

In each experiment, we first load the \rocksdb with 900 million record
generated from the above \rocksdb Performance Benchmark.
Each record has a 20-byte key and a 400-byte value.
The resulting \rocksdb is around $110$ GB.
\leco, \texttt{Baseline\_1}, \texttt{Baseline\_16}, and \texttt{Baseline\_128} achieve a compression ratio of 28.1\%, 71.3\%, 18.9\% and 15.9\%, respectively on the index blocks in \rocksdb. 
We then perform 200M non-empty \texttt{Seek} queries using 64 threads.
The query keys are generated using YCSB~\cite{cooper2010benchmarking}
with a skewed configuration where $80\%$ of the queries access
$20\%$ of the total keys.
We repeat each experiment three times and report the average
measurement.

\cref{fig:rocksdb_index} shows the system throughputs for \leco,
and the baselines with a varying block cache size.
\rocksdb with \leco consistently outperforms the three baseline configurations by up to $16\%$
compared to the second-best configuration.
The reasons are two-fold.
First, compared to \texttt{Baseline\_1} where no compression for the
index blocks are carried out (each compression unit only contains one entry),
\leco produces smaller index blocks so that more data blocks can fit in the block
cache to save I/Os.
Such a performance improvement is more recognizable with a smaller block cache.

Second, compared to \texttt{Baseline\_16} and \texttt{Baseline\_128}
where the index blocks are compressed using Delta Encoding.
Although \leco no longer exhibits an index-size advantage over
these baselines, it saves
a significant amount of computations.
Compared to \texttt{Baseline\_128} which need to decompress the entire
128-entry unit before it accesses a single entry, 
\leco only requires two memory probes to perform
a random access in the index block.

To sum up, applying \leco speeds up binary search in the index blocks.
Such a small change improved the
performance of a complex system (\rocksdb) noticeably.
We believe that other systems with similar ``zone-map'' structures
can benefit from \leco as well.

\section{Related Work}
\label{sec:related}

Many prior compression algorithms leverage repetitions in a data sequence.
Null suppression omits the leading zeros in the bit representation of an integer
and records the byte length of each value~\cite{abadi2006, schlegel2010fast, 2016Vectorized, googlevarint, stepanov2011simd}.
Dictionary~\cite{liu2019mostly,raman2013db2,li2015padded,antoshenkov1996order,binnig2009dictionary,boncz2020fsst,zhang2020order}
and entropy-based compression algorithms~\cite{huffman1952,witten1987arithmetic}
build a bijective map between the original values and the code words.
Block compression algorithms such as LZ77~\cite{lz77}, Gzip~\cite{gzip}, 
Snappy~\cite{snappy}, LZ4~\cite{lz4}, and zstd~\cite{zstd} 
 achieve compression by replacing
repeated bit patterns with shorter dictionary codes.
These approaches, however, miss the opportunity to exploit the serial
correlation between values to achieve a compressed size beyond Shannon’s
Entropy.

A pioneer work by Boffa et al.~\cite{boffa2021learned} proposed to use a similar linear model
as in the PGM-Index~\cite{ferragina2020pgm} with a customized partitioning algorithm (i.e., la\_vector)
to compress a specific data structure called the rank\&select dictionaries.
Their approach represents a specific design point in the \leco framework that is much more general
and extensible in model types and partitioning algorithms.
Also, \leco's default variable-length partitioning algorithm is shown to be more
efficient than la\_vector for compressing columnar data.

Semantic compression~\cite{jagadish1999semantic,gao2016squish,ilkhechi2020deepsqueeze} 
aims to compress tabular data by
exploiting correlations between columns
using complex models like Bayesian networks.
LFR\cite{LFR} and DFR\cite{DFR} use linear model or Delta-like model to compress data without partitioning.
Because their model parameters vary at each data point, they do not support quick random access.

Data partitioning plays an essential role in achieving a good compression ratio
for various algorithms.
Several prior work~\cite{ottaviano2014partitioned, pibiri2019optimally} targeting
inverted indexes proposed partitioning algorithms for specific compression schemes
like Elias-Fano~\cite{vigna2013quasi} and VByte~\cite{thiel1972program,williams1999compressing}.
The partitioning algorithms introduced in \cref{sec:partitioner} are applicable
to an arbitrary linear combination of regression models.
In terms of storage format, FastPFOR~\cite{zukowski2006super} and
NewPFD~\cite{yan2009inverted} stores outlier values separately in a
different format to improve the overall storage and query efficiency.

Time-series/IoT data compression field adopts a similar idea with \leco of approximating data distribution with models, but they target keeping the prediction error within a predetermined threshold and achieve \textbf{lossy} compression. 
Their optimization goal is to minimize the total space of model parameters.
Partitioning algorithms for linear models~\cite{elmeleegy2009online,luo2015piecewise,Xie2014Maximum} and constant value models~\cite{2003Capturing} are designed to minimize the segment number. 
Sim-Piece\cite{kitsiossim} introduces a more compact format to keep the output models. 
Eichinger et al.~\cite{2015AEichinger} consider utilizing higher order models 
but require additional computation effort in the approximation process.

Codec selection is critical in improving data compression performances. A common practice is to define a feature set and use machine learning classifiers for selection.
\citeauthor{abadi2006}~\cite{abadi2006} empirically analyzed the performance of different codecs and manually built a decision tree for selection. 
While the features introduced by
CodecDB~\cite{codecdb}  
overlook the chance to utilize distribution patterns, in contrast to our Regressor Selector.

Both learned indexes and learned compression use regression
to model data distributions.
RMI~\cite{kraska2018case} and RS~\cite{kipf2020radixspline}
apply hierarchical machine learning models to fit the CDFs,
while PGM-Index~\cite{ferragina2020pgm}, FITing-Tree~\cite{galakatos2019fiting}, and CARMI~\cite{CARMI}
put more effort into the partitioning strategies to reduce model prediction errors.
ALEX~\cite{ALEX} and Finedex~\cite{li2021finedex} proposed techniques such as a gapped array and non-blocking retraining to
improve the indexes' update efficiency.

Previous work~\cite{abadi2013design,zukowski2012vectorwise} have shown that
heavyweight compression algorithms~\cite{gzip,snappy,huffman1952} designed
for disk-oriented systems could incur notable computational overhead to the
overall system performance.
Algorithms such as FSST~\cite{boncz2020fsst} and PIDS~\cite{jiang2020pids},
therefore, emphasize low CPU usage besides a competitive compression ratio.
Other related work reduces the computational overhead by enabling direct
query execution on compressed formats~\cite{abadi2006,damme2020,codecdb},
including filter and aggregation/join pushdowns~\cite{graefe1990data,li2013bitweaving,feng2015byteslice,das2015query,lang2016data,christian2010speeding,lee2014joins}.
\vspace{-0.1cm}

\section{Conclusion}
\label{sec:conclusion}

This paper introduces \leco, a lightweight compression framework that
uses machine learning techniques to exploit serial correlation between
the values in a column.
We provide a complementary perspective besides Shannon's entropy
to the general data compression problem.
The \leco framework bridges data mining and data compression with a highly modular design.
Both our micro-benchmark and system evaluation show that \leco is able to
achieve better storage efficiency and faster query processing simultaneously.

\newpage

\bibliographystyle{ACM-Reference-Format}
\bibliography{ref}


\begin{thebibliography}{118}

%%% the Web address.

\ifx \showCODEN    \undefined \def \showCODEN     #1{\unskip}     \fi
\ifx \showDOI      \undefined \def \showDOI       #1{#1}\fi
\ifx \showISBNx    \undefined \def \showISBNx     #1{\unskip}     \fi
\ifx \showISBNxiii \undefined \def \showISBNxiii  #1{\unskip}     \fi
\ifx \showISSN     \undefined \def \showISSN      #1{\unskip}     \fi
\ifx \showLCCN     \undefined \def \showLCCN      #1{\unskip}     \fi
\ifx \shownote     \undefined \def \shownote      #1{#1}          \fi
\ifx \showarticletitle \undefined \def \showarticletitle #1{#1}   \fi
\ifx \showURL      \undefined \def \showURL       {\relax}        \fi
% invisible to TeX
\providecommand\bibfield[2]{#2}
\providecommand\bibinfo[2]{#2}
\providecommand\natexlab[1]{#1}
\providecommand\showeprint[2][]{arXiv:#2}

\bibitem[\protect\citeauthoryear{??}{goo}{2009}]%
        {googlevarint}
 \bibinfo{year}{2009}\natexlab{}.
\newblock \bibinfo{title}{{Google Varint}}.
\newblock \bibinfo{howpublished}{\url{https://static.googleusercontent.com/media/research.google.com/en//people/jeff/WSDM09-keynote.pdf}}.
\newblock


\bibitem[\protect\citeauthoryear{??}{ema}{2018}]%
        {emaildataset}
 \bibinfo{year}{2018}\natexlab{}.
\newblock \bibinfo{title}{{300 Million Email Database}}.
\newblock \bibinfo{howpublished}{\url{https://archive.org/details/300MillionEmailDatabase}}.
\newblock


\bibitem[\protect\citeauthoryear{??}{eng}{2020}]%
        {englishwords}
 \bibinfo{year}{2020}\natexlab{}.
\newblock \bibinfo{title}{{English Word Dataset in HOPE}}.
\newblock \bibinfo{howpublished}{\url{https://github.com/efficient/HOPE/blob/master/datasets/words.txt}}.
\newblock


\bibitem[\protect\citeauthoryear{??}{arr}{2022}]%
        {arrow}
 \bibinfo{year}{2022}\natexlab{}.
\newblock \bibinfo{title}{{Apache Arrow}}.
\newblock \bibinfo{howpublished}{\url{https://arrow.apache.org/}}.
\newblock


\bibitem[\protect\citeauthoryear{??}{orc}{2022}]%
        {orc}
 \bibinfo{year}{2022}\natexlab{}.
\newblock \bibinfo{title}{{Apache ORC}}.
\newblock \bibinfo{howpublished}{\url{https://orc.apache.org/}}.
\newblock


\bibitem[\protect\citeauthoryear{??}{par}{2022}]%
        {parquet}
 \bibinfo{year}{2022}\natexlab{}.
\newblock \bibinfo{title}{{Apache Parquet}}.
\newblock \bibinfo{howpublished}{\url{https://parquet.apache.org/}}.
\newblock


\bibitem[\protect\citeauthoryear{??}{gzi}{2022}]%
        {gzip}
 \bibinfo{year}{2022}\natexlab{}.
\newblock \bibinfo{title}{{GNU GZip}}.
\newblock \bibinfo{howpublished}{\url{https://www.gnu.org/software/gzip/}}.
\newblock


\bibitem[\protect\citeauthoryear{??}{sna}{2022}]%
        {snappy}
 \bibinfo{year}{2022}\natexlab{}.
\newblock \bibinfo{title}{{Google snappy}}.
\newblock \bibinfo{howpublished}{\url{http://google.github.io/snappy/}}.
\newblock


\bibitem[\protect\citeauthoryear{??}{mov}{2022}]%
        {movieiddataset}
 \bibinfo{year}{2022}\natexlab{}.
\newblock \bibinfo{title}{{Kaggle Movie ID dataset}}.
\newblock \bibinfo{howpublished}{\url{https://www.kaggle.com/datasets/grouplens/movielens-20m-dataset?select=rating.csv}}.
\newblock


\bibitem[\protect\citeauthoryear{??}{hou}{2022}]%
        {houseprice}
 \bibinfo{year}{2022}\natexlab{}.
\newblock \bibinfo{title}{{Kaggle USA Real Estate Dataset}}.
\newblock \bibinfo{howpublished}{\url{https://www.kaggle.com/datasets/ahmedshahriarsakib/usa-real-estate-dataset?select=realtor-dataset-100k.csv}}.
\newblock


\bibitem[\protect\citeauthoryear{??}{lz4}{2022}]%
        {lz4}
 \bibinfo{year}{2022}\natexlab{}.
\newblock \bibinfo{title}{{Lz4}}.
\newblock \bibinfo{howpublished}{\url{https://github.com/lz4/lz4}}.
\newblock


\bibitem[\protect\citeauthoryear{??}{per}{2022}]%
        {personal-comm}
 \bibinfo{year}{2022}\natexlab{}.
\newblock \bibinfo{title}{{Personal communication, anonymized for review.}}
\newblock \bibinfo{howpublished}{\url{}}.
\newblock


\bibitem[\protect\citeauthoryear{??}{hea}{2022}]%
        {heatwave2021}
 \bibinfo{year}{2022}\natexlab{}.
\newblock \bibinfo{title}{{Real-time Analytics for MySQL Database Service}}.
\newblock \bibinfo{howpublished}{\url{https://www.oracle.com/mysql/}}.
\newblock


\bibitem[\protect\citeauthoryear{??}{Roc}{2022a}]%
        {Rocksdgithub}
 \bibinfo{year}{2022}\natexlab{a}.
\newblock \bibinfo{title}{{Rocksdb Github}}.
\newblock \bibinfo{howpublished}{\url{https://github.com/facebook/rocksdb}}.
\newblock


\bibitem[\protect\citeauthoryear{??}{Roc}{2022b}]%
        {Rocksdbperf}
 \bibinfo{year}{2022}\natexlab{b}.
\newblock \bibinfo{title}{{Rocksdb Performance Benchmarks}}.
\newblock \bibinfo{howpublished}{\url{https://github.com/facebook/rocksdb/wiki/Performance-Benchmarks}}.
\newblock


\bibitem[\protect\citeauthoryear{??}{sam}{2022}]%
        {samsungssd}
 \bibinfo{year}{2022}\natexlab{}.
\newblock \bibinfo{title}{{Samsung 980 PRO 4.0 NVMe SSD}}.
\newblock \bibinfo{howpublished}{\url{https://www.samsung.com/us/computing/memory-storage/solid-state-drives/980-pro-pcie-4-0-nvme-ssd-1tb-mz-v8p1t0b-am/}}.
\newblock


\bibitem[\protect\citeauthoryear{??}{sin}{2022}]%
        {singlestore}
 \bibinfo{year}{2022}\natexlab{}.
\newblock \bibinfo{title}{{SingleStore}}.
\newblock \bibinfo{howpublished}{\url{https://www.singlestore.com/}}.
\newblock


\bibitem[\protect\citeauthoryear{??}{mlt}{2022}]%
        {mltimestamp}
 \bibinfo{year}{2022}\natexlab{}.
\newblock \bibinfo{title}{{UCI Machine Learning Repository: Timestamp in Bar Crawl: Detecting Heavy Drinking Data Set}}.
\newblock \bibinfo{howpublished}{\url{https://archive.ics.uci.edu/ml/datasets/Bar+Crawl\%3A+Detecting+Heavy+Drinking}}.
\newblock


\bibitem[\protect\citeauthoryear{??}{zst}{2022}]%
        {zstd}
 \bibinfo{year}{2022}\natexlab{}.
\newblock \bibinfo{title}{{Zstandard}}.
\newblock \bibinfo{howpublished}{\url{https://github.com/facebook/zstd}}.
\newblock


\bibitem[\protect\citeauthoryear{??}{geo}{2023}]%
        {geoname}
 \bibinfo{year}{2023}\natexlab{}.
\newblock \bibinfo{title}{{GeoNames Data}}.
\newblock \bibinfo{howpublished}{\url{https://www.geonames.org/export/}}.
\newblock


\bibitem[\protect\citeauthoryear{??}{his}{2023}]%
        {histdata}
 \bibinfo{year}{2023}\natexlab{}.
\newblock \bibinfo{title}{{HistData GRXEUR}}.
\newblock \bibinfo{howpublished}{\url{https://www.histdata.com/}}.
\newblock


\bibitem[\protect\citeauthoryear{??}{cou}{2023}]%
        {courseinfo}
 \bibinfo{year}{2023}\natexlab{}.
\newblock \bibinfo{title}{{Kaggle Udemy Courses}}.
\newblock \bibinfo{howpublished}{\url{https://www.kaggle.com/datasets/hossaingh/udemy-courses}}.
\newblock


\bibitem[\protect\citeauthoryear{??}{mlc}{2023}]%
        {mlcourse}
 \bibinfo{year}{2023}\natexlab{}.
\newblock \bibinfo{title}{{mlcourse.ai}}.
\newblock \bibinfo{howpublished}{\url{https://github.com/Yorko/mlcourse.ai/tree/main/data}}.
\newblock


\bibitem[\protect\citeauthoryear{??}{pub}{2023}]%
        {publicbi}
 \bibinfo{year}{2023}\natexlab{}.
\newblock \bibinfo{title}{{Public BI Benchmark}}.
\newblock \bibinfo{howpublished}{\url{https://homepages.cwi.nl/~boncz/PublicBIbenchmark/}}.
\newblock


\bibitem[\protect\citeauthoryear{??}{tec}{2023}]%
        {techrep}
 \bibinfo{year}{2023}\natexlab{}.
\newblock \bibinfo{title}{{Technical Report.}}
\newblock \bibinfo{howpublished}{\url{https://gitfront.io/r/Leco2023/Hk2zGFeQUSVw/Learn-to-Compress/blob/Leco_sigmod2024_techreport.pdf}}.
\newblock


\bibitem[\protect\citeauthoryear{??}{tpc}{2023a}]%
        {tpcds}
 \bibinfo{year}{2023}\natexlab{a}.
\newblock \bibinfo{title}{{TPC-DS Benchmark Standard Specification}}.
\newblock \bibinfo{howpublished}{\url{https://www.tpc.org/tpcds/}}.
\newblock


\bibitem[\protect\citeauthoryear{??}{tpc}{2023b}]%
        {tpch}
 \bibinfo{year}{2023}\natexlab{b}.
\newblock \bibinfo{title}{{TPC-H Benchmark Standard Specification}}.
\newblock \bibinfo{howpublished}{\url{https://www.tpc.org/tpch/}}.
\newblock


\bibitem[\protect\citeauthoryear{Abadi, Boncz, Amiato, Idreos, and Madden}{Abadi et~al\mbox{.}}{2013}]%
        {abadi2013design}
\bibfield{author}{\bibinfo{person}{Daniel Abadi}, \bibinfo{person}{Peter Boncz}, \bibinfo{person}{Stavros~Harizopoulos Amiato}, \bibinfo{person}{Stratos Idreos}, {and} \bibinfo{person}{Samuel Madden}.} \bibinfo{year}{2013}\natexlab{}.
\newblock \bibinfo{booktitle}{\emph{The design and implementation of modern column-oriented database systems}}.
\newblock \bibinfo{publisher}{Now Hanover, Mass.}
\newblock


\bibitem[\protect\citeauthoryear{Abadi, Madden, and Ferreira}{Abadi et~al\mbox{.}}{2006}]%
        {abadi2006}
\bibfield{author}{\bibinfo{person}{Daniel Abadi}, \bibinfo{person}{Samuel Madden}, {and} \bibinfo{person}{Miguel Ferreira}.} \bibinfo{year}{2006}\natexlab{}.
\newblock \showarticletitle{Integrating compression and execution in column-oriented database systems}. In \bibinfo{booktitle}{\emph{Proceedings of the 2006 ACM SIGMOD international conference on Management of data}}. \bibinfo{pages}{671--682}.
\newblock


\bibitem[\protect\citeauthoryear{Agrawal and Agrawal}{Agrawal and Agrawal}{2015}]%
        {agrawal2015survey}
\bibfield{author}{\bibinfo{person}{Shikha Agrawal} {and} \bibinfo{person}{Jitendra Agrawal}.} \bibinfo{year}{2015}\natexlab{}.
\newblock \showarticletitle{Survey on anomaly detection using data mining techniques}.
\newblock \bibinfo{journal}{\emph{Procedia Computer Science}}  \bibinfo{volume}{60} (\bibinfo{year}{2015}), \bibinfo{pages}{708--713}.
\newblock


\bibitem[\protect\citeauthoryear{Andrew~Lamb}{Andrew~Lamb}{2012}]%
        {lamb2012vertica}
\bibfield{author}{\bibinfo{person}{Ramakrishna Varadarajan Nga Tran Ben Vandiver Lyric Doshi Chuck~Bear Andrew~Lamb, Matt~Fuller}.} \bibinfo{year}{2012}\natexlab{}.
\newblock \showarticletitle{The Vertica Analytic Database: C-Store 7 Years Later}.
\newblock \bibinfo{journal}{\emph{Proceedings of the VLDB Endowment}} \bibinfo{volume}{5}, \bibinfo{number}{12} (\bibinfo{year}{2012}), \bibinfo{pages}{1790--1801}.
\newblock


\bibitem[\protect\citeauthoryear{Antoshenkov, Lomet, and Murray}{Antoshenkov et~al\mbox{.}}{1996}]%
        {antoshenkov1996order}
\bibfield{author}{\bibinfo{person}{Gennady Antoshenkov}, \bibinfo{person}{David Lomet}, {and} \bibinfo{person}{James Murray}.} \bibinfo{year}{1996}\natexlab{}.
\newblock \showarticletitle{Order preserving string compression}. In \bibinfo{booktitle}{\emph{Proceedings of the Twelfth International Conference on Data Engineering}}. IEEE, \bibinfo{pages}{655--663}.
\newblock


\bibitem[\protect\citeauthoryear{Armenatzoglou, Basu, Bhanoori, Cai, Chainani, Chinta, Govindaraju, Green, Gupta, Hillig, et~al\mbox{.}}{Armenatzoglou et~al\mbox{.}}{2022}]%
        {armenatzoglou2022}
\bibfield{author}{\bibinfo{person}{Nikos Armenatzoglou}, \bibinfo{person}{Sanuj Basu}, \bibinfo{person}{Naga Bhanoori}, \bibinfo{person}{Mengchu Cai}, \bibinfo{person}{Naresh Chainani}, \bibinfo{person}{Kiran Chinta}, \bibinfo{person}{Venkatraman Govindaraju}, \bibinfo{person}{Todd~J Green}, \bibinfo{person}{Monish Gupta}, \bibinfo{person}{Sebastian Hillig}, {et~al\mbox{.}}} \bibinfo{year}{2022}\natexlab{}.
\newblock \showarticletitle{Amazon Redshift Re-invented}. In \bibinfo{booktitle}{\emph{Proceedings of the 2022 ACM SIGMOD International Conference on Management of Data}}. \bibinfo{pages}{2205--2217}.
\newblock


\bibitem[\protect\citeauthoryear{Binnig, Hildenbrand, and F{\"a}rber}{Binnig et~al\mbox{.}}{2009}]%
        {binnig2009dictionary}
\bibfield{author}{\bibinfo{person}{Carsten Binnig}, \bibinfo{person}{Stefan Hildenbrand}, {and} \bibinfo{person}{Franz F{\"a}rber}.} \bibinfo{year}{2009}\natexlab{}.
\newblock \showarticletitle{Dictionary-based order-preserving string compression for main memory column stores}. In \bibinfo{booktitle}{\emph{Proceedings of the 2009 ACM SIGMOD International Conference on Management of data}}. \bibinfo{pages}{283--296}.
\newblock


\bibitem[\protect\citeauthoryear{Boffa, Ferragina, and Vinciguerra}{Boffa et~al\mbox{.}}{2021}]%
        {boffa2021learned}
\bibfield{author}{\bibinfo{person}{Antonio Boffa}, \bibinfo{person}{Paolo Ferragina}, {and} \bibinfo{person}{Giorgio Vinciguerra}.} \bibinfo{year}{2021}\natexlab{}.
\newblock \showarticletitle{A “Learned” Approach to Quicken and Compress Rank/Select Dictionaries}. In \bibinfo{booktitle}{\emph{2021 Proceedings of the Workshop on Algorithm Engineering and Experiments (ALENEX)}}. SIAM, \bibinfo{pages}{46--59}.
\newblock


\bibitem[\protect\citeauthoryear{Boncz, Neumann, and Leis}{Boncz et~al\mbox{.}}{2020}]%
        {boncz2020fsst}
\bibfield{author}{\bibinfo{person}{Peter Boncz}, \bibinfo{person}{Thomas Neumann}, {and} \bibinfo{person}{Viktor Leis}.} \bibinfo{year}{2020}\natexlab{}.
\newblock \showarticletitle{FSST: fast random access string compression}.
\newblock \bibinfo{journal}{\emph{Proceedings of the VLDB Endowment}} \bibinfo{volume}{13}, \bibinfo{number}{12} (\bibinfo{year}{2020}), \bibinfo{pages}{2649--2661}.
\newblock


\bibitem[\protect\citeauthoryear{Boncz, Zukowski, and Nes}{Boncz et~al\mbox{.}}{2005}]%
        {DBLP:conf/cidr/BonczZN05}
\bibfield{author}{\bibinfo{person}{Peter~A. Boncz}, \bibinfo{person}{Marcin Zukowski}, {and} \bibinfo{person}{Niels Nes}.} \bibinfo{year}{2005}\natexlab{}.
\newblock \showarticletitle{MonetDB/X100: Hyper-Pipelining Query Execution}. In \bibinfo{booktitle}{\emph{Second Biennial Conference on Innovative Data Systems Research, {CIDR}}}. \bibinfo{pages}{225--237}.
\newblock


\bibitem[\protect\citeauthoryear{Boniol, Paparrizos, Palpanas, and Franklin}{Boniol et~al\mbox{.}}{2021}]%
        {boniol2021sand}
\bibfield{author}{\bibinfo{person}{Paul Boniol}, \bibinfo{person}{John Paparrizos}, \bibinfo{person}{Themis Palpanas}, {and} \bibinfo{person}{Michael~J Franklin}.} \bibinfo{year}{2021}\natexlab{}.
\newblock \showarticletitle{SAND: streaming subsequence anomaly detection}.
\newblock \bibinfo{journal}{\emph{Proceedings of the VLDB Endowment}} \bibinfo{volume}{14}, \bibinfo{number}{10} (\bibinfo{year}{2021}), \bibinfo{pages}{1717--1729}.
\newblock


\bibitem[\protect\citeauthoryear{Borroni}{Borroni}{2013}]%
        {Borroni2013}
\bibfield{author}{\bibinfo{person}{C.~G. Borroni}.} \bibinfo{year}{2013}\natexlab{}.
\newblock \showarticletitle{A new rank correlation measure}.
\newblock \bibinfo{journal}{\emph{Statistical Papers}} \bibinfo{volume}{54}, \bibinfo{number}{2} (\bibinfo{year}{2013}), \bibinfo{pages}{255--270}.
\newblock


\bibitem[\protect\citeauthoryear{Cameron}{Cameron}{1966}]%
        {cameron1966piece}
\bibfield{author}{\bibinfo{person}{Scott~H Cameron}.} \bibinfo{year}{1966}\natexlab{}.
\newblock \bibinfo{booktitle}{\emph{Piece-wise linear approximations}}.
\newblock \bibinfo{type}{{T}echnical {R}eport}. \bibinfo{institution}{IIT RESEARCH INST CHICAGO IL COMPUTER SCIENCES DIV}.
\newblock


\bibitem[\protect\citeauthoryear{Christian, Kai-Uwe, Franz, and Alexander}{Christian et~al\mbox{.}}{2010}]%
        {christian2010speeding}
\bibfield{author}{\bibinfo{person}{Lemke Christian}, \bibinfo{person}{Sattler Kai-Uwe}, \bibinfo{person}{Faerber Franz}, {and} \bibinfo{person}{Zeier Alexander}.} \bibinfo{year}{2010}\natexlab{}.
\newblock \showarticletitle{Speeding up queries in column stores: a case for compression}.
\newblock \bibinfo{journal}{\emph{DaWaK (2010)}} (\bibinfo{year}{2010}), \bibinfo{pages}{117--129}.
\newblock


\bibitem[\protect\citeauthoryear{Cloud}{Cloud}{2017}]%
        {osm}
\bibfield{author}{\bibinfo{person}{Google Cloud}.} \bibinfo{year}{2017}\natexlab{}.
\newblock \bibinfo{title}{{OpenStreetMap(2017)}}.
\newblock \bibinfo{howpublished}{\url{https://console.cloud.google.com/marketplace/details/openstreetmap/geo-openstreetmap}}.
\newblock


\bibitem[\protect\citeauthoryear{Cooper, Silberstein, Tam, Ramakrishnan, and Sears}{Cooper et~al\mbox{.}}{2010}]%
        {cooper2010benchmarking}
\bibfield{author}{\bibinfo{person}{Brian~F Cooper}, \bibinfo{person}{Adam Silberstein}, \bibinfo{person}{Erwin Tam}, \bibinfo{person}{Raghu Ramakrishnan}, {and} \bibinfo{person}{Russell Sears}.} \bibinfo{year}{2010}\natexlab{}.
\newblock \showarticletitle{Benchmarking cloud serving systems with YCSB}. In \bibinfo{booktitle}{\emph{Proceedings of the 1st ACM symposium on Cloud computing}}. \bibinfo{pages}{143--154}.
\newblock


\bibitem[\protect\citeauthoryear{Dageville, Cruanes, Zukowski, Antonov, Avanes, Bock, Claybaugh, Engovatov, Hentschel, Huang, et~al\mbox{.}}{Dageville et~al\mbox{.}}{2016}]%
        {dageville2016}
\bibfield{author}{\bibinfo{person}{Benoit Dageville}, \bibinfo{person}{Thierry Cruanes}, \bibinfo{person}{Marcin Zukowski}, \bibinfo{person}{Vadim Antonov}, \bibinfo{person}{Artin Avanes}, \bibinfo{person}{Jon Bock}, \bibinfo{person}{Jonathan Claybaugh}, \bibinfo{person}{Daniel Engovatov}, \bibinfo{person}{Martin Hentschel}, \bibinfo{person}{Jiansheng Huang}, {et~al\mbox{.}}} \bibinfo{year}{2016}\natexlab{}.
\newblock \showarticletitle{The snowflake elastic data warehouse}. In \bibinfo{booktitle}{\emph{Proceedings of the 2016 International Conference on Management of Data}}. \bibinfo{pages}{215--226}.
\newblock


\bibitem[\protect\citeauthoryear{Damme, Ungeth{\"u}m, Pietrzyk, Krause, Habich, and Lehner}{Damme et~al\mbox{.}}{2020}]%
        {damme2020}
\bibfield{author}{\bibinfo{person}{Patrick Damme}, \bibinfo{person}{Annett Ungeth{\"u}m}, \bibinfo{person}{Johannes Pietrzyk}, \bibinfo{person}{Alexander Krause}, \bibinfo{person}{Dirk Habich}, {and} \bibinfo{person}{Wolfgang Lehner}.} \bibinfo{year}{2020}\natexlab{}.
\newblock \showarticletitle{Morphstore: Analytical query engine with a holistic compression-enabled processing model}.
\newblock \bibinfo{journal}{\emph{arXiv preprint arXiv:2004.09350}} (\bibinfo{year}{2020}).
\newblock


\bibitem[\protect\citeauthoryear{Das, Yan, Zait, Valluri, Vyas, Krishnamachari, Gaharwar, Kamp, and Mukherjee}{Das et~al\mbox{.}}{2015}]%
        {das2015query}
\bibfield{author}{\bibinfo{person}{Dinesh Das}, \bibinfo{person}{Jiaqi Yan}, \bibinfo{person}{Mohamed Zait}, \bibinfo{person}{Satyanarayana~R Valluri}, \bibinfo{person}{Nirav Vyas}, \bibinfo{person}{Ramarajan Krishnamachari}, \bibinfo{person}{Prashant Gaharwar}, \bibinfo{person}{Jesse Kamp}, {and} \bibinfo{person}{Niloy Mukherjee}.} \bibinfo{year}{2015}\natexlab{}.
\newblock \showarticletitle{Query optimization in Oracle 12c database in-memory}.
\newblock \bibinfo{journal}{\emph{Proceedings of the VLDB Endowment}} \bibinfo{volume}{8}, \bibinfo{number}{12} (\bibinfo{year}{2015}), \bibinfo{pages}{1770--1781}.
\newblock


\bibitem[\protect\citeauthoryear{Ding, Minhas, Yu, Wang, Do, Li, Zhang, Chandramouli, Gehrke, Kossmann, Lomet, and Kraska}{Ding et~al\mbox{.}}{2019}]%
        {ALEX}
\bibfield{author}{\bibinfo{person}{Jialin Ding}, \bibinfo{person}{Umar~Farooq Minhas}, \bibinfo{person}{Jia Yu}, \bibinfo{person}{Chi Wang}, \bibinfo{person}{Jaeyoung Do}, \bibinfo{person}{Yinan Li}, \bibinfo{person}{Hantian Zhang}, \bibinfo{person}{Badrish Chandramouli}, \bibinfo{person}{Johannes Gehrke}, \bibinfo{person}{Donald Kossmann}, \bibinfo{person}{David Lomet}, {and} \bibinfo{person}{Tim Kraska}.} \bibinfo{year}{2019}\natexlab{}.
\newblock \showarticletitle{ALEX: An Updatable Adaptive Learned Index}.
\newblock  (\bibinfo{year}{2019}).
\newblock
\urldef\tempurl%
\url{https://doi.org/10.1145/3318464.3389711}
\showDOI{\tempurl}
\showeprint{arXiv:1905.08898}


\bibitem[\protect\citeauthoryear{Dong, Kryczka, Jin, and Stumm}{Dong et~al\mbox{.}}{2021}]%
        {dong2021rocksdb}
\bibfield{author}{\bibinfo{person}{Siying Dong}, \bibinfo{person}{Andrew Kryczka}, \bibinfo{person}{Yanqin Jin}, {and} \bibinfo{person}{Michael Stumm}.} \bibinfo{year}{2021}\natexlab{}.
\newblock \showarticletitle{RocksDB: evolution of development priorities in a key-value store serving large-scale applications}.
\newblock \bibinfo{journal}{\emph{ACM Transactions on Storage (TOS)}} \bibinfo{volume}{17}, \bibinfo{number}{4} (\bibinfo{year}{2021}), \bibinfo{pages}{1--32}.
\newblock


\bibitem[\protect\citeauthoryear{Duda}{Duda}{2013}]%
        {duda2013asymmetric}
\bibfield{author}{\bibinfo{person}{Jarek Duda}.} \bibinfo{year}{2013}\natexlab{}.
\newblock \showarticletitle{Asymmetric numeral systems: entropy coding combining speed of Huffman coding with compression rate of arithmetic coding}.
\newblock \bibinfo{journal}{\emph{arXiv preprint arXiv:1311.2540}} (\bibinfo{year}{2013}).
\newblock


\bibitem[\protect\citeauthoryear{Eichinger, Efros, Karnouskos, and Böhm}{Eichinger et~al\mbox{.}}{2015}]%
        {2015AEichinger}
\bibfield{author}{\bibinfo{person}{Frank Eichinger}, \bibinfo{person}{Pavel Efros}, \bibinfo{person}{Stamatis Karnouskos}, {and} \bibinfo{person}{Klemens Böhm}.} \bibinfo{year}{2015}\natexlab{}.
\newblock \showarticletitle{A time-series compression technique and its application to the smart grid}.
\newblock \bibinfo{journal}{\emph{The VLDB Journal}} (\bibinfo{year}{2015}).
\newblock


\bibitem[\protect\citeauthoryear{Elmeleegy, Elmagarmid, Cecchet, Aref, and Zwaenepoel}{Elmeleegy et~al\mbox{.}}{2009}]%
        {elmeleegy2009online}
\bibfield{author}{\bibinfo{person}{Hazem Elmeleegy}, \bibinfo{person}{Ahmed Elmagarmid}, \bibinfo{person}{Emmanuel Cecchet}, \bibinfo{person}{Walid~G Aref}, {and} \bibinfo{person}{Willy Zwaenepoel}.} \bibinfo{year}{2009}\natexlab{}.
\newblock \showarticletitle{Online piece-wise linear approximation of numerical streams with precision guarantees}.
\newblock  (\bibinfo{year}{2009}).
\newblock


\bibitem[\protect\citeauthoryear{Faloutsos and Megalooikonomou}{Faloutsos and Megalooikonomou}{2007}]%
        {faloutsos2007data}
\bibfield{author}{\bibinfo{person}{Christos Faloutsos} {and} \bibinfo{person}{Vasileios Megalooikonomou}.} \bibinfo{year}{2007}\natexlab{}.
\newblock \showarticletitle{On data mining, compression, and kolmogorov complexity}.
\newblock \bibinfo{journal}{\emph{Data mining and knowledge discovery}} \bibinfo{volume}{15}, \bibinfo{number}{1} (\bibinfo{year}{2007}), \bibinfo{pages}{3--20}.
\newblock


\bibitem[\protect\citeauthoryear{F{\"a}rber, Cha, Primsch, Bornh{\"o}vd, Sigg, and Lehner}{F{\"a}rber et~al\mbox{.}}{2012}]%
        {farber2012}
\bibfield{author}{\bibinfo{person}{Franz F{\"a}rber}, \bibinfo{person}{Sang~Kyun Cha}, \bibinfo{person}{J{\"u}rgen Primsch}, \bibinfo{person}{Christof Bornh{\"o}vd}, \bibinfo{person}{Stefan Sigg}, {and} \bibinfo{person}{Wolfgang Lehner}.} \bibinfo{year}{2012}\natexlab{}.
\newblock \showarticletitle{SAP HANA database: data management for modern business applications}.
\newblock \bibinfo{journal}{\emph{ACM Sigmod Record}} \bibinfo{volume}{40}, \bibinfo{number}{4} (\bibinfo{year}{2012}), \bibinfo{pages}{45--51}.
\newblock


\bibitem[\protect\citeauthoryear{Feng, Lo, Kao, and Xu}{Feng et~al\mbox{.}}{2015}]%
        {feng2015byteslice}
\bibfield{author}{\bibinfo{person}{Ziqiang Feng}, \bibinfo{person}{Eric Lo}, \bibinfo{person}{Ben Kao}, {and} \bibinfo{person}{Wenjian Xu}.} \bibinfo{year}{2015}\natexlab{}.
\newblock \showarticletitle{Byteslice: Pushing the envelop of main memory data processing with a new storage layout}. In \bibinfo{booktitle}{\emph{Proceedings of the 2015 ACM SIGMOD International Conference on Management of Data}}. \bibinfo{pages}{31--46}.
\newblock


\bibitem[\protect\citeauthoryear{Ferragina and Vinciguerra}{Ferragina and Vinciguerra}{2020}]%
        {ferragina2020pgm}
\bibfield{author}{\bibinfo{person}{Paolo Ferragina} {and} \bibinfo{person}{Giorgio Vinciguerra}.} \bibinfo{year}{2020}\natexlab{}.
\newblock \showarticletitle{The PGM-index: a fully-dynamic compressed learned index with provable worst-case bounds}.
\newblock \bibinfo{journal}{\emph{Proceedings of the VLDB Endowment}} \bibinfo{volume}{13}, \bibinfo{number}{8} (\bibinfo{year}{2020}), \bibinfo{pages}{1162--1175}.
\newblock


\bibitem[\protect\citeauthoryear{Flamm}{Flamm}{2019}]%
        {flamm2019measuring}
\bibfield{author}{\bibinfo{person}{Kenneth Flamm}.} \bibinfo{year}{2019}\natexlab{}.
\newblock \showarticletitle{Measuring Moore’s law: evidence from price, cost, and quality indexes}.
\newblock In \bibinfo{booktitle}{\emph{Measuring and Accounting for Innovation in the 21st Century}}. \bibinfo{publisher}{University of Chicago Press}.
\newblock


\bibitem[\protect\citeauthoryear{Galakatos, Markovitch, Binnig, Fonseca, and Kraska}{Galakatos et~al\mbox{.}}{2019}]%
        {galakatos2019fiting}
\bibfield{author}{\bibinfo{person}{Alex Galakatos}, \bibinfo{person}{Michael Markovitch}, \bibinfo{person}{Carsten Binnig}, \bibinfo{person}{Rodrigo Fonseca}, {and} \bibinfo{person}{Tim Kraska}.} \bibinfo{year}{2019}\natexlab{}.
\newblock \showarticletitle{Fiting-tree: A data-aware index structure}. In \bibinfo{booktitle}{\emph{Proceedings of the 2019 International Conference on Management of Data}}. \bibinfo{pages}{1189--1206}.
\newblock


\bibitem[\protect\citeauthoryear{Gao and Parameswaran}{Gao and Parameswaran}{2016}]%
        {gao2016squish}
\bibfield{author}{\bibinfo{person}{Yihan Gao} {and} \bibinfo{person}{Aditya Parameswaran}.} \bibinfo{year}{2016}\natexlab{}.
\newblock \showarticletitle{Squish: Near-optimal compression for archival of relational datasets}. In \bibinfo{booktitle}{\emph{Proceedings of the 22nd ACM SIGKDD International Conference on Knowledge Discovery and Data Mining}}. \bibinfo{pages}{1575--1584}.
\newblock


\bibitem[\protect\citeauthoryear{Goldstein, Ramakrishnan, and Shaft}{Goldstein et~al\mbox{.}}{1998}]%
        {goldstein1998}
\bibfield{author}{\bibinfo{person}{Jonathan Goldstein}, \bibinfo{person}{Raghu Ramakrishnan}, {and} \bibinfo{person}{Uri Shaft}.} \bibinfo{year}{1998}\natexlab{}.
\newblock \showarticletitle{Compressing relations and indexes}. In \bibinfo{booktitle}{\emph{Proceedings 14th International Conference on Data Engineering}}. IEEE, \bibinfo{pages}{370--379}.
\newblock


\bibitem[\protect\citeauthoryear{Graefe and Shapiro}{Graefe and Shapiro}{1990}]%
        {graefe1990data}
\bibfield{author}{\bibinfo{person}{Goetz Graefe} {and} \bibinfo{person}{Leonard~D Shapiro}.} \bibinfo{year}{1990}\natexlab{}.
\newblock \bibinfo{booktitle}{\emph{Data compression and database performance}}.
\newblock \bibinfo{publisher}{University of Colorado, Boulder, Department of Computer Science}.
\newblock


\bibitem[\protect\citeauthoryear{Gupta, Agarwal, Tan, Kulesza, Pathak, Stefani, and Srinivasan}{Gupta et~al\mbox{.}}{2015}]%
        {gupta2015}
\bibfield{author}{\bibinfo{person}{Anurag Gupta}, \bibinfo{person}{Deepak Agarwal}, \bibinfo{person}{Derek Tan}, \bibinfo{person}{Jakub Kulesza}, \bibinfo{person}{Rahul Pathak}, \bibinfo{person}{Stefano Stefani}, {and} \bibinfo{person}{Vidhya Srinivasan}.} \bibinfo{year}{2015}\natexlab{}.
\newblock \showarticletitle{Amazon redshift and the case for simpler data warehouses}. In \bibinfo{booktitle}{\emph{Proceedings of the 2015 ACM SIGMOD international conference on management of data}}. \bibinfo{pages}{1917--1923}.
\newblock


\bibitem[\protect\citeauthoryear{Huang, Liu, Cui, Fang, Ma, Xu, Shen, Tang, Zhou, Huang, et~al\mbox{.}}{Huang et~al\mbox{.}}{2020}]%
        {huang2020tidb}
\bibfield{author}{\bibinfo{person}{Dongxu Huang}, \bibinfo{person}{Qi Liu}, \bibinfo{person}{Qiu Cui}, \bibinfo{person}{Zhuhe Fang}, \bibinfo{person}{Xiaoyu Ma}, \bibinfo{person}{Fei Xu}, \bibinfo{person}{Li Shen}, \bibinfo{person}{Liu Tang}, \bibinfo{person}{Yuxing Zhou}, \bibinfo{person}{Menglong Huang}, {et~al\mbox{.}}} \bibinfo{year}{2020}\natexlab{}.
\newblock \showarticletitle{TiDB: a Raft-based HTAP database}.
\newblock \bibinfo{journal}{\emph{Proceedings of the VLDB Endowment}} \bibinfo{volume}{13}, \bibinfo{number}{12} (\bibinfo{year}{2020}), \bibinfo{pages}{3072--3084}.
\newblock


\bibitem[\protect\citeauthoryear{Huffman}{Huffman}{1952}]%
        {huffman1952}
\bibfield{author}{\bibinfo{person}{David~A Huffman}.} \bibinfo{year}{1952}\natexlab{}.
\newblock \showarticletitle{A method for the construction of minimum-redundancy codes}.
\newblock \bibinfo{journal}{\emph{Proceedings of the IRE}} \bibinfo{volume}{40}, \bibinfo{number}{9} (\bibinfo{year}{1952}), \bibinfo{pages}{1098--1101}.
\newblock


\bibitem[\protect\citeauthoryear{Hung, Jeung, and Aberer}{Hung et~al\mbox{.}}{2012}]%
        {hung2012evaluation}
\bibfield{author}{\bibinfo{person}{Nguyen Quoc~Viet Hung}, \bibinfo{person}{Hoyoung Jeung}, {and} \bibinfo{person}{Karl Aberer}.} \bibinfo{year}{2012}\natexlab{}.
\newblock \showarticletitle{An evaluation of model-based approaches to sensor data compression}.
\newblock \bibinfo{journal}{\emph{IEEE Transactions on Knowledge and Data Engineering}} \bibinfo{volume}{25}, \bibinfo{number}{11} (\bibinfo{year}{2012}), \bibinfo{pages}{2434--2447}.
\newblock


\bibitem[\protect\citeauthoryear{Ilkhechi, Crotty, Galakatos, Mao, Fan, Shi, and Cetintemel}{Ilkhechi et~al\mbox{.}}{2020}]%
        {ilkhechi2020deepsqueeze}
\bibfield{author}{\bibinfo{person}{Amir Ilkhechi}, \bibinfo{person}{Andrew Crotty}, \bibinfo{person}{Alex Galakatos}, \bibinfo{person}{Yicong Mao}, \bibinfo{person}{Grace Fan}, \bibinfo{person}{Xiran Shi}, {and} \bibinfo{person}{Ugur Cetintemel}.} \bibinfo{year}{2020}\natexlab{}.
\newblock \showarticletitle{DeepSqueeze: deep semantic compression for tabular data}. In \bibinfo{booktitle}{\emph{Proceedings of the 2020 ACM SIGMOD international conference on management of data}}. \bibinfo{pages}{1733--1746}.
\newblock


\bibitem[\protect\citeauthoryear{Jagadish, Madar, and Ng}{Jagadish et~al\mbox{.}}{1999}]%
        {jagadish1999semantic}
\bibfield{author}{\bibinfo{person}{HV Jagadish}, \bibinfo{person}{Jason Madar}, {and} \bibinfo{person}{Raymond~T Ng}.} \bibinfo{year}{1999}\natexlab{}.
\newblock \showarticletitle{Semantic compression and pattern extraction with fascicles}. In \bibinfo{booktitle}{\emph{VLDB}}, Vol.~\bibinfo{volume}{99}. \bibinfo{pages}{186--97}.
\newblock


\bibitem[\protect\citeauthoryear{Jiang, Liu, Jin, Paparrizos, and Elmore}{Jiang et~al\mbox{.}}{2020}]%
        {jiang2020pids}
\bibfield{author}{\bibinfo{person}{Hao Jiang}, \bibinfo{person}{Chunwei Liu}, \bibinfo{person}{Qi Jin}, \bibinfo{person}{John Paparrizos}, {and} \bibinfo{person}{Aaron~J Elmore}.} \bibinfo{year}{2020}\natexlab{}.
\newblock \showarticletitle{PIDS: attribute decomposition for improved compression and query performance in columnar storage}.
\newblock \bibinfo{journal}{\emph{Proceedings of the VLDB Endowment}} \bibinfo{volume}{13}, \bibinfo{number}{6} (\bibinfo{year}{2020}), \bibinfo{pages}{925--938}.
\newblock


\bibitem[\protect\citeauthoryear{Jiang, Liu, Paparrizos, Chien, Ma, and Elmore}{Jiang et~al\mbox{.}}{2021}]%
        {codecdb}
\bibfield{author}{\bibinfo{person}{Hao Jiang}, \bibinfo{person}{Chunwei Liu}, \bibinfo{person}{John Paparrizos}, \bibinfo{person}{Andrew~A Chien}, \bibinfo{person}{Jihong Ma}, {and} \bibinfo{person}{Aaron~J Elmore}.} \bibinfo{year}{2021}\natexlab{}.
\newblock \showarticletitle{Good to the Last Bit: Data-Driven Encoding with CodecDB}. In \bibinfo{booktitle}{\emph{Proceedings of the 2021 International Conference on Management of Data}}. \bibinfo{pages}{843--856}.
\newblock


\bibitem[\protect\citeauthoryear{Kemper and Neumann}{Kemper and Neumann}{2011}]%
        {kemper2011hyper}
\bibfield{author}{\bibinfo{person}{Alfons Kemper} {and} \bibinfo{person}{Thomas Neumann}.} \bibinfo{year}{2011}\natexlab{}.
\newblock \showarticletitle{HyPer: A hybrid OLTP\&OLAP main memory database system based on virtual memory snapshots}. In \bibinfo{booktitle}{\emph{2011 IEEE 27th International Conference on Data Engineering}}. IEEE, \bibinfo{pages}{195--206}.
\newblock


\bibitem[\protect\citeauthoryear{Kipf, Marcus, Renen, Stoian, Kemper, Kraska, and Neumann}{Kipf et~al\mbox{.}}{2019}]%
        {2019SOSD}
\bibfield{author}{\bibinfo{person}{A. Kipf}, \bibinfo{person}{R Marcus}, \bibinfo{person}{A~Van Renen}, \bibinfo{person}{M. Stoian}, \bibinfo{person}{A. Kemper}, \bibinfo{person}{T. Kraska}, {and} \bibinfo{person}{T. Neumann}.} \bibinfo{year}{2019}\natexlab{}.
\newblock \showarticletitle{SOSD: A Benchmark for Learned Indexes}.
\newblock  (\bibinfo{year}{2019}).
\newblock


\bibitem[\protect\citeauthoryear{Kipf, Marcus, van Renen, Stoian, Kemper, Kraska, and Neumann}{Kipf et~al\mbox{.}}{2020}]%
        {kipf2020radixspline}
\bibfield{author}{\bibinfo{person}{Andreas Kipf}, \bibinfo{person}{Ryan Marcus}, \bibinfo{person}{Alexander van Renen}, \bibinfo{person}{Mihail Stoian}, \bibinfo{person}{Alfons Kemper}, \bibinfo{person}{Tim Kraska}, {and} \bibinfo{person}{Thomas Neumann}.} \bibinfo{year}{2020}\natexlab{}.
\newblock \showarticletitle{RadixSpline: a single-pass learned index}. In \bibinfo{booktitle}{\emph{Proceedings of the Third International Workshop on Exploiting Artificial Intelligence Techniques for Data Management}}. \bibinfo{pages}{1--5}.
\newblock


\bibitem[\protect\citeauthoryear{Kitsios, Liakos, Papakonstantinopoulou, and Kotidis}{Kitsios et~al\mbox{.}}{[n.d.]}]%
        {kitsiossim}
\bibfield{author}{\bibinfo{person}{Xenophon Kitsios}, \bibinfo{person}{Panagiotis Liakos}, \bibinfo{person}{Katia Papakonstantinopoulou}, {and} \bibinfo{person}{Yannis Kotidis}.} \bibinfo{year}{[n.d.]}\natexlab{}.
\newblock \showarticletitle{Sim-Piece: Highly Accurate Piecewise Linear Approximation through Similar Segment Merging}.
\newblock  (\bibinfo{year}{[n.\,d.]}).
\newblock


\bibitem[\protect\citeauthoryear{Kraska, Beutel, Chi, Dean, and Polyzotis}{Kraska et~al\mbox{.}}{2018}]%
        {kraska2018case}
\bibfield{author}{\bibinfo{person}{Tim Kraska}, \bibinfo{person}{Alex Beutel}, \bibinfo{person}{Ed~H Chi}, \bibinfo{person}{Jeffrey Dean}, {and} \bibinfo{person}{Neoklis Polyzotis}.} \bibinfo{year}{2018}\natexlab{}.
\newblock \showarticletitle{The case for learned index structures}. In \bibinfo{booktitle}{\emph{Proceedings of the 2018 international conference on management of data}}. \bibinfo{pages}{489--504}.
\newblock


\bibitem[\protect\citeauthoryear{Lahiri, Chavan, Colgan, Das, Ganesh, Gleeson, Hase, Holloway, Kamp, Lee, et~al\mbox{.}}{Lahiri et~al\mbox{.}}{2015}]%
        {lahiri2015}
\bibfield{author}{\bibinfo{person}{Tirthankar Lahiri}, \bibinfo{person}{Shasank Chavan}, \bibinfo{person}{Maria Colgan}, \bibinfo{person}{Dinesh Das}, \bibinfo{person}{Amit Ganesh}, \bibinfo{person}{Mike Gleeson}, \bibinfo{person}{Sanket Hase}, \bibinfo{person}{Allison Holloway}, \bibinfo{person}{Jesse Kamp}, \bibinfo{person}{Teck-Hua Lee}, {et~al\mbox{.}}} \bibinfo{year}{2015}\natexlab{}.
\newblock \showarticletitle{Oracle database in-memory: A dual format in-memory database}. In \bibinfo{booktitle}{\emph{2015 IEEE 31st International Conference on Data Engineering}}. IEEE, \bibinfo{pages}{1253--1258}.
\newblock


\bibitem[\protect\citeauthoryear{Lang, M{\"u}hlbauer, Funke, Boncz, Neumann, and Kemper}{Lang et~al\mbox{.}}{2016}]%
        {lang2016data}
\bibfield{author}{\bibinfo{person}{Harald Lang}, \bibinfo{person}{Tobias M{\"u}hlbauer}, \bibinfo{person}{Florian Funke}, \bibinfo{person}{Peter~A Boncz}, \bibinfo{person}{Thomas Neumann}, {and} \bibinfo{person}{Alfons Kemper}.} \bibinfo{year}{2016}\natexlab{}.
\newblock \showarticletitle{Data blocks: Hybrid OLTP and OLAP on compressed storage using both vectorization and compilation}. In \bibinfo{booktitle}{\emph{Proceedings of the 2016 International Conference on Management of Data}}. \bibinfo{pages}{311--326}.
\newblock


\bibitem[\protect\citeauthoryear{Larson, Birka, Hanson, Huang, Nowakiewicz, and Papadimos}{Larson et~al\mbox{.}}{2015}]%
        {larson2015real}
\bibfield{author}{\bibinfo{person}{Per-{\AA}ke Larson}, \bibinfo{person}{Adrian Birka}, \bibinfo{person}{Eric~N Hanson}, \bibinfo{person}{Weiyun Huang}, \bibinfo{person}{Michal Nowakiewicz}, {and} \bibinfo{person}{Vassilis Papadimos}.} \bibinfo{year}{2015}\natexlab{}.
\newblock \showarticletitle{Real-time analytical processing with SQL server}.
\newblock \bibinfo{journal}{\emph{Proceedings of the VLDB Endowment}} \bibinfo{volume}{8}, \bibinfo{number}{12} (\bibinfo{year}{2015}), \bibinfo{pages}{1740--1751}.
\newblock


\bibitem[\protect\citeauthoryear{Lazaridis and Mehrotra}{Lazaridis and Mehrotra}{2003}]%
        {2003Capturing}
\bibfield{author}{\bibinfo{person}{I. Lazaridis} {and} \bibinfo{person}{S. Mehrotra}.} \bibinfo{year}{2003}\natexlab{}.
\newblock \showarticletitle{Capturing sensor-generated time series with quality guarantees}. In \bibinfo{booktitle}{\emph{Data Engineering, 2003. Proceedings. 19th International Conference on}}.
\newblock


\bibitem[\protect\citeauthoryear{Lee, Moon, Kim, Kim, Cha, and Han}{Lee et~al\mbox{.}}{2017}]%
        {lee2017parallel}
\bibfield{author}{\bibinfo{person}{Juchang Lee}, \bibinfo{person}{SeungHyun Moon}, \bibinfo{person}{Kyu~Hwan Kim}, \bibinfo{person}{Deok~Hoe Kim}, \bibinfo{person}{Sang~Kyun Cha}, {and} \bibinfo{person}{Wook-Shin Han}.} \bibinfo{year}{2017}\natexlab{}.
\newblock \showarticletitle{Parallel replication across formats in SAP HANA for scaling out mixed OLTP/OLAP workloads}.
\newblock \bibinfo{journal}{\emph{Proceedings of the VLDB Endowment}} \bibinfo{volume}{10}, \bibinfo{number}{12} (\bibinfo{year}{2017}), \bibinfo{pages}{1598--1609}.
\newblock


\bibitem[\protect\citeauthoryear{Lee, Attaluri, Barber, Chainani, Draese, Ho, Idreos, Kim, Lightstone, Lohman, et~al\mbox{.}}{Lee et~al\mbox{.}}{2014}]%
        {lee2014joins}
\bibfield{author}{\bibinfo{person}{Jae-Gil Lee}, \bibinfo{person}{Gopi Attaluri}, \bibinfo{person}{Ronald Barber}, \bibinfo{person}{Naresh Chainani}, \bibinfo{person}{Oliver Draese}, \bibinfo{person}{Frederick Ho}, \bibinfo{person}{Stratos Idreos}, \bibinfo{person}{Min-Soo Kim}, \bibinfo{person}{Sam Lightstone}, \bibinfo{person}{Guy Lohman}, {et~al\mbox{.}}} \bibinfo{year}{2014}\natexlab{}.
\newblock \showarticletitle{Joins on encoded and partitioned data}.
\newblock \bibinfo{journal}{\emph{Proceedings of the VLDB Endowment}} \bibinfo{volume}{7}, \bibinfo{number}{13} (\bibinfo{year}{2014}), \bibinfo{pages}{1355--1366}.
\newblock


\bibitem[\protect\citeauthoryear{Lemire and Boytsov}{Lemire and Boytsov}{2015}]%
        {lemire2015decoding}
\bibfield{author}{\bibinfo{person}{Daniel Lemire} {and} \bibinfo{person}{Leonid Boytsov}.} \bibinfo{year}{2015}\natexlab{}.
\newblock \showarticletitle{Decoding billions of integers per second through vectorization}.
\newblock \bibinfo{journal}{\emph{Software: Practice and Experience}} \bibinfo{volume}{45}, \bibinfo{number}{1} (\bibinfo{year}{2015}), \bibinfo{pages}{1--29}.
\newblock


\bibitem[\protect\citeauthoryear{Li, Vit{\'a}nyi, et~al\mbox{.}}{Li et~al\mbox{.}}{2008}]%
        {li2008introduction}
\bibfield{author}{\bibinfo{person}{Ming Li}, \bibinfo{person}{Paul Vit{\'a}nyi}, {et~al\mbox{.}}} \bibinfo{year}{2008}\natexlab{}.
\newblock \bibinfo{booktitle}{\emph{An introduction to Kolmogorov complexity and its applications}}. Vol.~\bibinfo{volume}{3}.
\newblock \bibinfo{publisher}{Springer}.
\newblock


\bibitem[\protect\citeauthoryear{Li, Hua, Jia, and Zuo}{Li et~al\mbox{.}}{2021}]%
        {li2021finedex}
\bibfield{author}{\bibinfo{person}{Pengfei Li}, \bibinfo{person}{Yu Hua}, \bibinfo{person}{Jingnan Jia}, {and} \bibinfo{person}{Pengfei Zuo}.} \bibinfo{year}{2021}\natexlab{}.
\newblock \showarticletitle{FINEdex: a fine-grained learned index scheme for scalable and concurrent memory systems}.
\newblock \bibinfo{journal}{\emph{Proceedings of the VLDB Endowment}} \bibinfo{volume}{15}, \bibinfo{number}{2} (\bibinfo{year}{2021}), \bibinfo{pages}{321--334}.
\newblock


\bibitem[\protect\citeauthoryear{Li, Chasseur, and Patel}{Li et~al\mbox{.}}{2015}]%
        {li2015padded}
\bibfield{author}{\bibinfo{person}{Yinan Li}, \bibinfo{person}{Craig Chasseur}, {and} \bibinfo{person}{Jignesh~M Patel}.} \bibinfo{year}{2015}\natexlab{}.
\newblock \showarticletitle{A padded encoding scheme to accelerate scans by leveraging skew}. In \bibinfo{booktitle}{\emph{Proceedings of the 2015 ACM SIGMOD International Conference on Management of Data}}. \bibinfo{pages}{1509--1524}.
\newblock


\bibitem[\protect\citeauthoryear{Li and Patel}{Li and Patel}{2013}]%
        {li2013bitweaving}
\bibfield{author}{\bibinfo{person}{Yinan Li} {and} \bibinfo{person}{Jignesh~M Patel}.} \bibinfo{year}{2013}\natexlab{}.
\newblock \showarticletitle{Bitweaving: Fast scans for main memory data processing}. In \bibinfo{booktitle}{\emph{Proceedings of the 2013 ACM SIGMOD International Conference on Management of Data}}. \bibinfo{pages}{289--300}.
\newblock


\bibitem[\protect\citeauthoryear{Libraries.io}{Libraries.io}{2017}]%
        {libio}
\bibfield{author}{\bibinfo{person}{Libraries.io}.} \bibinfo{year}{2017}\natexlab{}.
\newblock \bibinfo{title}{{Repository ID in Libraries.io}}.
\newblock \bibinfo{howpublished}{\url{https://libraries.io/data}}.
\newblock


\bibitem[\protect\citeauthoryear{Liu, Umbenhower, Jiang, Subramaniam, Ma, and Elmore}{Liu et~al\mbox{.}}{2019}]%
        {liu2019mostly}
\bibfield{author}{\bibinfo{person}{Chunwei Liu}, \bibinfo{person}{McKade Umbenhower}, \bibinfo{person}{Hao Jiang}, \bibinfo{person}{Pranav Subramaniam}, \bibinfo{person}{Jihong Ma}, {and} \bibinfo{person}{Aaron~J Elmore}.} \bibinfo{year}{2019}\natexlab{}.
\newblock \showarticletitle{Mostly order preserving dictionaries}. In \bibinfo{booktitle}{\emph{2019 IEEE 35th International Conference on Data Engineering (ICDE)}}. IEEE, \bibinfo{pages}{1214--1225}.
\newblock


\bibitem[\protect\citeauthoryear{Luo, Yi, Cheng, Li, Fan, He, and Mu}{Luo et~al\mbox{.}}{2015}]%
        {luo2015piecewise}
\bibfield{author}{\bibinfo{person}{Ge Luo}, \bibinfo{person}{Ke Yi}, \bibinfo{person}{Siu-Wing Cheng}, \bibinfo{person}{Zhenguo Li}, \bibinfo{person}{Wei Fan}, \bibinfo{person}{Cheng He}, {and} \bibinfo{person}{Yadong Mu}.} \bibinfo{year}{2015}\natexlab{}.
\newblock \showarticletitle{Piecewise linear approximation of streaming time series data with max-error guarantees}. In \bibinfo{booktitle}{\emph{2015 IEEE 31st international conference on data engineering}}. IEEE, \bibinfo{pages}{173--184}.
\newblock


\bibitem[\protect\citeauthoryear{Ottaviano and Venturini}{Ottaviano and Venturini}{2014}]%
        {ottaviano2014partitioned}
\bibfield{author}{\bibinfo{person}{Giuseppe Ottaviano} {and} \bibinfo{person}{Rossano Venturini}.} \bibinfo{year}{2014}\natexlab{}.
\newblock \showarticletitle{Partitioned elias-fano indexes}. In \bibinfo{booktitle}{\emph{Proceedings of the 37th international ACM SIGIR conference on Research \& development in information retrieval}}. \bibinfo{pages}{273--282}.
\newblock


\bibitem[\protect\citeauthoryear{{\"O}zcan, Tian, and T{\"o}z{\"u}n}{{\"O}zcan et~al\mbox{.}}{2017}]%
        {ozcan2017hybrid}
\bibfield{author}{\bibinfo{person}{Fatma {\"O}zcan}, \bibinfo{person}{Yuanyuan Tian}, {and} \bibinfo{person}{Pinar T{\"o}z{\"u}n}.} \bibinfo{year}{2017}\natexlab{}.
\newblock \showarticletitle{Hybrid transactional/analytical processing: A survey}. In \bibinfo{booktitle}{\emph{Proceedings of the 2017 ACM International Conference on Management of Data}}. \bibinfo{pages}{1771--1775}.
\newblock


\bibitem[\protect\citeauthoryear{Pezzini, Feinberg, Rayner, and Edjlali}{Pezzini et~al\mbox{.}}{2014}]%
        {pezzini2014hybrid}
\bibfield{author}{\bibinfo{person}{Massimo Pezzini}, \bibinfo{person}{Donald Feinberg}, \bibinfo{person}{Nigel Rayner}, {and} \bibinfo{person}{Roxane Edjlali}.} \bibinfo{year}{2014}\natexlab{}.
\newblock \showarticletitle{Hybrid transaction/analytical processing will foster opportunities for dramatic business innovation}.
\newblock \bibinfo{journal}{\emph{Gartner (2014, January 28) Available at https://www. gartner. com/doc/2657815/hybrid-transactionanalyticalprocessing-foster-opportunities}} (\bibinfo{year}{2014}), \bibinfo{pages}{4--20}.
\newblock


\bibitem[\protect\citeauthoryear{Pibiri and Venturini}{Pibiri and Venturini}{2019}]%
        {pibiri2019optimally}
\bibfield{author}{\bibinfo{person}{Giulio~Ermanno Pibiri} {and} \bibinfo{person}{Rossano Venturini}.} \bibinfo{year}{2019}\natexlab{}.
\newblock \showarticletitle{On optimally partitioning variable-byte codes}.
\newblock \bibinfo{journal}{\emph{IEEE Transactions on Knowledge and Data Engineering}} \bibinfo{volume}{32}, \bibinfo{number}{9} (\bibinfo{year}{2019}), \bibinfo{pages}{1812--1823}.
\newblock


\bibitem[\protect\citeauthoryear{Plaisance, Kurz, and Lemire}{Plaisance et~al\mbox{.}}{2016}]%
        {2016Vectorized}
\bibfield{author}{\bibinfo{person}{J. Plaisance}, \bibinfo{person}{N. Kurz}, {and} \bibinfo{person}{D Lemire}.} \bibinfo{year}{2016}\natexlab{}.
\newblock \showarticletitle{Vectorized VByte Decoding}.
\newblock \bibinfo{journal}{\emph{Computerence}} (\bibinfo{year}{2016}).
\newblock


\bibitem[\protect\citeauthoryear{Plattner}{Plattner}{2009}]%
        {plattner2009common}
\bibfield{author}{\bibinfo{person}{Hasso Plattner}.} \bibinfo{year}{2009}\natexlab{}.
\newblock \showarticletitle{A common database approach for OLTP and OLAP using an in-memory column database}. In \bibinfo{booktitle}{\emph{Proceedings of the 2009 ACM SIGMOD International Conference on Management of data}}. \bibinfo{pages}{1--2}.
\newblock


\bibitem[\protect\citeauthoryear{Raman, Attaluri, Barber, Chainani, Kalmuk, KulandaiSamy, Leenstra, Lightstone, Liu, Lohman, et~al\mbox{.}}{Raman et~al\mbox{.}}{2013}]%
        {raman2013db2}
\bibfield{author}{\bibinfo{person}{Vijayshankar Raman}, \bibinfo{person}{Gopi Attaluri}, \bibinfo{person}{Ronald Barber}, \bibinfo{person}{Naresh Chainani}, \bibinfo{person}{David Kalmuk}, \bibinfo{person}{Vincent KulandaiSamy}, \bibinfo{person}{Jens Leenstra}, \bibinfo{person}{Sam Lightstone}, \bibinfo{person}{Shaorong Liu}, \bibinfo{person}{Guy~M Lohman}, {et~al\mbox{.}}} \bibinfo{year}{2013}\natexlab{}.
\newblock \showarticletitle{DB2 with BLU acceleration: So much more than just a column store}.
\newblock \bibinfo{journal}{\emph{Proceedings of the VLDB Endowment}} \bibinfo{volume}{6}, \bibinfo{number}{11} (\bibinfo{year}{2013}), \bibinfo{pages}{1080--1091}.
\newblock


\bibitem[\protect\citeauthoryear{Raman and Swart}{Raman and Swart}{2006}]%
        {raman2006wring}
\bibfield{author}{\bibinfo{person}{Vijayshankar Raman} {and} \bibinfo{person}{Garret Swart}.} \bibinfo{year}{2006}\natexlab{}.
\newblock \showarticletitle{How to wring a table dry: Entropy compression of relations and querying of compressed relations}. In \bibinfo{booktitle}{\emph{Proceedings of the 32nd international conference on Very large data bases}}. \bibinfo{pages}{858--869}.
\newblock


\bibitem[\protect\citeauthoryear{Schlegel, Gemulla, and Lehner}{Schlegel et~al\mbox{.}}{2010}]%
        {schlegel2010fast}
\bibfield{author}{\bibinfo{person}{Benjamin Schlegel}, \bibinfo{person}{Rainer Gemulla}, {and} \bibinfo{person}{Wolfgang Lehner}.} \bibinfo{year}{2010}\natexlab{}.
\newblock \showarticletitle{Fast integer compression using SIMD instructions}. In \bibinfo{booktitle}{\emph{Proceedings of the Sixth International Workshop on Data Management on New Hardware}}. \bibinfo{pages}{34--40}.
\newblock


\bibitem[\protect\citeauthoryear{Seidel}{Seidel}{1991}]%
        {seidel1991small}
\bibfield{author}{\bibinfo{person}{Raimund Seidel}.} \bibinfo{year}{1991}\natexlab{}.
\newblock \showarticletitle{Small-dimensional linear programming and convex hulls made easy}.
\newblock \bibinfo{journal}{\emph{Discrete \& Computational Geometry}}  \bibinfo{volume}{6} (\bibinfo{year}{1991}), \bibinfo{pages}{423--434}.
\newblock


\bibitem[\protect\citeauthoryear{Shannon}{Shannon}{1948}]%
        {shannon1948}
\bibfield{author}{\bibinfo{person}{Claude~Elwood Shannon}.} \bibinfo{year}{1948}\natexlab{}.
\newblock \showarticletitle{A mathematical theory of communication}.
\newblock \bibinfo{journal}{\emph{The Bell system technical journal}} \bibinfo{volume}{27}, \bibinfo{number}{3} (\bibinfo{year}{1948}), \bibinfo{pages}{379--423}.
\newblock


\bibitem[\protect\citeauthoryear{Silvestri and Venturini}{Silvestri and Venturini}{2010}]%
        {silvestri2010vsencoding}
\bibfield{author}{\bibinfo{person}{Fabrizio Silvestri} {and} \bibinfo{person}{Rossano Venturini}.} \bibinfo{year}{2010}\natexlab{}.
\newblock \showarticletitle{Vsencoding: efficient coding and fast decoding of integer lists via dynamic programming}. In \bibinfo{booktitle}{\emph{Proceedings of the 19th ACM international conference on Information and knowledge management}}. \bibinfo{pages}{1219--1228}.
\newblock


\bibitem[\protect\citeauthoryear{Stepanov, Gangolli, Rose, Ernst, and Oberoi}{Stepanov et~al\mbox{.}}{2011}]%
        {stepanov2011simd}
\bibfield{author}{\bibinfo{person}{Alexander~A Stepanov}, \bibinfo{person}{Anil~R Gangolli}, \bibinfo{person}{Daniel~E Rose}, \bibinfo{person}{Ryan~J Ernst}, {and} \bibinfo{person}{Paramjit~S Oberoi}.} \bibinfo{year}{2011}\natexlab{}.
\newblock \showarticletitle{SIMD-based decoding of posting lists}. In \bibinfo{booktitle}{\emph{Proceedings of the 20th ACM international conference on Information and knowledge management}}. \bibinfo{pages}{317--326}.
\newblock


\bibitem[\protect\citeauthoryear{Taylor, Griffiths, Xu, and Mouzakitis}{Taylor et~al\mbox{.}}{2019}]%
        {taylor2018data}
\bibfield{author}{\bibinfo{person}{Phillip~M Taylor}, \bibinfo{person}{Nathan Griffiths}, \bibinfo{person}{Zhou Xu}, {and} \bibinfo{person}{Alexandros Mouzakitis}.} \bibinfo{year}{2019}\natexlab{}.
\newblock \showarticletitle{Data mining and compression: where to apply it and what are the effects?}. In \bibinfo{booktitle}{\emph{Proceedings of the 8th SIGKDD International Workshop on Urban Computing}}. ACM.
\newblock


\bibitem[\protect\citeauthoryear{Thiel and Heaps}{Thiel and Heaps}{1972}]%
        {thiel1972program}
\bibfield{author}{\bibinfo{person}{Larry~H Thiel} {and} \bibinfo{person}{HS Heaps}.} \bibinfo{year}{1972}\natexlab{}.
\newblock \showarticletitle{Program design for retrospective searches on large data bases}.
\newblock \bibinfo{journal}{\emph{Information Storage and Retrieval}} \bibinfo{volume}{8}, \bibinfo{number}{1} (\bibinfo{year}{1972}), \bibinfo{pages}{1--20}.
\newblock


\bibitem[\protect\citeauthoryear{Vigna}{Vigna}{2013}]%
        {vigna2013quasi}
\bibfield{author}{\bibinfo{person}{Sebastiano Vigna}.} \bibinfo{year}{2013}\natexlab{}.
\newblock \showarticletitle{Quasi-succinct indices}. In \bibinfo{booktitle}{\emph{Proceedings of the sixth ACM international conference on Web search and data mining}}. \bibinfo{pages}{83--92}.
\newblock


\bibitem[\protect\citeauthoryear{Welton, Kimpe, Cope, Patrick, Iskra, and Ross}{Welton et~al\mbox{.}}{2011}]%
        {welton2011}
\bibfield{author}{\bibinfo{person}{Benjamin Welton}, \bibinfo{person}{Dries Kimpe}, \bibinfo{person}{Jason Cope}, \bibinfo{person}{Christina~M Patrick}, \bibinfo{person}{Kamil Iskra}, {and} \bibinfo{person}{Robert Ross}.} \bibinfo{year}{2011}\natexlab{}.
\newblock \showarticletitle{Improving i/o forwarding throughput with data compression}. In \bibinfo{booktitle}{\emph{2011 IEEE International Conference on Cluster Computing}}. IEEE, \bibinfo{pages}{438--445}.
\newblock


\bibitem[\protect\citeauthoryear{Williams and Zobel}{Williams and Zobel}{1999}]%
        {williams1999compressing}
\bibfield{author}{\bibinfo{person}{Hugh~E Williams} {and} \bibinfo{person}{Justin Zobel}.} \bibinfo{year}{1999}\natexlab{}.
\newblock \showarticletitle{Compressing integers for fast file access}.
\newblock \bibinfo{journal}{\emph{Comput. J.}} \bibinfo{volume}{42}, \bibinfo{number}{3} (\bibinfo{year}{1999}), \bibinfo{pages}{193--201}.
\newblock


\bibitem[\protect\citeauthoryear{Witten, Neal, and Cleary}{Witten et~al\mbox{.}}{1987}]%
        {witten1987arithmetic}
\bibfield{author}{\bibinfo{person}{Ian~H Witten}, \bibinfo{person}{Radford~M Neal}, {and} \bibinfo{person}{John~G Cleary}.} \bibinfo{year}{1987}\natexlab{}.
\newblock \showarticletitle{Arithmetic coding for data compression}.
\newblock \bibinfo{journal}{\emph{Commun. ACM}} \bibinfo{volume}{30}, \bibinfo{number}{6} (\bibinfo{year}{1987}), \bibinfo{pages}{520--540}.
\newblock


\bibitem[\protect\citeauthoryear{Wongkham, Lu, Liu, Zhong, Lo, and Wang}{Wongkham et~al\mbox{.}}{2022}]%
        {wongkham2022updatable}
\bibfield{author}{\bibinfo{person}{Chaichon Wongkham}, \bibinfo{person}{Baotong Lu}, \bibinfo{person}{Chris Liu}, \bibinfo{person}{Zhicong Zhong}, \bibinfo{person}{Eric Lo}, {and} \bibinfo{person}{Tianzheng Wang}.} \bibinfo{year}{2022}\natexlab{}.
\newblock \showarticletitle{Are updatable learned indexes ready?}
\newblock \bibinfo{journal}{\emph{Proceedings of the VLDB Endowment}} \bibinfo{volume}{15}, \bibinfo{number}{11} (\bibinfo{year}{2022}), \bibinfo{pages}{3004--3017}.
\newblock


\bibitem[\protect\citeauthoryear{Xie, Qing, Zhang, Xiangliang, Zhou, Xiaofang, Deng, Ke, Pang, and Chaoyi}{Xie et~al\mbox{.}}{2014}]%
        {Xie2014Maximum}
\bibfield{author}{\bibinfo{person}{Xie}, \bibinfo{person}{Qing}, \bibinfo{person}{Zhang}, \bibinfo{person}{Xiangliang}, \bibinfo{person}{Zhou}, \bibinfo{person}{Xiaofang}, \bibinfo{person}{Deng}, \bibinfo{person}{Ke}, \bibinfo{person}{Pang}, {and} \bibinfo{person}{Chaoyi}.} \bibinfo{year}{2014}\natexlab{}.
\newblock \showarticletitle{Maximum error-bounded Piecewise Linear Representation for online stream approximation}.
\newblock \bibinfo{journal}{\emph{VLDB journal: The international journal of very large data bases}} (\bibinfo{year}{2014}).
\newblock


\bibitem[\protect\citeauthoryear{Xu, Siyamwala, Ghosh, Suri, Awasthi, Guz, Shayesteh, and Balakrishnan}{Xu et~al\mbox{.}}{2015}]%
        {xu2015performance}
\bibfield{author}{\bibinfo{person}{Qiumin Xu}, \bibinfo{person}{Huzefa Siyamwala}, \bibinfo{person}{Mrinmoy Ghosh}, \bibinfo{person}{Tameesh Suri}, \bibinfo{person}{Manu Awasthi}, \bibinfo{person}{Zvika Guz}, \bibinfo{person}{Anahita Shayesteh}, {and} \bibinfo{person}{Vijay Balakrishnan}.} \bibinfo{year}{2015}\natexlab{}.
\newblock \showarticletitle{Performance analysis of NVMe SSDs and their implication on real world databases}. In \bibinfo{booktitle}{\emph{Proceedings of the 8th ACM International Systems and Storage Conference}}. \bibinfo{pages}{1--11}.
\newblock


\bibitem[\protect\citeauthoryear{Xuejun, Dingyi, and Xiaojiang}{Xuejun et~al\mbox{.}}{2011}]%
        {DFR}
\bibfield{author}{\bibinfo{person}{Ren Xuejun}, \bibinfo{person}{Fang Dingyi}, {and} \bibinfo{person}{Chen Xiaojiang}.} \bibinfo{year}{2011}\natexlab{}.
\newblock \showarticletitle{A Difference Fitting Residuals algorithm for lossless data compression in wireless sensor nodes}. In \bibinfo{booktitle}{\emph{2011 IEEE 3rd International Conference on Communication Software and Networks}}. \bibinfo{pages}{481--485}.
\newblock
\urldef\tempurl%
\url{https://doi.org/10.1109/ICCSN.2011.6013638}
\showDOI{\tempurl}


\bibitem[\protect\citeauthoryear{Xuejun and Zhongyuan}{Xuejun and Zhongyuan}{2018}]%
        {LFR}
\bibfield{author}{\bibinfo{person}{Ren Xuejun} {and} \bibinfo{person}{Ren Zhongyuan}.} \bibinfo{year}{2018}\natexlab{}.
\newblock \showarticletitle{A Sensor Node Lossless Compression Algorithm Based on Linear Fitting Residuals Coding}. In \bibinfo{booktitle}{\emph{Proceedings of the 10th International Conference on Computer Modeling and Simulation}} \emph{(\bibinfo{series}{ICCMS '18})}. \bibinfo{publisher}{Association for Computing Machinery}, \bibinfo{address}{New York, NY, USA}, \bibinfo{pages}{62–66}.
\newblock
\showISBNx{9781450363396}
\urldef\tempurl%
\url{https://doi.org/10.1145/3177457.3177482}
\showDOI{\tempurl}


\bibitem[\protect\citeauthoryear{Yan, Ding, and Suel}{Yan et~al\mbox{.}}{2009}]%
        {yan2009inverted}
\bibfield{author}{\bibinfo{person}{Hao Yan}, \bibinfo{person}{Shuai Ding}, {and} \bibinfo{person}{Torsten Suel}.} \bibinfo{year}{2009}\natexlab{}.
\newblock \showarticletitle{Inverted index compression and query processing with optimized document ordering}. In \bibinfo{booktitle}{\emph{Proceedings of the 18th international conference on World wide web}}. \bibinfo{pages}{401--410}.
\newblock


\bibitem[\protect\citeauthoryear{Zhang, Lim, Leis, Andersen, Kaminsky, Keeton, and Pavlo}{Zhang et~al\mbox{.}}{2018}]%
        {zhang2018surf}
\bibfield{author}{\bibinfo{person}{Huanchen Zhang}, \bibinfo{person}{Hyeontaek Lim}, \bibinfo{person}{Viktor Leis}, \bibinfo{person}{David~G Andersen}, \bibinfo{person}{Michael Kaminsky}, \bibinfo{person}{Kimberly Keeton}, {and} \bibinfo{person}{Andrew Pavlo}.} \bibinfo{year}{2018}\natexlab{}.
\newblock \showarticletitle{Surf: Practical range query filtering with fast succinct tries}. In \bibinfo{booktitle}{\emph{Proceedings of the 2018 International Conference on Management of Data}}. \bibinfo{pages}{323--336}.
\newblock


\bibitem[\protect\citeauthoryear{Zhang, Liu, Andersen, Kaminsky, Keeton, and Pavlo}{Zhang et~al\mbox{.}}{2020}]%
        {zhang2020order}
\bibfield{author}{\bibinfo{person}{Huanchen Zhang}, \bibinfo{person}{Xiaoxuan Liu}, \bibinfo{person}{David~G Andersen}, \bibinfo{person}{Michael Kaminsky}, \bibinfo{person}{Kimberly Keeton}, {and} \bibinfo{person}{Andrew Pavlo}.} \bibinfo{year}{2020}\natexlab{}.
\newblock \showarticletitle{Order-preserving key compression for in-memory search trees}. In \bibinfo{booktitle}{\emph{Proceedings of the 2020 ACM SIGMOD International Conference on Management of Data}}. \bibinfo{pages}{1601--1615}.
\newblock


\bibitem[\protect\citeauthoryear{Zhang and Gao}{Zhang and Gao}{2022}]%
        {CARMI}
\bibfield{author}{\bibinfo{person}{Jiaoyi Zhang} {and} \bibinfo{person}{Yihan Gao}.} \bibinfo{year}{2022}\natexlab{}.
\newblock \showarticletitle{CARMI: A Cache-Aware Learned Index with a Cost-based Construction Algorithm}.
\newblock \bibinfo{journal}{\emph{Proceedings of the VLDB Endowment}} \bibinfo{volume}{15}, \bibinfo{number}{11} (\bibinfo{year}{2022}), \bibinfo{pages}{2679 -- 2691}.
\newblock


\bibitem[\protect\citeauthoryear{Ziv and Lempel}{Ziv and Lempel}{1977}]%
        {lz77}
\bibfield{author}{\bibinfo{person}{J. Ziv} {and} \bibinfo{person}{A. Lempel}.} \bibinfo{year}{1977}\natexlab{}.
\newblock \showarticletitle{A universal algorithm for data compression}.
\newblock \bibinfo{journal}{\emph{IEEE Transactions on Information Theory}} \bibinfo{volume}{23}, \bibinfo{number}{3} (\bibinfo{year}{1977}), \bibinfo{pages}{337--343}.
\newblock


\bibitem[\protect\citeauthoryear{Zukowski, Heman, Nes, and Boncz}{Zukowski et~al\mbox{.}}{2006}]%
        {zukowski2006super}
\bibfield{author}{\bibinfo{person}{Marcin Zukowski}, \bibinfo{person}{Sandor Heman}, \bibinfo{person}{Niels Nes}, {and} \bibinfo{person}{Peter Boncz}.} \bibinfo{year}{2006}\natexlab{}.
\newblock \showarticletitle{Super-scalar RAM-CPU cache compression}. In \bibinfo{booktitle}{\emph{22nd International Conference on Data Engineering (ICDE'06)}}. IEEE, \bibinfo{pages}{59--59}.
\newblock


\bibitem[\protect\citeauthoryear{Zukowski, Van~de Wiel, and Boncz}{Zukowski et~al\mbox{.}}{2012}]%
        {zukowski2012vectorwise}
\bibfield{author}{\bibinfo{person}{Marcin Zukowski}, \bibinfo{person}{Mark Van~de Wiel}, {and} \bibinfo{person}{Peter Boncz}.} \bibinfo{year}{2012}\natexlab{}.
\newblock \showarticletitle{Vectorwise: A vectorized analytical DBMS}. In \bibinfo{booktitle}{\emph{2012 IEEE 28th International Conference on Data Engineering}}. IEEE, \bibinfo{pages}{1349--1350}.
\newblock


\end{thebibliography}

\end{document}